\documentclass[12pt,american]{article}
\usepackage{lmodern}

\usepackage{babel}
\usepackage{soul}
\usepackage{xcolor}
\usepackage{geometry}
\usepackage{color}
\usepackage{titlesec}
\titleformat{\section}{\large\bfseries}{\thesection}{1em}{}
\usepackage{verbatim}
\usepackage{amsmath}
\usepackage{array}
\usepackage{multirow}
\usepackage{caption}
\usepackage{subcaption}
\usepackage{graphicx}
\usepackage{setspace}
\usepackage{esint}
\usepackage{hhline}
\usepackage[colorlinks=true,citecolor=gray,linkcolor=black,urlcolor=gray]{hyperref}
\usepackage{dcolumn}
\usepackage{float}
\usepackage[normalem]{ulem}
\usepackage{natbib}

\usepackage[title]{appendix}

\usepackage{todonotes}

\makeatletter
\usepackage[utf8]{inputenc}

\begin{document}

\title{Fires and Local Labor Markets\thanks{We would like to thank David Childers, Serge Coulombe, Jacob Gellman, Obie Porteous, Brigitte Roth Tran, Nikodem Szumilo,  Margaret Walls, Matthew Wibbenmeyer, and participants at the Applied Time Series Econometrics Workshop at the Federal Reserve Bank of St. Louis (2022), SEA Conference (2022), AERE Summer Conference (2023), and the 49th State Economics Workshop (2023) for their helpful comments and suggestions. We also thank Breanna Guo, Krithika Vasireddy, Sam Madden, Raeanne Smith, Masataka Mori, and Tai Nguyen for their excellent research assistance.}
}

\author{Raphaelle G. Coulombe\footnote{Department of Economics, Middlebury College. E-mail: rgauvincoulombe@middlebury.edu} \and Akhil Rao\footnote{Department of Economics, Middlebury College. E-mail: akhilr@middlebury.edu. Corresponding author.}}

\date{\today}

\maketitle
\thispagestyle{empty}

\begin{abstract}
\begin{doublespace}
We study the dynamic effects of fires on county labor markets in the US using a novel geophysical measure of fire exposure based on satellite imagery. We find increased fire exposure causes lower employment growth in the short and medium run, with medium-run effects being linked to migration. We also document heterogeneous effects across counties by education and industrial concentration levels, states of the business cycle, and fire size. By overcoming challenges in measuring fire impacts, we identify vulnerable places and economic states, offering guidance on tailoring relief efforts and contributing to a broader understanding of natural disasters' economic impacts.


\end{doublespace}

\end{abstract}
\begin{onehalfspace}
\addtocounter{page}{-1}

\thispagestyle{empty} 
\end{onehalfspace}

\newpage{}

\section{Introduction}

On April 6, 2022, forest managers in the Santa Fe National Forest initiated a controlled burn.\footnote{Such burns are a vital part of wildfire prevention and forest maintenance \citep{wagle1979controlled}.} Unfortunately, seasonal winds that afternoon caused the fire to spread. The once-controlled burn quickly became the largest wildfire in the contiguous United States, not contained until August 21, 2022 \citep{davis2022, reed2022}. People were evacuated; property was destroyed; and, though no one died, smoke from the fire likely triggered underlying health issues in surrounding and far-off areas.\footnote{In addition to its other demerits, smoke does not respect political boundaries \citep{langmann2009vegetation, burke2023smoke}. Fires burning in Quebec and western Canada recently caused people in New York---typically insulated from fire activity---to briefly experience the worst air quality on the planet \citep{bilefsky2023wildfires,
hauser2023wildfires}.} Over April-July 2022, San Miguel County saw a 9\% reduction in employment; employment in Taos County took eight months to return to its pre-fire level. As fire risk increases due to climate change and the expansion of wildlife-urban interfaces, understanding the economic impacts of fires---of all sizes---is increasingly important.\footnote{See \citet{westerling2006,jolly2015,coronese2019,bhatia2019,vecchi2021changes} on the connections between climate change and fires, or \citet{radeloff2018rapid, schug2023global} on the connections between expanding wildlife-urban interfaces and fire risk.} How have fires tended to affect economic activity in US counties? Are they short, sharp shocks with quick recoveries, or slower, more drawn-out affairs? Are their effects worse in particular places or times?

In this paper we analyze the dynamic effects of fires on local labor markets in the US. We construct a novel geophysical measure of fire exposure at the county-month level using hourly satellite imagery for the entire US over Nov 2000-May 2022. This measure allows us to identify the monthly economic effects of marginal increases in burn area. We use local projections (LPs) \citep{jorda2005} to estimate the time path of these effects over the 3 years following a fire, exploring heterogeneity across counties, states of the business cycle, and fire size. A key advantage of the LP method is its flexibility in allowing for state dependence in the estimated impulse responses and persistence in the shock.

We find that greater fire exposure decreases county employment growth in both the short run (1-7 months after fire) and the medium run (2-3 years after fire). While the short-run effects likely reflect labor supply and demand responses, we find the medium-run effects are consistent with estimated increases in net out-migration. The effects of a 13 km$^2$ fire impulse---the mean monthly burn area in counties that experienced fires---accumulate to 0.26 percentage points of lost monthly employment growth over the subsequent 3 years. For context, the average county-level monthly employment growth rate is 0.03\%, implying the cumulative 3-year effect of a fire impulse is on the order of 25\% of employment growth over that horizon. We document significant heterogeneity in employment growth responses across county-level labor market characteristics, state of the business cycle, and fire size.

There are three challenges in studying the effects of fires on employment growth. First, fire activity estimates based on disaster declarations may not accurately reflect the event's timing and intensity, an issue we detail in section \ref{sec:data}. Second, while fires may indirectly affect national economic metrics via trade and migration, their direct impacts are typically localized, reducing their visibility in national or regional economic statistics. Third, fires may only last days or weeks; in some exceptional cases they may last months. While some disasters can have long-lasting effects visible in quarterly, annual, or decadal data \citep{long2018, boustan2020}, due to their brevity the impacts of individual fires may not be visible over such long intervals. To overcome these challenges, we combine satellite imagery from the National Aeronautics and Space Administration (NASA) of the universe of fires affecting the US since November of 2001, compiled and hosted by the National Oceanic and Atmospheric Administration (NOAA) and US Geological Survey (USGS) with monthly employment data from the Bureau of Labor Statistics.

Recent analyses of the effects of fires on labor market outcomes in the US have found mixed effects across time and space. For instance, \citet{nielsen2013effects} report that large wildfires can temporarily boost local wage growth and employment, but also fuel economic volatility. \citet{walls2023econwildfires} and \citet{meier2023eurofire} offer complementary perspectives on the economic effects of wildfires from different settings. \citet{walls2023econwildfires} focus on Western US labor markets at both the county and establishment levels and identify an immediate increase in employment, specifically in construction, which tapers off in the long run. In contrast, \citet{meier2023eurofire} focus on Southern Europe and identify a decline in annual GDP growth in regions affected by wildfires, suggestive of a longer-lasting economic effect. \citet{meier2023eurofire} also document a sectoral shift in employment, with tourism-related jobs declining and jobs in construction and real estate-related services increasing. Considering wildfires and natural disasters more generally, \citet{tran2021} document that wildfires in the US initially generate declines in personal income and wages followed by eventual increases, albeit with longer-run negative impacts at the state level. Returning to the Western US, \citet{borgschulte2022air} find that smoke exposure can reduce earnings and labor force participation, and increase Social Security claims. This mechanism may explain some of the short-run negative effects found in other studies.

Our analysis contributes to the growing literature on fires and economic activity, as well as the broader literature on the economic impacts of natural disasters, in three ways. First, we study fires using geophysical measures of fire exposure. Both official disaster declarations and dollar value measures may be endogenous to economic conditions and not synced with actual fire timings, an issue which satellite measurements of burn area avoid.\footnote{For instance, dollar-damage figures might be higher in counties better equipped economically for fire recovery, skewing impulse response estimates towards quicker recoveries. Also, two identical fires with the same employment effects might be recorded with varying intensities based on the relative wealth of the areas.} Our use of geophysical measures of disaster activity mirrors \citet{walls2023econwildfires} and \citet{meier2023eurofire}'s analyses of fires and \citet{strobl2011} and \citet{felbermayr2014growthnd}'s analyses of hurricanes and other disasters. Our estimates also provide a tool to decompose the impacts of past fire exposure on labor markets. We execute such a decomposition to gauge the historical fluctuations in regional employment due to fire exposure during the sample period.

Second, while there is much evidence regarding the long-term growth effects of natural disasters broadly (e.g. \citet{noy2009, cavallo2013, lackner2018, boustan2020}), less is known about the impacts of fires, especially at the county-month level. Our paper parallels the county-level analysis by \citet{borgschulte2022air}, who use high-frequency geophysical measures of wind flows and smoke exposure to study employment impacts. We incorporate similar measures of fire-driven smoke exposure along with spatial lags of fire exposure to flexibly control for potential spatial spillovers due to fires in other regions. To our knowledge this study is the first to use geophysical measures of fire \emph{and} smoke exposure at the county-month level for the entire US.

Third, the national scale of our dataset enables identification of heterogeneous impacts of fires across county characteristics such as average education levels, industrial concentration, and slack states of the local labor market. We find that counties with high industrial concentration or lower average educational attainment are particularly vulnerable to fires. We also find that the labor market effects of fires are amplified during high-slack periods. These findings can guide policymakers in tailoring relief efforts more effectively, reducing the total cost of fires.

The paper is organized as follows. In section \ref{sec:mech} we describe potential mechanisms through which fires may affect local employment, motivating our empirical analysis. In section \ref{sec:data} we describe our data, present summary statistics, and highlight some issues in measuring economic impacts of fires without geophysical measures. In section \ref{sec:empirical-framework} we present our empirical framework and identification strategy. We show our main results in Section \ref{sec:results}, including robustness checks against spatial spillovers. We conclude in Section \ref{sec:conclusion}. Additional summary statistics and robustness checks are shown in the Appendix.

\section{Why and how fires may affect local employment}
\label{sec:mech}

Fires---particularly wildfires, but also prescribed or controlled burns---can affect both demand and supply sides of a local labor market. Fires may also have different effects over time, as different mechanisms become operational \citep{nielsen2013effects}.

On the labor supply side, local barriers can impede workers from reaching their places of work in the immediate aftermath of a fire. This is especially true in cases where a fire results in road closures, or where health issues triggered by smoke exposure prevent individuals from working. Labor force characteristics also play a role in modulating the effects of fires. For example, regions with a higher share of white-collar employment may see less disruption if these workers are able to switch to remote work \citep{yasenov2020can}. Over longer horizons, households may migrate in response to changing fire risk perceptions. Here, too, responses may vary by labor force characteristics. For example, workers with lower educational attainment may be less likely to migrate in response to changing labor market conditions \citep{wozniak2010college, balgova2018don}.

On the labor demand side, a fire can dampen local labor demand due to a reduction in consumer demand and the destruction of physical capital. This is particularly true for regions heavily reliant on tourism, where fires can deter visitors \citep{otrachshenko2022fire}, further impacting hospitality and other service industries. Fires are also a source of unsystematic risk affecting specific industries like services and agriculture, so regional diversification within a region might increase its resilience against fires---an implication of portfolio theory. Outside the context of fires, the regional economic literature has found evidence that regional industrial diversification supports employment stability (e.g. \citet{Barth1975, Nourse1968, Sherwood1990}, and \citet{Dissart2003}), and that more-diversified economies are more robust to economic shocks (e.g. \citet{kluge2018sectoral, coulson2020economic}). The industrial concentration of a region is therefore likely to be a key determinant of the extent of fire's impact on labor demand. 

On the other side, financial aid and reconstruction efforts in the wake of large or unexpected fires can boost local employment, offsetting or reversing some of the negative effects. The time horizon at which such positive forces operate is unclear: while funds may not be disbursed immediately, forward-looking agents without binding credit constraints may be able to borrow against the expectation of relief funds. Further, all of these effects may be modulated by the broader state of the economy. In a state of high labor market slack, for instance, where the number of unemployed people per job opening is high, the negative economic effects of a fire could be magnified as the fire exacerbates existing underemployment or unemployment. Similarly, more businesses might be on the brink of bankruptcy, again amplifying the negative effects. The fire and resulting disruption can therefore add additional burdens to an already-struggling labor market. In a low slack environment---closer to full employment---the economy may exhibit more resilience to such shocks.

Given these features, the overall relationship between fires and economic activity is likely to be heterogeneous across counties and time. Labor market characteristics, the business cycle, the size of the shock, and time since the shock are all likely to play a role. We therefore explore these factors systematically in what follows.

\section{Data}
\label{sec:data}
\paragraph{Fire exposure measurements.}

We obtain fire exposure data from the NASA Earthdata LPDAAC MCD64A1 product, hosted by USGS \citep{usgs_fires}. The dataset divides the Earth into a set of tiles which are further split into 500 m$^2$ grid cells. Each cell contains information about burn status when sampled, the date of the detected fire, and burn measurement uncertainty. The underlying data comes from the MODIS satellite products, hosted by NOAA, and contains hourly 1 km$^2$-pixel measurements of fire activity captured by the MODIS satellite \citep{noaa_fires}. The LPDAAC MCD64A1 data product identifies burn status within 500 m$^2$ grid cells by using information on cell characteristics such as burn-sensitive vegetation and reflectance \citep{giglio2018collection}. We obtain the latitude-longitude coordinates of each grid cell using the inverse mapping described in \citep{giglio2020collection}, then link each coordinate to a county FIPS code using the FCC's Census Block Conversions API \citep{fcc_api}, and finally aggregate from hourly to monthly frequency. The dataset covers the entire US over the period from November of 2000 to May of 2022.

Table \ref{tab:burn_summary} shows summary statistics of burn areas in km$^2$ for the US as a whole and within Census regions over the Nov 2000-May 2022 period. Most counties do not experience fires in most months, but those which do experience an average monthly burn area of roughly 13.1 km$^2$. This varies widely across regions which experience fires: counties in the West observe roughly 26.3 km$^2$ of burn area per month on average, while counties in the Northeast see burn areas of roughly 2.2 km$^2$ per month. Even among these counties however fire exposure is highly unequal: the median exposure is 2 km$^2$ but the maximum is 7,069 km$^2$. 
  
\begin{table}[!htbp]
\begin{center}
\caption{Summary statistics for county-level burn areas in km$^2$ per month over Nov 2000-May 2022}
\label{tab:burn_summary}
\begin{tabular}{|l|c|c|c|}
\hline
 & Mean & Mean $|$ burn & Obs\\
 \hline \hline
US & 1.0 & 13.1 & 824,589\\
\hline
Midwest & 0.5 & 10.8 & 268,419\\
Northeast & 0 & 2.2 & 55,120\\
South & 0.8 & 7.9 & 362,719\\
West & 4.0 & 26.3 & 115,217\\
 \hline
\end{tabular}    
\end{center}
\raggedright
\begin{singlespace}
\footnotesize{The first column shows the unconditional mean and the second column shows the mean for counties which have fires. The final column shows the unconditional sample size for each region.} \end{singlespace}
\end{table}

\begin{figure}[!h] 
\caption{Total burned area in US counties over time} \label{fig:fire_maps}
\centering
\begin{subfigure}[b]{0.49\textwidth}
\includegraphics[height=4.5cm,width=\textwidth]{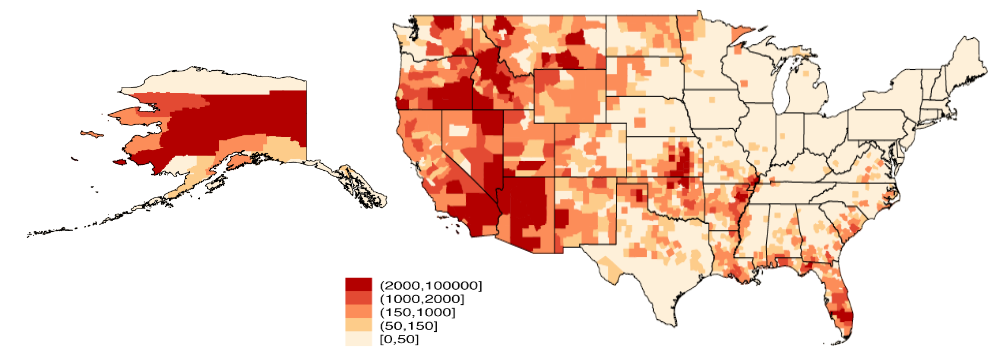}  
\caption{2002--2006}
\end{subfigure}
\begin{subfigure}[b]{0.49\textwidth}
\includegraphics[height=4.5cm,width=\textwidth]{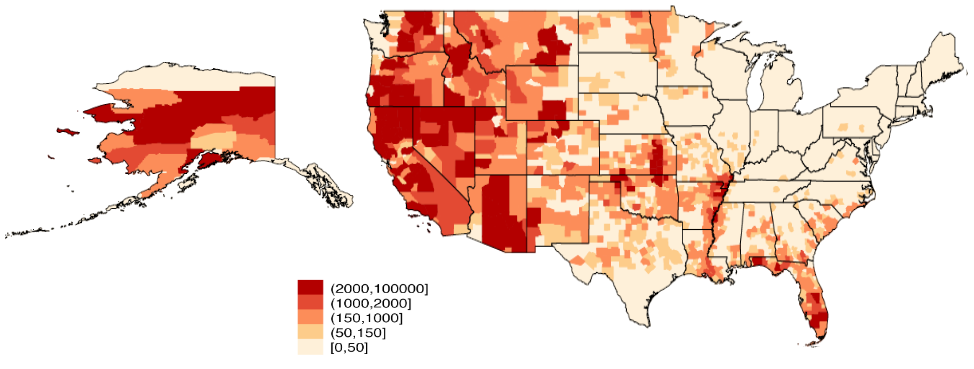}  
\caption{2017--2021}
\end{subfigure}
\raggedright
\begin{singlespace}
    \footnotesize{Darker colors indicate larger burned areas within counties.}
\end{singlespace}
\end{figure}

Figure \ref{fig:fire_maps} shows changes in total burn area in US counties over the sample period. Panels (a) and (b) show that fires are generally concentrated in the West and South. Still, more counties experienced fire activity in 2017-2021 than in 2002-2006, particularly in the Midwest, South, and along the Canadian border.  

\begin{figure}[!h] 
\caption{Total burned area for selected fires and FEMA declarations} \label{fig:burnfema}
\centering
\includegraphics[height=7cm,width=\textwidth]{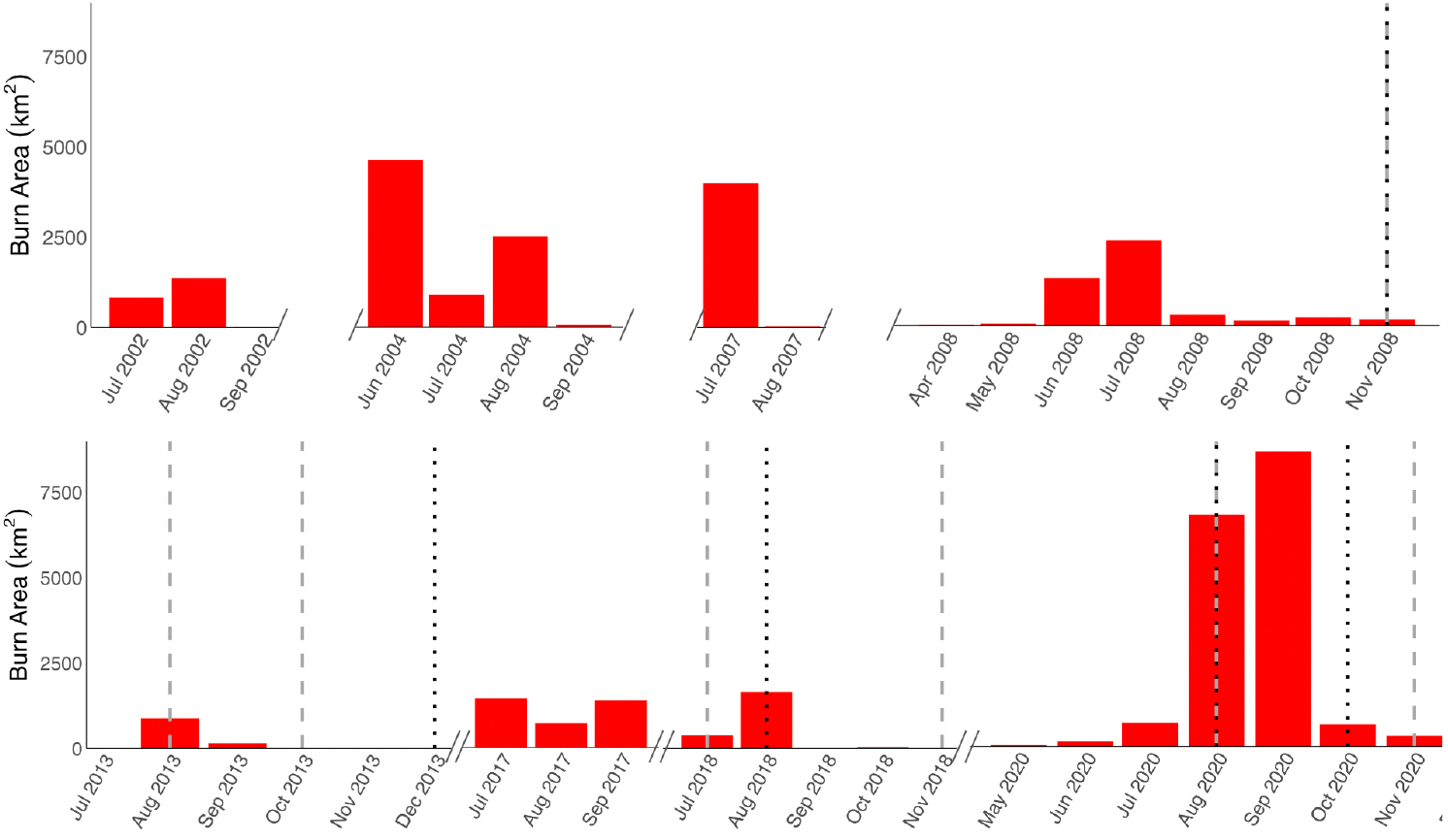}
\raggedright
\begin{singlespace}
    \footnotesize{The fires represented are the Biscuit Fire in Oregon (Jul 2002-Nov 2002), the Taylor Complex Fire in Alaska (Jun 2004-Sep 2004), the Murphy Complex Fire in Nevada and Idaho (Jul 2007-Aug 2007), the 2008 California Wildfires (Apr-Nov), the Rim Fire in California (Aug 2013-Sep 2013), the 2017 Montana Wildfires (Jun-Sep), the Mendocino Complex Fire in California (Jul 2018-Dec 2018), and the 2020 California Wildfires (Feb 2020-Jan 2021). The grey dashed lines represent the months FEMA records as the major disaster period, while the black dotted line is the date that FEMA declared the major disaster.}
\end{singlespace}
\end{figure}

Our burn area measure has a low correlation (0.03) at the county-month level with FEMA fire declarations that received a ``Major Disaster'' status. This is partly driven by the inclusion of small fires in our data and partly by the difference between FEMA declaration and actual burn timing. While our measure captures 84\% of FEMA-declared county-months (i.e. 84\% of county-months that FEMA declares a fire are also recorded as having fires in our data), only 67\% of FEMA-declared county-months occur at the time of the fire as detected by satellite.\footnote{The FEMA fire declarations also include events that should not be included in our analysis. 9/11 is one notable example where our fire exposure measure shows zero burn area while FEMA declared a fire-related emergency in most New York counties.} Consequently, FEMA declarations cover only 8\% of the largest 1\% of county-month burn areas. Figure \ref{fig:burnfema} shows total burn area for eight of the largest fires in our dataset.\footnote{These eight were selected because they were in the top 1\% of fires by area and had specific names, indicative of their significance.} The gray dashed lines represent the final range of dates FEMA recorded as ``Major Disaster'', while the black dotted line is the date that FEMA declared the major disaster. The first three fires shown burned a total of 2,220 km$^2$, 8,132 km$^2$, and 4,041 km$^2$ respectively but never received a ``Major Disaster'' status. The 2008 California Wildfires started in April 2008 and reached a peak in June and July but were only declared a ``Major Disaster'' by FEMA in November 2008. The bottom row shows more examples of large fires that were declared months after the fires were first detected. Overall, the figure highlights the large differences in both event timing classification and intensive margin measurements between our geophysical measure of burn area and FEMA declaration/event dates. 

\paragraph{Labor force statistics.} We obtain monthly data on the number of people employed at the county level from the Quarterly Census of Employment and Wages (QCEW) program available from the BLS. The number of people employed in a county is constructed from employer reports. Workers are therefore included in the county where they work---not where they live. Monthly employment data is available since January 1990 and includes both part-time and full-time workers as well as workers who are on paid vacations. We use the X-12 algorithm for seasonal adjustment as the number of employed workers shows strong seasonal patterns. While we use the full labor force statistics dataset for seasonal adjustment, we restrict our sample of labor force statistics to the period after November of 2000 to match our fire data. 

Table \ref{tab:empg_summary} shows some summary statistics of monthly employment growth for the US as a whole and within Census regions over the Nov 2000-May 2022 period. The mean value of employment growth rates nationally is 0.03\%, but with marked geographical differences: counties in the South and West with fire exposure appear to have higher (0.10\% and 0.14\%, respectively) and more widely varied employment growth. Interestingly, counties that did experience fires have higher employment growth across all regions compared to counties without fire exposure. 

\begin{table}[!htbp]
\centering
\caption{Summary statistics for county-level monthly employment growth (\%) over Nov 2000-May 2022}
\begin{tabular}{|l|c|c|c|c|c|c|}
\hline
 & Mean & Mean $|$ burn & 5th percentile & Median & 95th percentile & Obs \\
\hline
\hline
US & 0.03 & 0.09 & $-1.87$ & 0.06 & 1.93 & 824,589 \\
\hline
Midwest & 0.01 & 0.08 & -1.87 & 0.10 & 2.92 & 268,419 \\
Northeast & 0.00 & 0.05 & -1.03 & 0.03 & 1.09 & 55,120 \\
South & 0.04 & 0.10 & -1.78 & 0.08 & 1.83 & 362,719 \\
West & 0.07 & 0.14 & -2.82 & 0.10 & 2.92 & 115,217 \\
\hline
\end{tabular}
\vspace{0.2cm}
\raggedright
\begin{singlespace}
\footnotesize{The first column shows the unconditional mean, the second column shows the mean for counties which have fires, and the third, fourth, and fifth columns show the 5th, 50th, and 95th percentiles across all counties. The final column shows the unconditional sample size for each region.}\end{singlespace}
\label{tab:empg_summary}
\end{table}

\begin{figure}[htpb] 
 \caption{Fire-driven smoke-related PM$_{2.5}$ in US counties over time} \label{fig:smoke-maps}
 \centering
 \begin{subfigure}[b]{0.49\textwidth}
\includegraphics[height=5cm,width=\textwidth]{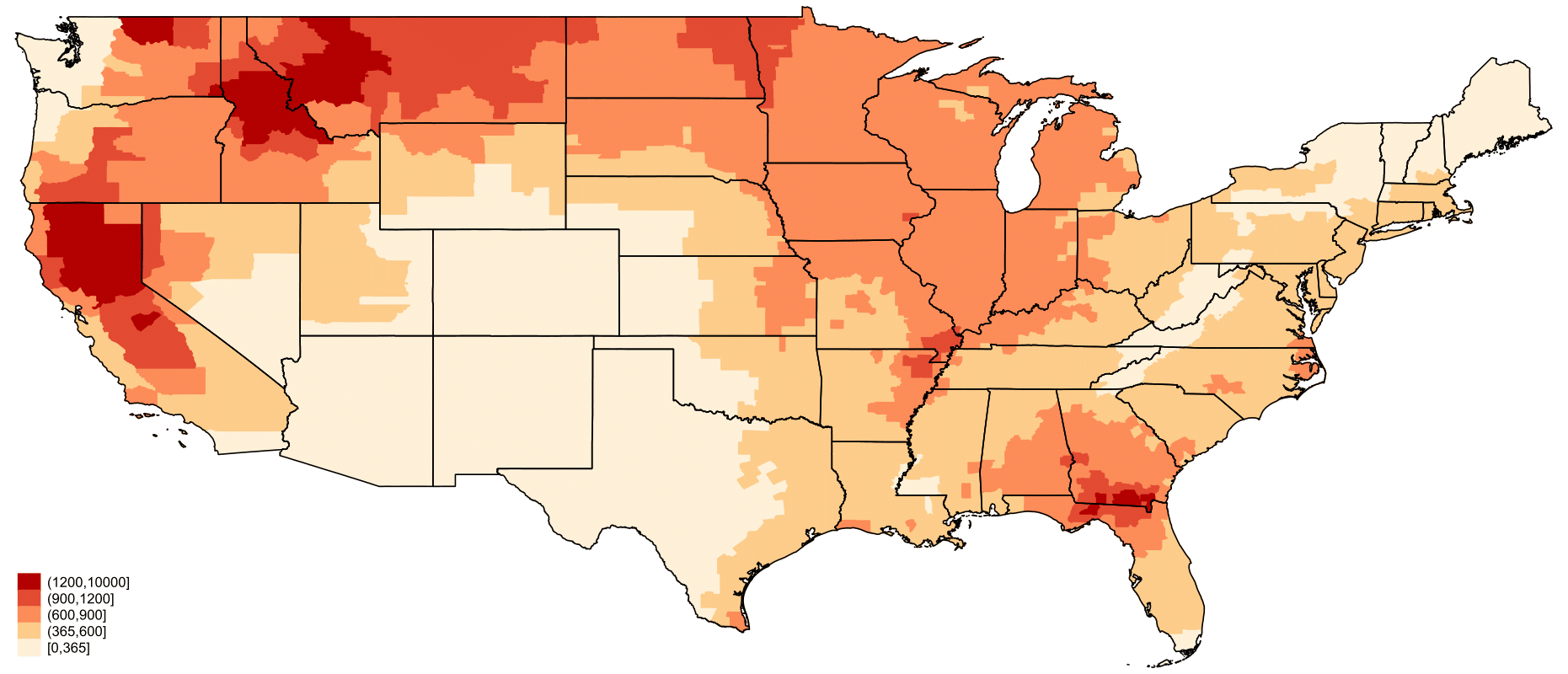}  
 \caption{2006--2010}
 \end{subfigure}
 \begin{subfigure}[b]{0.49\textwidth}
\includegraphics[height=5cm,width=\textwidth]{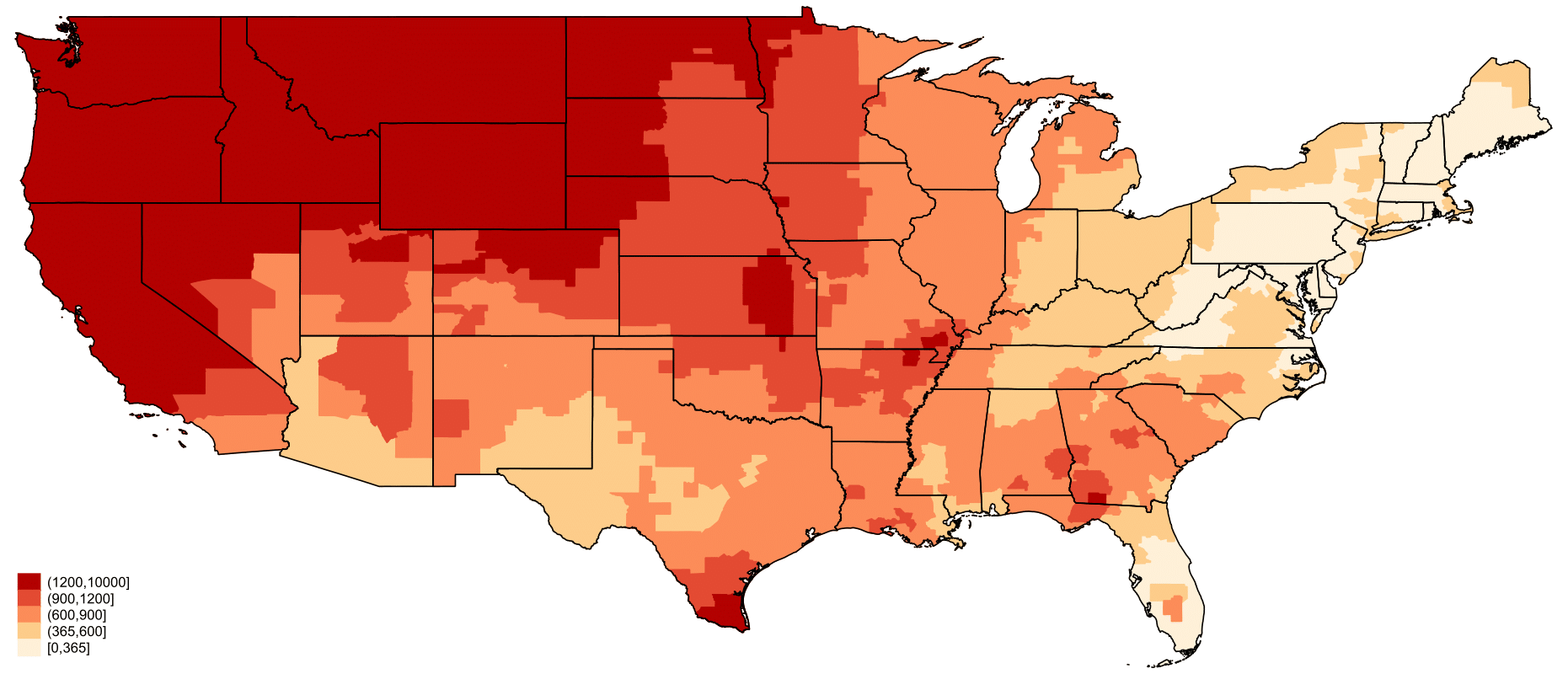}  
 \caption{2016--2020}
 \end{subfigure}
 \raggedright
 \begin{singlespace}
 \footnotesize{Darker colors indicate higher PM$_{2.5}$ levels within counties.} 
  \end{singlespace}
 \end{figure}

\paragraph{Smoke exposure measurements.} 
We use fire-related smoke exposure data from \citet{childs2022}. The dataset contains hourly estimates of smoke-related PM$_{2.5}$ over 10 km$^2$ grid cells for the contiguous US from 2006 onwards, which we aggregate to the county-month level. The dataset uses HMS smoke plume data along with particle transport and machine learning models to propagate smoke trajectories and estimate smoke exposure within grid cells. Figure \ref{fig:smoke-maps} shows changes in smoke exposure in US counties over the 2006--2020 period. Smoke exposure across the country increases over this period, even for counties that are less exposed to fire. 

Comparing the maps of fire-driven smoke and burn area reveals that while many counties are not affected by fire, many of them experience fire-driven smoke. Table \ref{tab:smoke_summary} presents summary statistics for county-level smoke-related PM$_{2.5}$. The West continues to dominate with an average of 20.3 $\mu g/m^3$ of fire-driven smoke exposure per county-month. 

\begin{table}[!htbp]
\begin{center}
\caption{Summary statistics for county-level smoke-related PM$_{2.5}$ ($\mu g/m^3$) per month over 2006-2021} \label{tab:smoke_summary}
\begin{tabular}{|l|c|c|c|}
\hline
 & Mean & Mean $|$ burn & Obs\\
 \hline \hline
US & 13.9 & 19.8 & 490,998\\
\hline
Midwest & 14.9 & 23.0 & 177,120\\
Northeast & 7.4 & 15.4 & 36,540\\
South & 11.1 & 14.4 & 212,178\\
West & 20.3 & 37.0 & 64,260\\
 \hline
\end{tabular}   
\end{center}
\raggedright
\begin{singlespace}
\footnotesize{The first column shows the unconditional mean, the second column shows the mean for counties which have fires, and the third shows the unconditional sample size for each region.} \end{singlespace}
\end{table}

\paragraph{County-to-county migration.} We obtain county-to-county migration data from the IRS Tax Stats Database. Migration patterns in this dataset are computed from year-to-year address changes as reported on individual income tax returns filed with the IRS. County-level migration data from this source is only available at an annual frequency. Our sample covers the 2003-2019 period. Since migration data is constructed from tax returns, this dataset does not include individuals who do not file a tax return and those who file late. 

Table \ref{tab:nexmpg_summary} shows some summary statistics of county-level net out-migration (out-migration minus in-migration) as a share of county population over the 2003-2019 period. The mean growth in net out-migration tends to be small but varies widely across regions. People appear to be leaving the Midwest, especially counties with fire exposure. The West has the largest annual increase in net migration despite being the most exposed to fires. 

\begin{table}[htpb]
\centering
\caption{Summary statistics for county-level annual net out-migration as a share of the population (\%) over 2003-2019} 
\begin{tabular}[t]{|l|c|c|c|c|c|c|}
 \hline
 & Mean & Mean $|$ burn & 5th percentile & Median & 95th percentile & Obs\\
 \hline
 \hline
US & -0.07 & -0.04 & -1.84 & 0.01 & 1.38 & 53,348\\
 \hline
Midwest & 0.12 & 0.28 & -1.31 & 0.15 & 1.42 & 17,901\\
Northeast & 0.12 & -0.04 & -0.75 & 0.14 & 0.89 & 3,689\\
South & -0.22 & -0.11 & -2.13 & -0.13 & 1.27 & 24,162\\
West & -0.18 & -0.22 & -0.95 & -0.10 & 1.06 & 7,596\\
 \hline
\end{tabular}
\vspace{0.2cm}
\raggedright
\begin{singlespace} \footnotesize{County-level net out-migration is calculated as the difference between a county's out-migration and in-migration as a share of the county's population in percent. The first column shows the unconditional mean, the second column shows the mean for counties which have fires, and the third, fourth, and fifth columns show the 5th, 50th, and 95th percentiles across all counties. The final column shows the unconditional sample size for each region.}\end{singlespace}
\label{tab:nexmpg_summary}
\end{table}

Our final dataset contains county-month observations with county characteristics, the number of people employed, the net out-migration rate as a fraction of the county's population the area burned in fires, and the amount of smoke exposure due to fires. 

\section{Empirical framework} 
\label{sec:empirical-framework}

We apply \citet{jorda2005}'s local projection method, modified for panel data, to estimate the dynamic response of labor market outcomes to fire activity. Let $t$ index months and $c$ index counties. Let $y_{c,t+h}$ denote the outcome of interest at period $t+h$,  $D_{c,t}$ county-level fire exposure, and $X'_{c,t}$ a vector of control variables. We estimate the following equation for a series of horizons $h\geq 0$:
\begin{equation} 
    y_{c,t+h} - y_{c,t-1} = \beta_h D_{c,t} + X'_{c,t} \gamma_h + \alpha_{c,h} + \mu_{t,h} + \epsilon_{c,t+h}. \label{eqn:base-model}
\end{equation} 
We consider two outcomes of interest: (log) employment and the net out-migration rate (as a share of county population). The dependent variable $y_{c,t+h} - y_{c,t-1}$ therefore represents the percent change in employment or the net out-migration rate between times $t-1$ and $t+h$. The slope parameter $\beta_h$ measures the change in the outcome between periods $t-1$ and $t+h$ due to an increase in fire exposure at time $t$. We rescale $\beta_h$ in all figures and tables so it has percentage point units. For example, letting $y_{c,t}$ be the natural log of employment in $c$ at $t$, a value of $\beta_3=-0.2$ means that 3 months after a fire impulse the growth rate of employment is 0.2 pp lower than a similar county which did not experience the fire impulse.  

We control for the county's fire exposure in the previous twenty four months ($\sum_{s=1}^{24} D_{c,t-s}$) as fire exposure tends to be serially correlated and affect the labor market over several months. We also include twenty four lags of the county's labor market outcome ($\sum_{s=1}^{24} y_{c,t-s}$) to account for serial correlation in employment.\footnote{We obtain very similar results when we control for twelve or thirty six lags of fire exposure and labor market outcomes.} We include horizon-specific county and year-month fixed effects, $\alpha_{c,h}$ and $\mu_{t,h}$, to control for county-specific time-invariant and common time-varying factors.\footnote{While the combination of county fixed effects and lagged dependent variable can create a Nickell bias, our results are almost identical when we exclude county fixed effects in equation \ref{eqn:base-model}. The size of our time dimension---251 periods in total---also mitigates Nickell bias.} We use Driscoll-Kraay (DK) standard errors for all point estimates to account for unmodeled spatiotemporal dependence.

\subsection{Identification}
We rely on variation in fire exposure across counties and over time to identify the causal effect of marginal fire exposure at month $t$ on a county's labor market outcome in month $t+h$, i.e. $\beta_h$.\footnote{In the language of \citet{rambachan2021common}, we are interested in the path of ``marginal impulse causal effects'' of fire activity on labor market outcomes.} The causal effect we study includes all subsequent responses to the fire impulse at time $t$, i.e. we are interested in the marginal impulse causal effect of fire on employment inclusive of aid and reconstruction, labor market adaptation, and other direct responses to the fire impulse.\footnote{This is a distinct estimand from what is identified by distributed lead/lag (DL) methods \citep{alloza2020dynamic, dube2022local}. Local projection (LP) and DL methods can be made equivalent, but by default they handle persistence (e.g. induced by fuel depletion or other fire dynamics) differently. DL models recover the marginal causal effect \emph{assuming a counterfactual shock with no persistence}. LP models recover the marginal causal effect \emph{allowing the shock to evolve as it would}. The latter is the most likely dynamic response of the system to the shock---the object of our interest in this analysis.} 

There are two key identifying assumptions. First, conditional on the control variables, fire exposure at time $t$ must be exogenous to county-level labor market outcomes at times $t, ..., t+h$. This is a ``no anticipation'' assumption. Since non-prescribed fires are generally unpredictable events---though the probability of a fire at any location-time pair can be calculated the occurrence and extent of a fire itself is random---and changing jobs or leaving the workforce is often a costly action, it seems unlikely that people adjust their employment or workforce status in anticipation of fire events over short horizons. The reverse causality channel---where people anticipating certain labor market outcomes deliberately create or prevent fires---similarly seems unlikely.\footnote{Employment at local fire departments may respond to labor market outcomes or fire activity, e.g. workers who are marginally attached to the labor force may join a volunteer fire department to assist in firefighting activities during a fire event. Such actions are part of the effect we are studying but are unlikely to be quantitatively significant. See \citet{walls2023econwildfires} for more details.} 

Second, conditional on controls, counties exposed to fires should have similar labor market trends as counties not exposed to fires---a ``parallel trends'' assumption. Our control variables and fixed effects likely hold relevant dimensions constant across counties. In particular, lagged labor market outcomes account for heterogeneous labor market trends, while lagged fire exposures account for heterogeneous fire propensity (e.g. availability of fuel following recent fires, location-specific fire seasonality). Controlling for these lags also addresses a backwards-looking Stable Unit Treatment Value Assumption (SUTVA) condition---conditional on controls, counties experiencing fires at month $t$ are unaffected by fires in the same county at months prior to $t$.

However, we still require a spatial SUTVA condition: that employment outcomes in counties other than $c$ are unaffected by fire exposure in county $c$, conditional on controls. We conduct two robustness checks to determine the extent to which such spatial spillovers affect our results. First, we control for smoke exposure; and second, we control for county-level spatial lags of fire exposure.

Smoke from fires in county $c$ may depress economic activity in counties other than $c$, contaminating the control group used to estimate the effect of fire in $c$. We control for contemporaneous and lagged smoke exposure in all counties, finding qualitatively-similar results as our baseline models without smoke. To flexibly control for regional spatial spillovers, we augment our baseline specification in equation \ref{eqn:base-model} with county-level spatial lags of fire exposure \citep{halleck2015slx}. We again find similar results, with little evidence of fire-driven spatial spillovers. We discuss these results further in section \ref{sec:spatial}. 

While our fire exposure measure allows us to study the effects of fires in unprecedented granularity, measurement error is a concern. The satellite data-generating process is informative about the types of measurement error we might expect. The MODIS satellite captures images of the Earth's surface, which are then fed into a contextual classification algorithm to determine which pixels contain fires \citep{giglio2020collection}. Though the algorithm contains processing steps to remove false positives due to factors like sun glints and water reflections, it may still miss small or low-intensity fires (incorrectly labeling such pixels as ``unburned''). Such small fires may not be representative of employment-affecting fire activity and so their exclusion may be appropriate. A related source of measurement error in our fire exposure measure comes from the discreteness of fire detection. The use of discrete measurement levels (i.e. burned/unburned at the 500 m$^2$) level induces measurement error when burned areas are not divisible by 500, e.g. a ``true'' burn area of 2600 m$^2$ will be recorded as 2500 m$^2$. Such measurement error is likely to attenuate our estimates toward zero. 

Finally, our estimation framework has similarities with difference-in-differences (DiD) frameworks in that we consider the causal impact of fires on US counties and identify these effects through comparison to counties experiencing less or no fire exposure. One may thus wonder about bias arising from ``forbidden comparisons'' (e.g. counties which have recently experienced fires being used as controls). The use of such controls has recently been shown to create negative weights in popular two-way fixed-effects (TWFE) estimators, particularly when treatment effects change over time and vary across treated units \citep{de2020two, sun2021estimating, callaway2021difference, goodman2021difference}. 

To see if negative weighting may present issues we run our analysis restricted the sample to treated counties and ``clean controls'' \citep{cengiz2019effect, dube2022local}. To account for the fact that counties may experience fires multiple times over our sample period, we construct samples with ``clean'' control groups using only counties which have no fire activity for 3 years before and after the impulse at $t$. Our results comparing counties experiencing fires to only clean controls closely match our results from the full sample (appendix \ref{apdx:robustness}, figure \ref{fig:clean_control}).

\section{Economic effects of fires}
\label{sec:results}

\begin{figure}[!h] 
\begin{center}
\caption{Response of employment growth to an increase in burn area} \label{fig:base-model-result}
\includegraphics[height=6cm,width=0.8\textwidth]{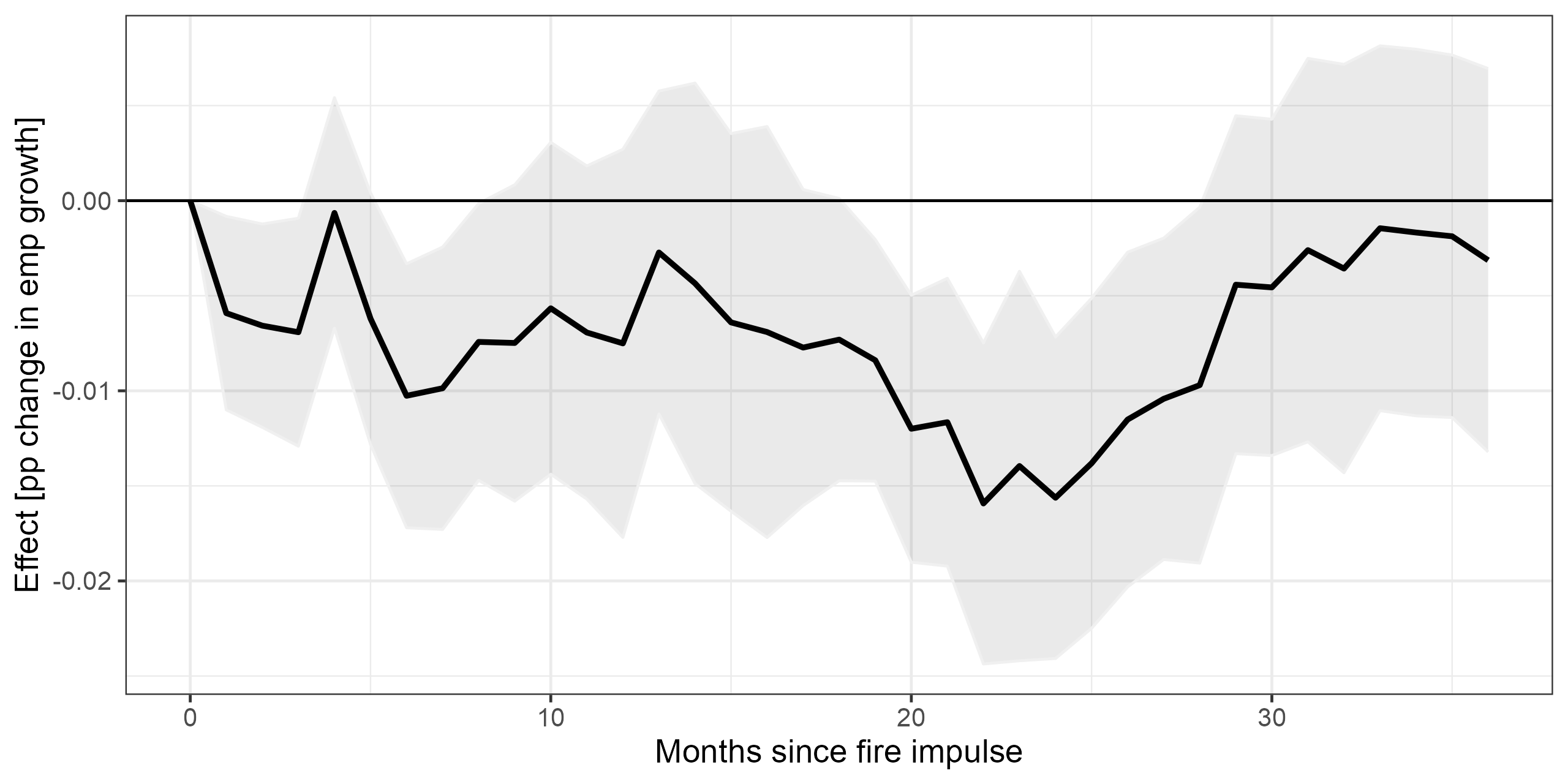}
\end{center}
\vspace{-0.4cm}
\raggedright
\begin{singlespace}
\footnotesize{The $y$ axis shows percentage point changes in employment growth in response to a burn area impulse of 13.1 km$^2$---the mean burn area in counties that experienced fires. Shaded areas indicate 95\% CIs computed using DK standard errors. Covariates include county and year-month fixed effects and 24 monthly lags of county employment and burn area.} \end{singlespace}
\end{figure}

We estimate equation \ref{eqn:base-model} with the percent change in the total number of people employed between periods $t-1$ and $t+h$ as the dependent variable (i.e. $\log(emp_{t+h}) - \log(emp_{t-1})$). The impulse response and cumulative effect over the horizon are computed from the estimated $\{\beta_h\}_0^{36}$. Figure \ref{fig:base-model-result} shows the impulse response of employment growth following a 13.1 km$^2$ impulse of additional fire exposure with 95\% confidence intervals computed using DK standard errors.\footnote{The size of fire impulse is set to 13.1 km$^2$---the average monthly burn area across counties experiencing fires---throughout the paper unless otherwise indicated.} 

The initial decline in employment growth in the month following an increase in fire activity is about 0.006 pp (or 6.7\% of the monthly employment growth), growing to about 0.015 pp (or 16.7\% of the monthly employment growth) two years after the fire occurs and fading a little around the one year mark. By three years after the fire impulse the response is centered on zero and not statistically significant. Over 3 years, this effect accumulates to 0.26 pp (standard error 0.07 pp) of monthly lost employment growth due to the fire impulse.\footnote{We describe the calculation of the cumulative effect and its standard error in Appendix \ref{apdx:cumulative-effect}} To put the effect magnitude in context, Table \ref{tab:empg_summary} shows that the average monthly employment growth rate in counties which experience fires is about 0.09\%. While the decrease in employment in the short-run (1-7 months after impulse) likely reflects labor supply and demand responses due to smoke-induced health issues, property and infrastructure damage, and associated reductions in consumer demand, the medium-run (2-3 years after impulse) decrease in employment likely reflects other factors. To investigate the medium-run effects we next turn to the impulse response of net out-migration to fire activity.

\subsection{Migration and medium-run responses}
\label{sec:migration}
While migration is often infeasible in the months immediately following a local shock, households may eventually respond to fires and similar shocks by migrating. \citet{sheldon2022mig} find that households' (particularly higher-income households') propensity to migrate following natural disasters is increasing in the amount of FEMA aid. To assess the degree to which net out-migration may help explain the observed decline in employment following an increase in fire exposure, we estimate equation \ref{eqn:base-model} with the change in the net out-migration rate as the dependent variable. 

\begin{figure}[!h]
\begin{center}
\caption{Response of net out-migration as a share of population to an increase in burn area}  \label{fig:out_mig} 
\includegraphics[height=6cm,width=0.8\textwidth]{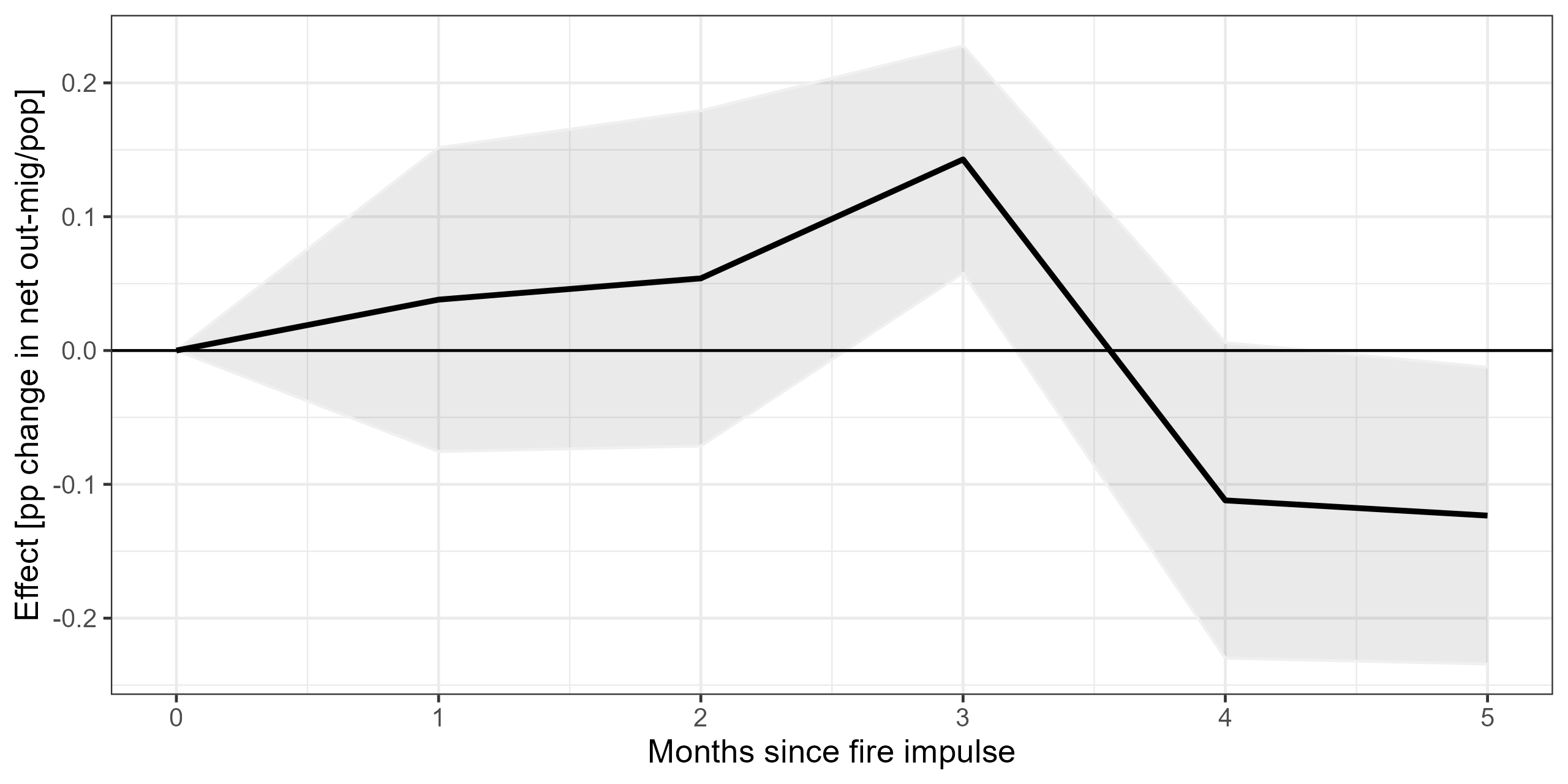}
\end{center}
\vspace{-0.4cm}
\raggedright
\begin{singlespace}
\footnotesize{The $y$ axis shows percentage point changes in net out-migration as a percentage of population in response to a burn area impulse of about 13 km$^2$---the mean burn area in counties that experienced fires. Shaded areas indicate 95\% CIs computed using DK standard errors. Covariates include county and year fixed effects and 2 yearly lags of net out-migration and burn area.}\end{singlespace}
\end{figure}

Figure \ref{fig:out_mig} shows the response of net out-migration to a fire impulse over the subsequent 5 years. Three years after the impulse, net out-migration increases by about 0.15 pp---a substantial effect given that the average net out-migration rate is -0.07\%. The timing of the effect is consistent with the timing at which the decline in employment reaches its peak in figure \ref{fig:base-model-result}. This suggests that part of the medium-run decline in employment is driven by an increase in net out-migration. Note that someone who migrates mid-year will only show up in the IRS Tax Stats Database the following year, so the decline in employment observed between 2 and 2.5 years after a fire impulse aligns with the increase in net out-migration observed at 3 years.\footnote{Though the effect at 5 years afterward is negative and statistically significant at the 5\% level, this may be a result of excessive variability in LP estimates at longer horizons. See \citet{ramey2012comment} for more details.}

\subsection{Labor supply and demand factors}
\label{sec:chars}

As discussed in section \ref{sec:mech}, fires can affect employment through both labor supply and labor demand channels. Our framework recovers effect paths inclusive of equilibrium responses. We therefore conduct a series of tests with sample restrictions to better understand the role of county-specific labor market factors in employment responses to fires.

\subsubsection{Industrial concentration}

``Industrial concentration'' measures the degree to which a few industries employ a large proportion of the population. Both portfolio theory and the regional economics literature support the hypothesis that lower industrial concentration (i.e. greater industrial diversification) supports greater employment stability in the face of shocks \citep{Nourse1968, Barth1975, Sherwood1990, Dissart2003}. We divide the sample into two groups of counties based on their degree of industrial concentration. We calculate concentration using the Herfindahl-Hirschman Index (HHI), which captures the distribution of employment across a set of industries. The HHI is formulated as:
\begin{align} \label{eq_herfin}
    H_c = \sum_{s=1}^{S_c}  \left( \frac{e_{s,c}}{e_c} \right)^2
\end{align}
where $S_c$ is the total number of industries in county $c$, $e_{s,c}$ denotes employment in industry $s$ in county $c$, and $e_c$ the total employment in county $c$. High HHIs indicate high industrial concentration. In contrast, a county where employment is dispersed across a large number of industries would have a low HHI. We use 2-digit NAICS codes as our industry classification. Data on the share of the population employed in each industry is from the American Community Survey.\footnote{The share of the population employed in each industry at the county level are calculated from the ACS 5-Year estimates in 2019.}  

\begin{figure}[!htbp]
\begin{center}
\caption{Response of employment growth to an increase in burn area split by industrial concentration level} \label{fig:fire_indconcentration}
\begin{subfigure}[b]{0.49\textwidth}
\includegraphics[width=\textwidth,height=5.5cm]{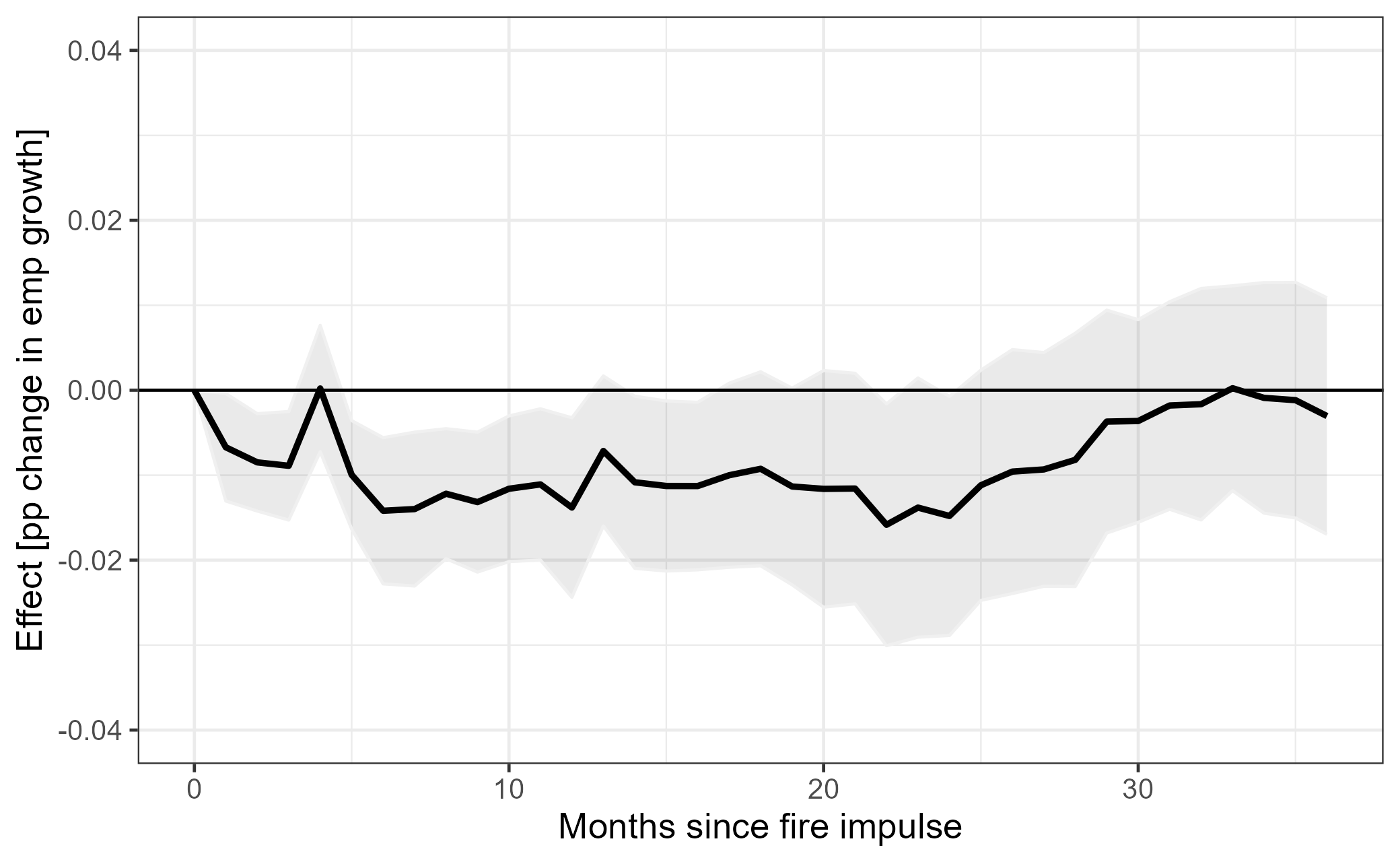}
\caption{High concentration}
\label{fig_highconc}
\end{subfigure}
\begin{subfigure}[b]{0.49\textwidth}
\includegraphics[width=\textwidth,height=5.5cm]{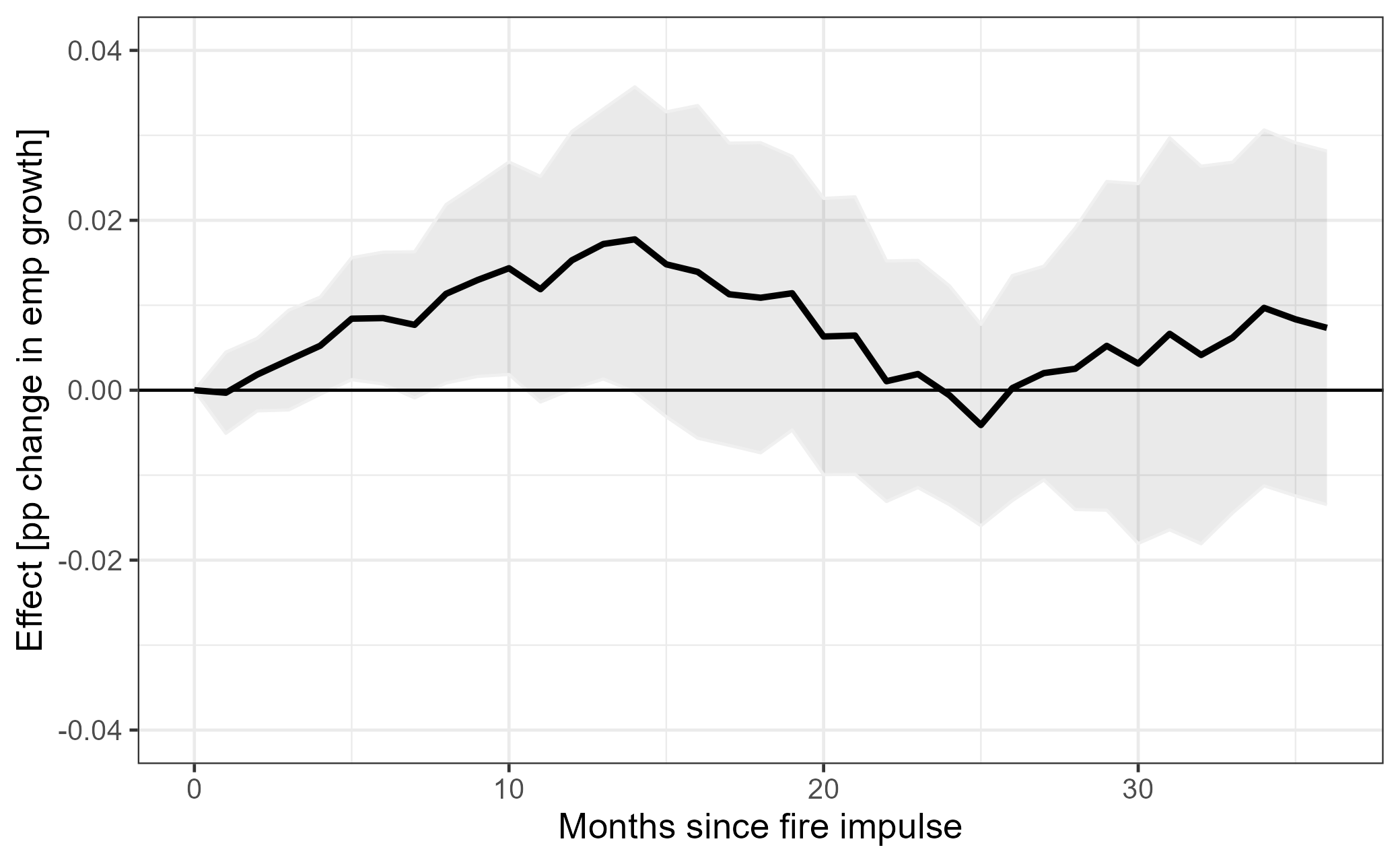}
\caption{Low concentration}
\end{subfigure}
\raggedright
\begin{singlespace}
 \footnotesize{The level of industrial concentration of a county is calculated from the Herfindahl-Hirschman Index. The $y$ axis shows percentage point changes in employment growth in response to a burn area impulse of about 13 km$^2$---the mean burn area in counties that experienced fires. Shaded areas indicate 95\% CIs computed using DK standard errors. Covariates include county and year-month fixed effects and 24 monthly lags of county employment and burn area.} \end{singlespace}
\end{center}
\end{figure}

We estimate equation \ref{eqn:base-model} separately for counties that are above and below the median level of industrial concentration. Figure \ref{fig:fire_indconcentration} shows the resulting impulse responses to an increase in fire burn area at $t$ for the counties that are above (left panel) and below (right panel) median HHI. The employment growth responses are strikingly different across county groups. Counties with lower industrial concentration experience positive economic effects from fires for about one year, which could reflect reconstruction efforts or government support. In contrast, counties with high industrial concentration see a decrease in employment growth of about 0.01 pp in the months following an increase in fire activity. The effect remains negative throughout the horizon and grows to about 0.15 pp two years after the fire occurs, though it is no longer statistically significant at the 5\% level. 
Over three years, the effect accumulates to 0.46 pp (standard error 0.05) of monthly employment growth lost due to the fire impulse for high-concentration counties. In contrast, over three years after the fire impulse low concentration counties see an \emph{increase} in employment growth of 0.21 pp (standard error 0.09). These results support the hypothesis that county-level industrial diversification increases a county's resilience against fire shocks specifically. 

Figure \ref{fig:fire_maps_hhi} in Appendix \ref{apdx:summary-stats} shows the total burn area in US counties by industrial concentration grouping. Coastal areas that experience fires tend to have more-diversified economies (low concentration) while counties with less-diversified economies (high concentration) that experience fires are distributed across the West, Midwest, and Southwest. Table \ref{table:indcomp} in Appendix \ref{apdx:summary-stats} shows the industrial composition across concentration groupings. While the overall industry composition is broadly similar across the two groups, counties that are more concentrated employ on average a larger share of the population in the natural resources sector (9.4\% vs. 5.3\%), which might be especially vulnerable to fires. Table \ref{tab:HHI-summary} in Appendix \ref{apdx:summary-stats} shows the top and bottom ten counties in terms of HHI. Petroleum County in Montana is the most concentrated (HHI 0.886), while Storey County in Nevada is the least concentrated (HHI 0.110). 

These two groups of counties also differ along other dimensions. Table \ref{table:descstat} in Appendix \ref{apdx:summary-stats} shows county-level summary statistics across the two groups. Interestingly, counties with more diversified economies also experience more fires, although both groups have significant fire exposure. The employment-to-population ratio is almost identical across the two groups. Counties that are more concentrated tend to have smaller populations and lower income per capita on average, although industrial concentration is only weakly correlated with population and income (-0.07 and -0.08, respectively, see Table \ref{table:correlation}). Finally, counties that are highly concentrated are also likely to be more rural than counties that are less concentrated. This aligns with \citet{krugman1991}'s core-periphery analysis of distribution of economic activity across space---urban areas are more likely to support a larger set of economic activities.

\subsubsection{Education}

As discussed in Section \ref{sec:mech}, the educational composition of a region's workforce is likely an important determinant of the economic effects of fires. Existing literature indicates that workers with more education are more mobile \citep{wozniak2010college, balgova2018don}, and may also work jobs that are less sensitive to fire-related disruptions \citep{yasenov2020can}. For each county, we calculate the average percentage of the population with a high school diploma since 2003 using data from the Census Bureau. Table \ref{tab:edu-summary} in Appendix \ref{apdx:summary-stats} shows that education levels vary greatly across counties. Kalawao County in Hawaii has the largest share of population without a high school diploma (60.5\%), while Douglas County in Colorado has the lowest (3.1\%). 

\begin{figure}[!htbp]
\begin{center}
\caption{Response of employment growth to an increase in fire exposure split by education level} \label{fig:fire_education}
\begin{subfigure}[b]{0.49\textwidth}
\includegraphics[width=\textwidth,height=5.5cm]{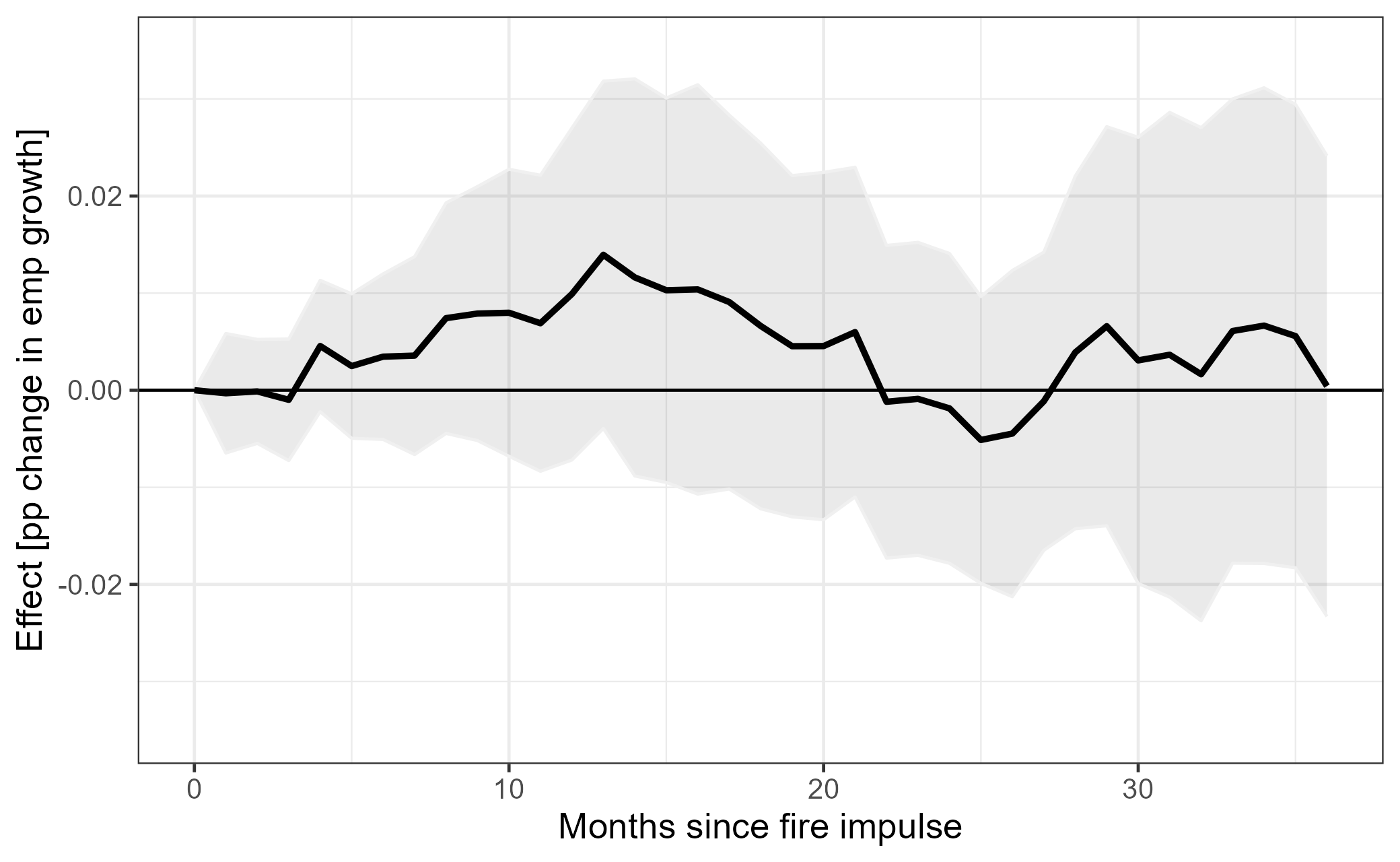}
\caption{High education}
\label{fig_highedu}
\end{subfigure}
\begin{subfigure}[b]{0.49\textwidth}
\includegraphics[width=\textwidth,height=5.5cm]{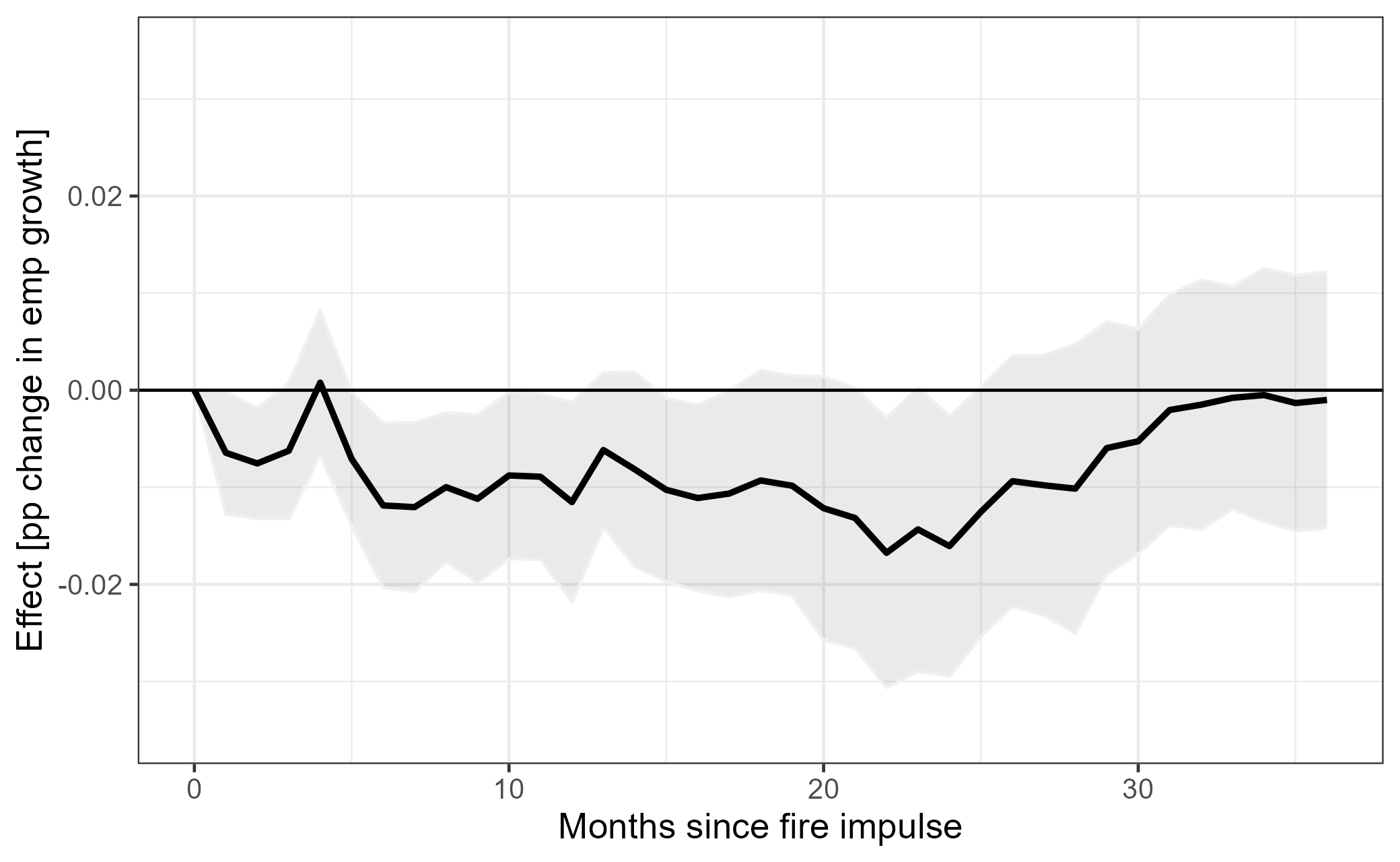}
\caption{Low education}
\label{fig:lowedu}
\end{subfigure}
\raggedright
\begin{singlespace}
 \footnotesize{County education level is calculated as the share of the population that does not have a high school diploma. The $y$ axis shows percentage point changes in employment growth in response to a burn area impulse of about 13 km$^2$---the mean burn area in counties that experienced fires. Shaded areas indicate 95\% CIs computed using DK standard errors. Covariates include county and year-month fixed effects and 24 monthly lags of county employment and burn area.} \end{singlespace}
\end{center}
\end{figure}

We estimate equation \ref{eqn:base-model} separately for counties that are above and below the median level of high school diploma attainment. Figure \ref{fig:fire_education} shows the resulting impulse responses to an increase in burn area for counties that are above (left panel) and below (right panel) the median level of high school diploma attainment. The level of education appears to matter a great deal. In particular, counties with below-median high school diploma attainment see a decrease in employment following fires that lasts for about two years. Over three years, the effect accumulates to 0.43 pp (standard error 0.08) of monthly employment growth lost due to the fire impulse for low-education counties. For counties with above-median high school diploma attainment, there is no statistically significant effect of fires on employment at any time horizon. 

Figure \ref{fig:fire_maps_edu} shows that less-educated counties are largely concentrated in the South with a smaller number of counties distributed over the West and in Alaska (with the exception of Anchorage).\footnote{Some of this gap may be attributable to persistent legacies of discriminatory and suppressive policies, e.g. \citet{williams2021persistence}. Exploring this connection is a promising direction for future research.} While Table \ref{table:descstat} shows that counties with below-median high school diploma attainment, like highly concentrated counties, tend to be in rural areas and have smaller population, it is worth noting that the correlation between education and industrial concentration across counties is essentially zero (0.0003) and not statistically significant. Not surprisingly, the share of population that is employed is lower in counties with below-median high school diploma attainment (42.2\% vs. 48.8\%).  
Overall, these results show that economies that are more diversified and have a more-educated workforce tend to be more economically robust to fires. 

\subsection{The state of the business cycle}
\label{sec:state-dependent-lp}

In addition to county-level labor supply and labor demand characteristics, temporally-varying conditions like labor market slack can also modulate the effects of fires on labor markets. Following the approach taken in \citet{auerbach2012slack}, \citet{owyang2013government}, and \citet{ramey2018} in the context of fiscal multipliers, we modify the baseline model to allow the effect of marginal fire exposure to depend on the state of the county's business cycle before the fire occurs. We use county-specific unemployment rates to measure the level of labor market slack in a county, and assess the county's economic state at the time the impulse occurs relative to its own historic slack levels. Let $I_{c,t}$ denote an indicator variable that equals 1 if the unemployment rate in county $c$ at month $t$ is greater than the 70$^{th}$ percentile of county $c$'s unemployment rate and 0 otherwise.\footnote{We find very similar results when we vary this threshold.} We estimate the following state-dependent LP model:
\begin{align} \label{eqn:slack-model}
y_{c,t+h} - y_{c,t-1} =  I_{c,t} \left[\beta^H_h D_{c,t}   \right]  + (1-I_{c,t})\left[\beta^L_h D_{c,t} \right] + X'_{c,t} \gamma_h + \alpha_{c,h} + \mu_{t,h} +  \epsilon_{c,t+h}
\end{align}
where $H$ denotes coefficients specific to the high-slack state and $L$ the low-slack state. The impulse response of employment growth in a high-slack state is computed from the estimated $\{\beta^H_h\}^{36}_0$, while the response in the low-slack state is computed from the estimated $\{\beta^L_h\}^{36}_0$. 

\begin{figure}[!t]
\caption{Response of employment and net out-migration to an increase in burn area split by level of slack} \label{fig:fire_emp_slack} 
\begin{subfigure}[b]{0.49\textwidth}
\includegraphics[width=\textwidth,height=5cm]{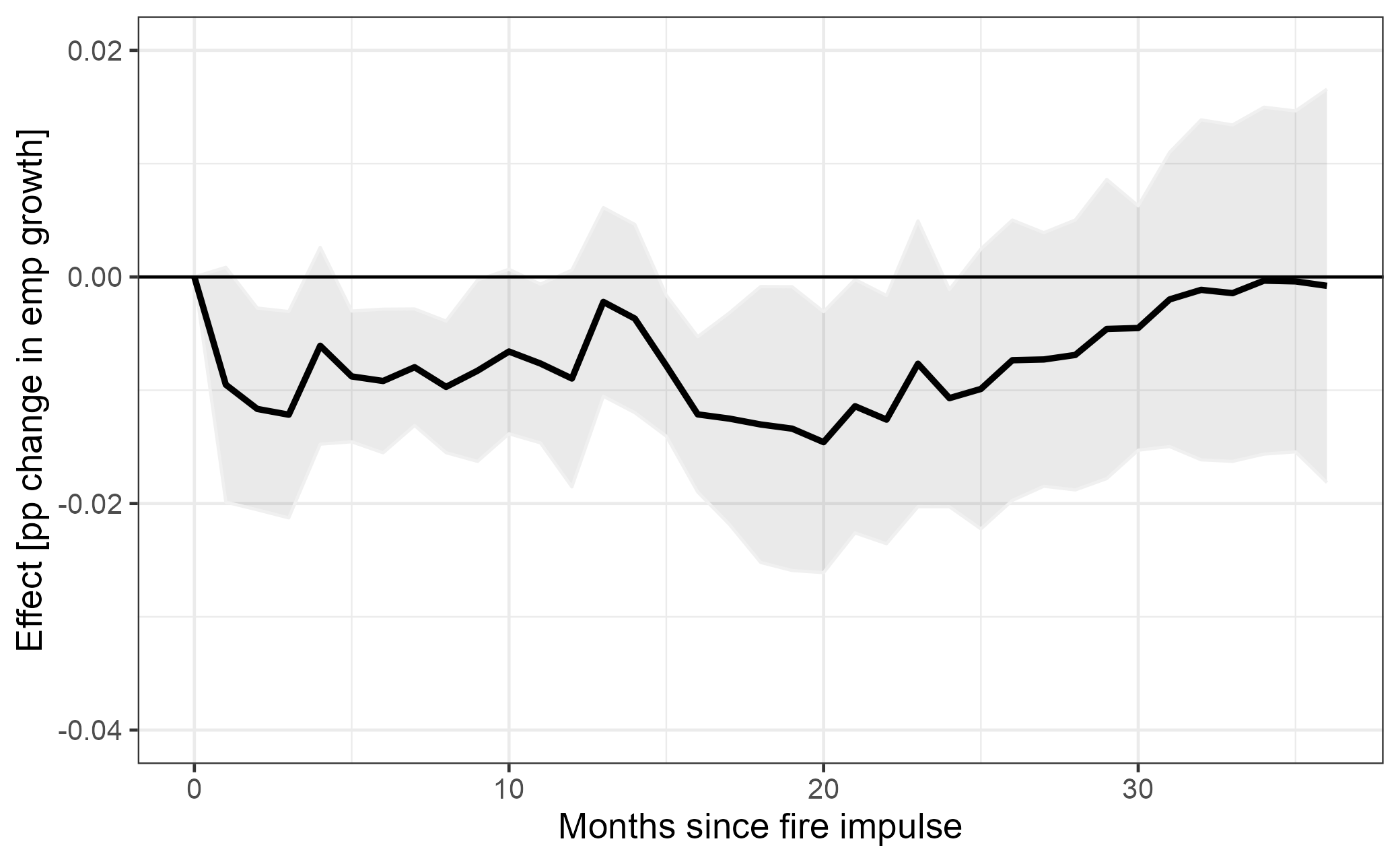}
\caption{Employment -- High slack}
\label{fig_highslack}
\end{subfigure}
\begin{subfigure}[b]{0.49\textwidth}
\includegraphics[width=\textwidth,height=5cm]{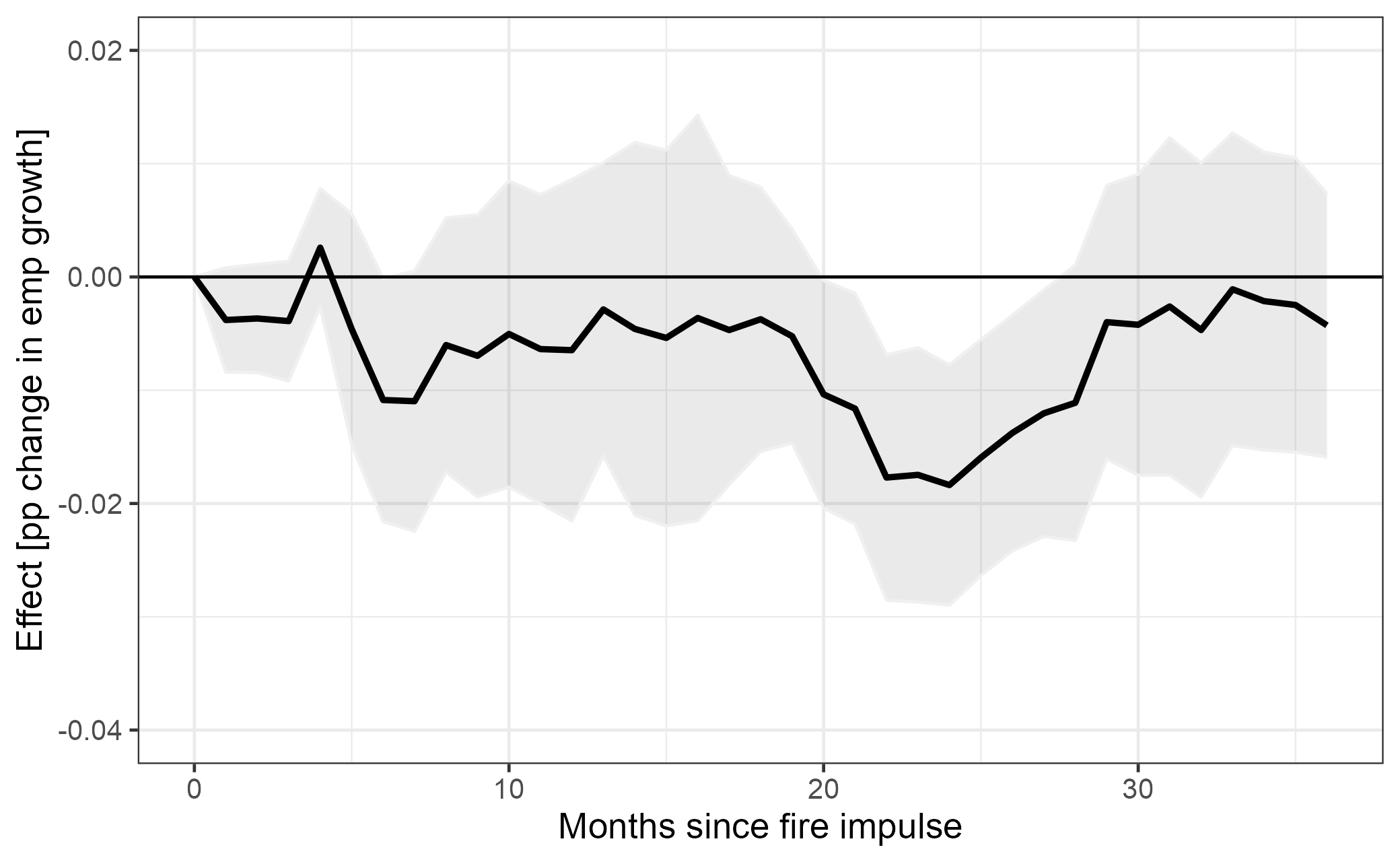}
\caption{Employment -- Low slack}
\label{fig_lowslack}
\end{subfigure}
\begin{subfigure}[b]{0.49\textwidth}
\includegraphics[width=\textwidth,height=5cm]{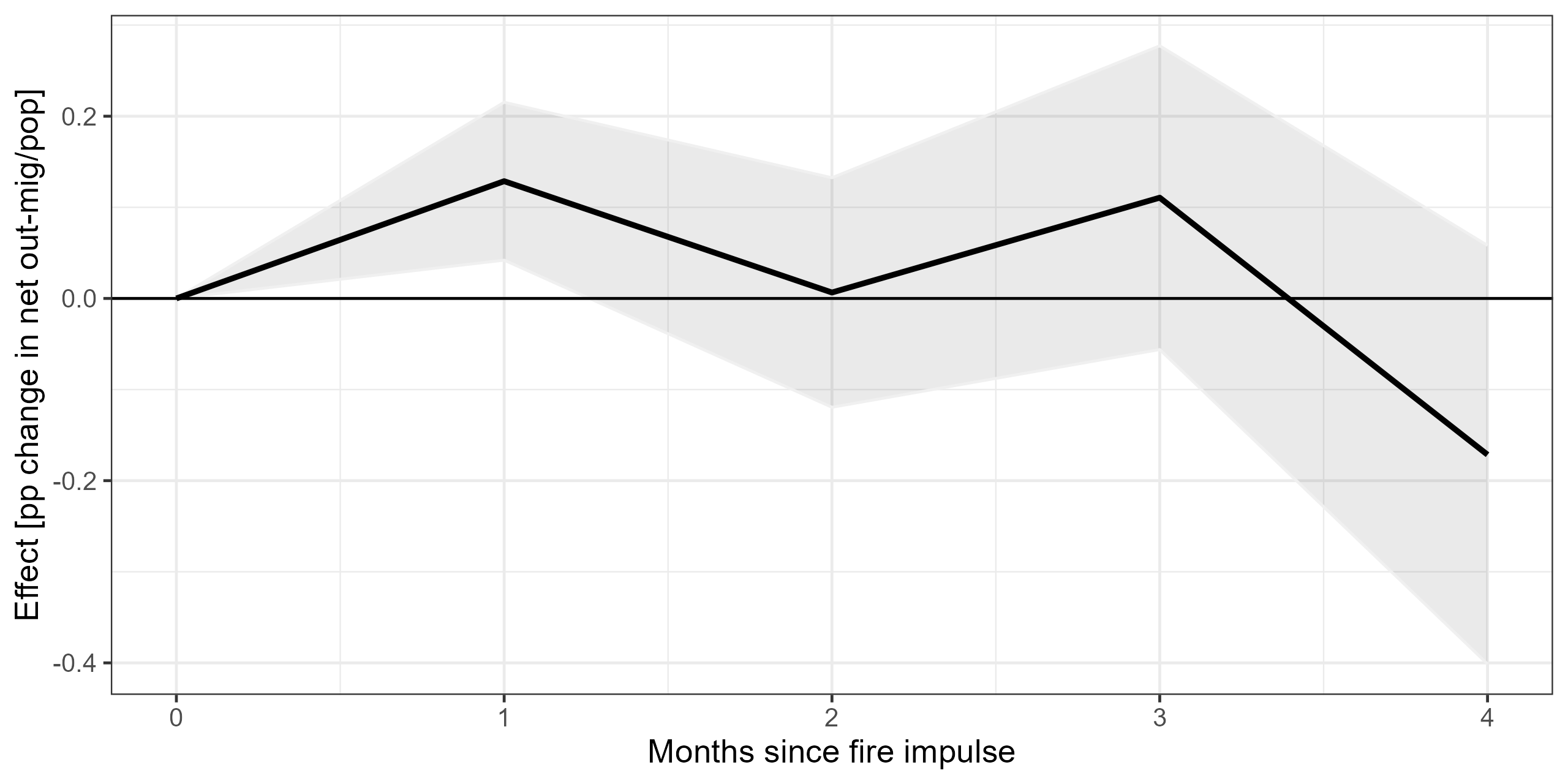}
\caption{Net out-migration -- High slack}
\label{fig_highslack_mig}
\end{subfigure}
\begin{subfigure}[b]{0.49\textwidth}
\includegraphics[width=\textwidth,height=5cm]{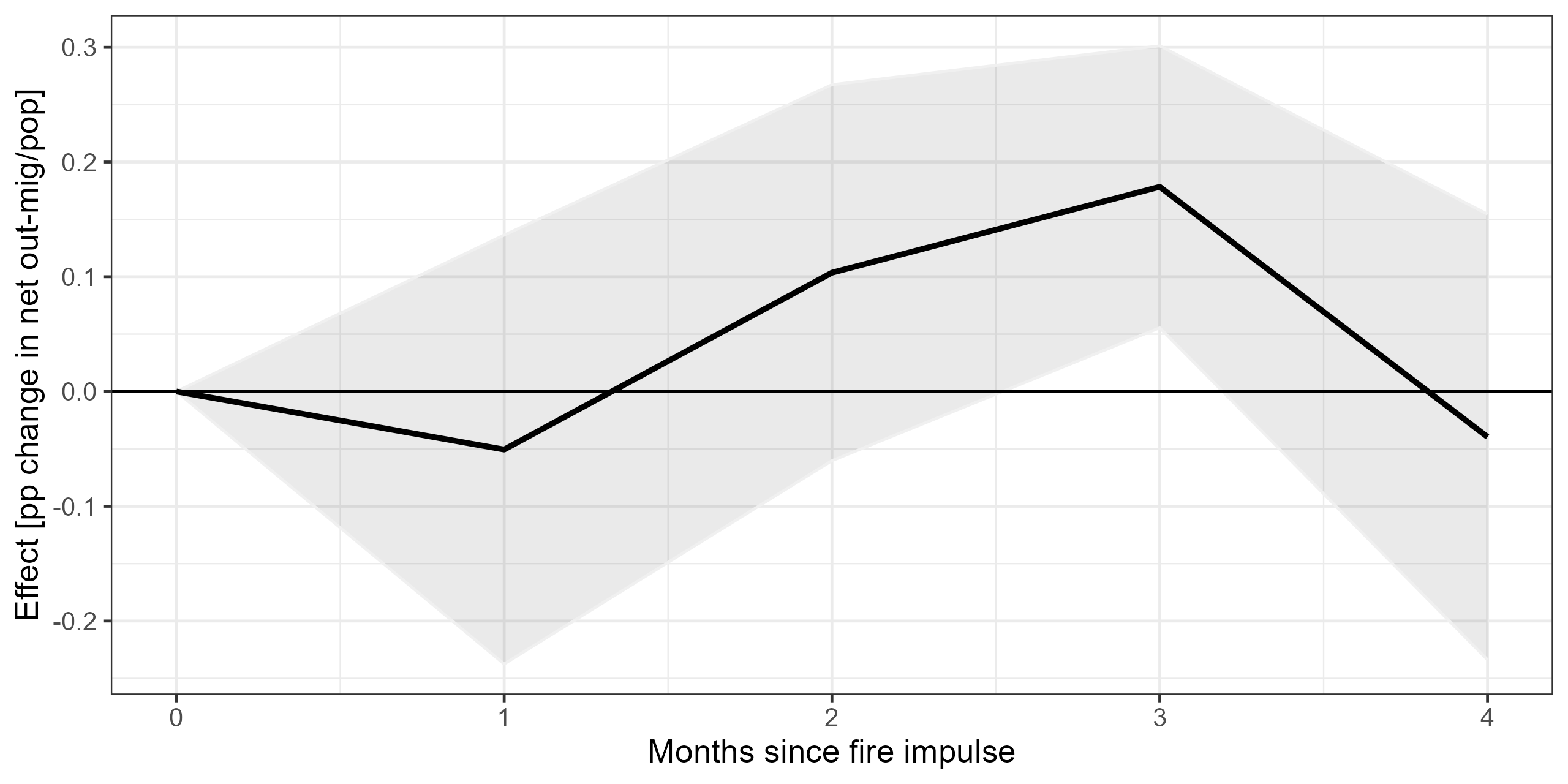}
\caption{Net out-migration -- Low slack}
\label{fig_lowslack_mig}
\end{subfigure}
\raggedright
\begin{singlespace}
 \footnotesize{The level of slack before the fire occurs is based on the county's unemployment rate. We consider a high slack state if the unemployment rate in county $c$ is greater than the $70^{th}$ percentile of county $c$'s unemployment rate. The $y$ axis shows percentage point changes in employment growth in response to a burn area impulse of about 13 km$^2$---the mean burn area in counties that experienced fires. Shaded areas indicate 95\% CIs computed using DK standard errors. Covariates include county and year-month fixed effects and 24 monthly lags of county employment and burn area.} \end{singlespace}
\end{figure}

Figure \ref{fig:fire_emp_slack} presents these impulse responses. The effect of fires on employment in a low-slack state, i.e. when the county's unemployment rate is below its historical 70$^{th}$-percentile level at period $t$, is shown in panel (a), while panel (b) shows the effect of a fire impulse during a high-slack state. The state of the economy appears to matter a great deal. During the first 20 months following a fire impulse, the decrease in employment growth is statistically significant at the 5\% level only if the fire impulse occurred in a high-slack (i.e. high unemployment) state. It takes about 20 months for the decrease in employment to become statistically significant at the 5\% level when the impulse occurs during a low-slack state. 

Panels (c) and (d) show the same state-dependent response for net out-migration. Workers are much more likely to leave immediately following a fire impulse if it occurs during a high-slack state. By contrast, fires impulses during low-slack states only induce net out-migration about 3 years after the impulse. These responses are consistent with the interpretation that the main labor market effect of a fire impulse during a low-slack state is a medium-run increase in net out-migration.

\newpage
\subsection{Heterogeneity in fires} 

\paragraph{Size and fire type.} As noted earlier, the distribution of fire exposure among counties is positively skewed: relatively few county-months have fires, and a smaller number have very large burn areas. The heterogeneity in fire sizes reflects underlying differences in fire types, e.g. some are uncontrolled wildfires while others are intentional prescribed burns. Could the impulse response of employment to fires vary by fire type? 

\begin{figure}[!htpb] 
\caption{Response of employment growth to an increase in burn area by fire size} \label{fig:fire-size-heterogeneity-small}
\begin{subfigure}[b]{0.49\textwidth}
\includegraphics[width=\textwidth,height=5.5cm]{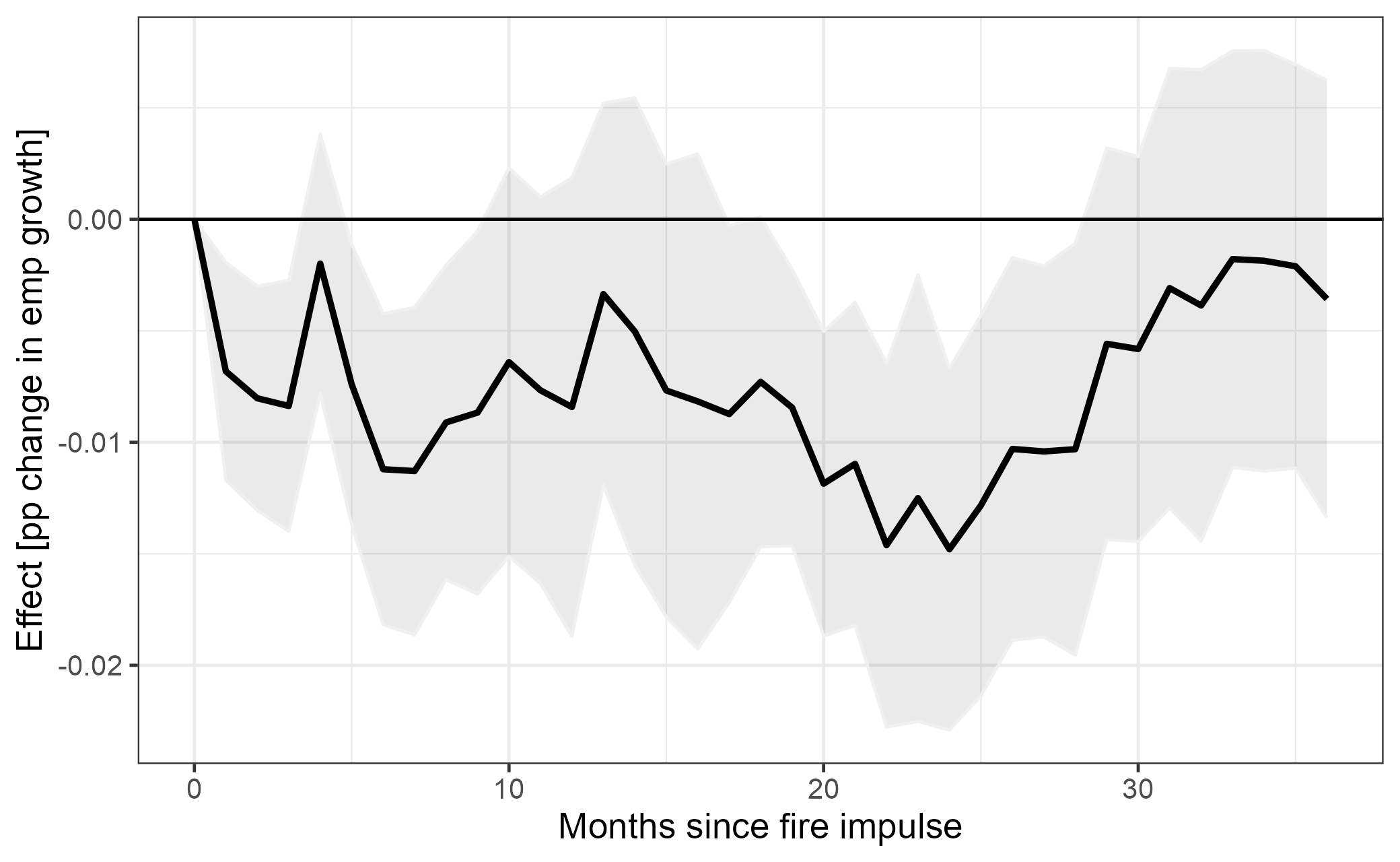}
\caption{Top 1\%}
\end{subfigure}
\begin{subfigure}[b]{0.49\textwidth}
\includegraphics[width=\textwidth,height=5.5cm]{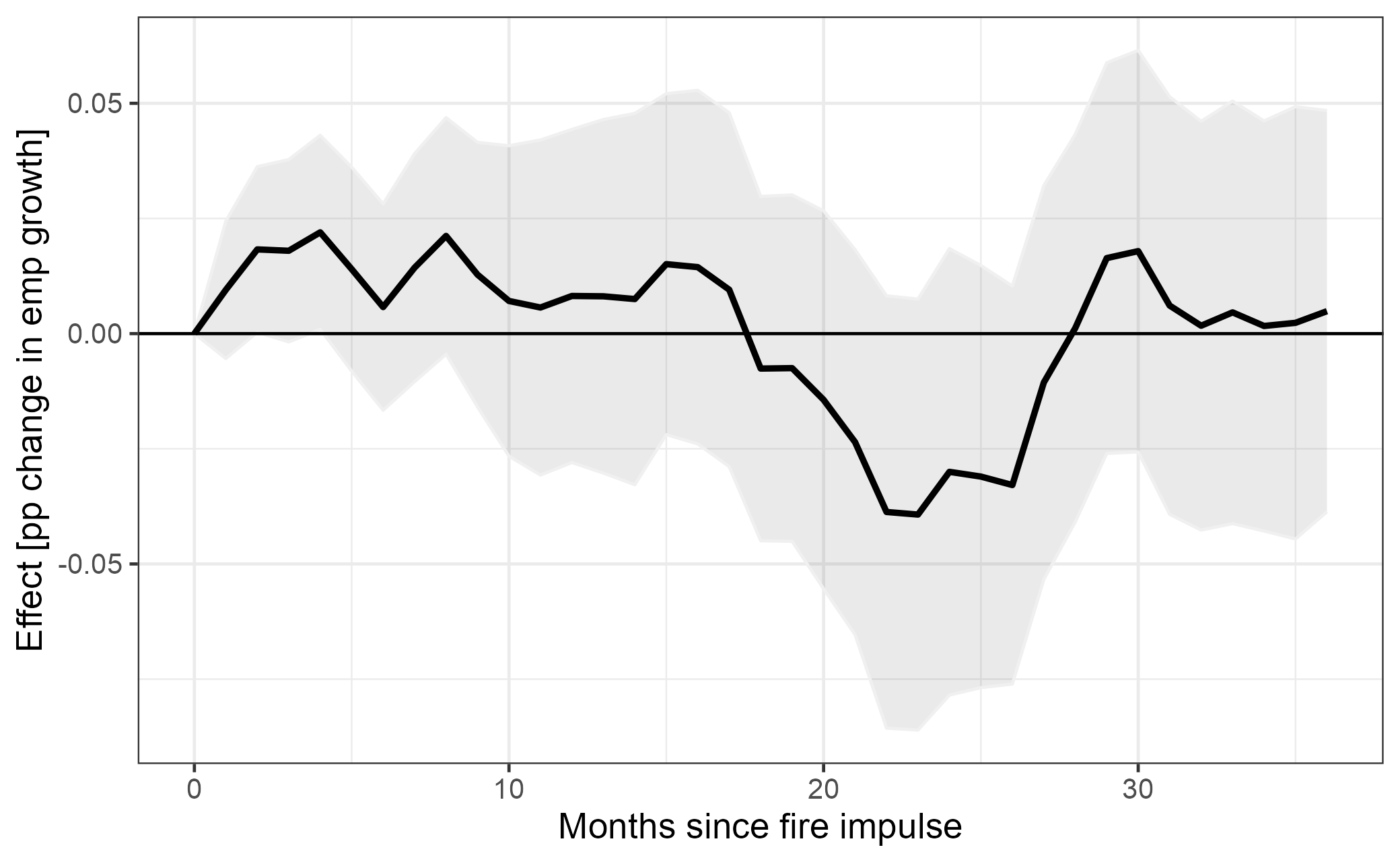}
\caption{Bottom 99\%}
\end{subfigure}
\raggedright
\begin{singlespace}
\footnotesize{The $y$ axis shows percentage changes in employment growth in response to a burn area impulse of about 13 km$^2$---the mean burn area in counties that experienced fires. Shaded areas indicate 95\% CIs computed DK standard errors. Covariates include county and year-month fixed effects and 24 monthly lags of county employment and burn area.} \end{singlespace}
\end{figure}

To investigate this, we modify the baseline model to allow fires of different sizes to have different effects. We use a state-dependent local projections model, similar to equation \ref{eqn:slack-model}, where the indicator $I_{c,t}$ is 1 if the burn area in county $c$ at time $t$ is greater than the $n$th percentile of burn areas in the sample. Figure \ref{fig:fire-size-heterogeneity-small} shows the response of employment growth to a 13 km$^2$ burn area impulse for $n=99$.\footnote{While the size of impulse is kept constant to facilitate comparison, it is worth noting that the average burn area for the bottom 90\% is 6.53 km$^2$. This is larger than the threshold for fire classification under the MTBS scheme described in \citet{eidenshink2007project} (roughly 4 km$^2$ in the West and 2 km$^2$ in the East).} 

The impulse response from the largest fires is virtually identical to the baseline model, with two distinct reductions in employment growth---one in the short run, one in the medium run---for the largest fires and a large but statistically-insignificant medium-run reduction for all other fire sizes. Figure \ref{fig:fire-size-heterogeneity-model} in Appendix \ref{apdx:additional-figures-tables} shows results are similar for $n \in \{ 90, 95, 99 \}$. These results suggest that the largest fires have the most notable effects. The low correlation with FEMA fire declaration measures suggest not all of these events may receive similar recovery aid. Differences between our results and those in the literature, e.g. \citet{tran2021} and \citet{walls2023econwildfires}, thus likely reflect differences in event timing classification and intensive margin measurements.\footnote{\citet{walls2023econwildfires} use the MTBS dataset to identify fire events, though their analysis focuses on quarterly employment outcomes. Our fire measure may provide greater power on the intensive margin, in addition to capturing more temporally-detailed dynamic responses. However, \citet{walls2023econwildfires} use a DL approach to identifying employment impacts of fires, thus focusing on a different estimand than we do (see footnote 7).} 

\paragraph{Regional differences.} As shown in table \ref{tab:burn_summary}, the average burn area varies substantially across the country. While the West experiences substantial fire activity over our sample period, the Northeast experiences virtually none. Additionally, the types of fires also differ substantially by region: while the West sees regular wildfires, the South sees a mix of wildfires and prescribed burns for agriculture (e.g. sugarcane burning before harvest in Florida). Does the impulse response of employment to fires vary by region? 

\begin{figure}[!h] 
\caption{Response of employment growth to an increase in burn area by region} \label{fig:region-IRFs}
\begin{subfigure}[b]{0.49\textwidth}
\includegraphics[width=\textwidth,height=5cm]{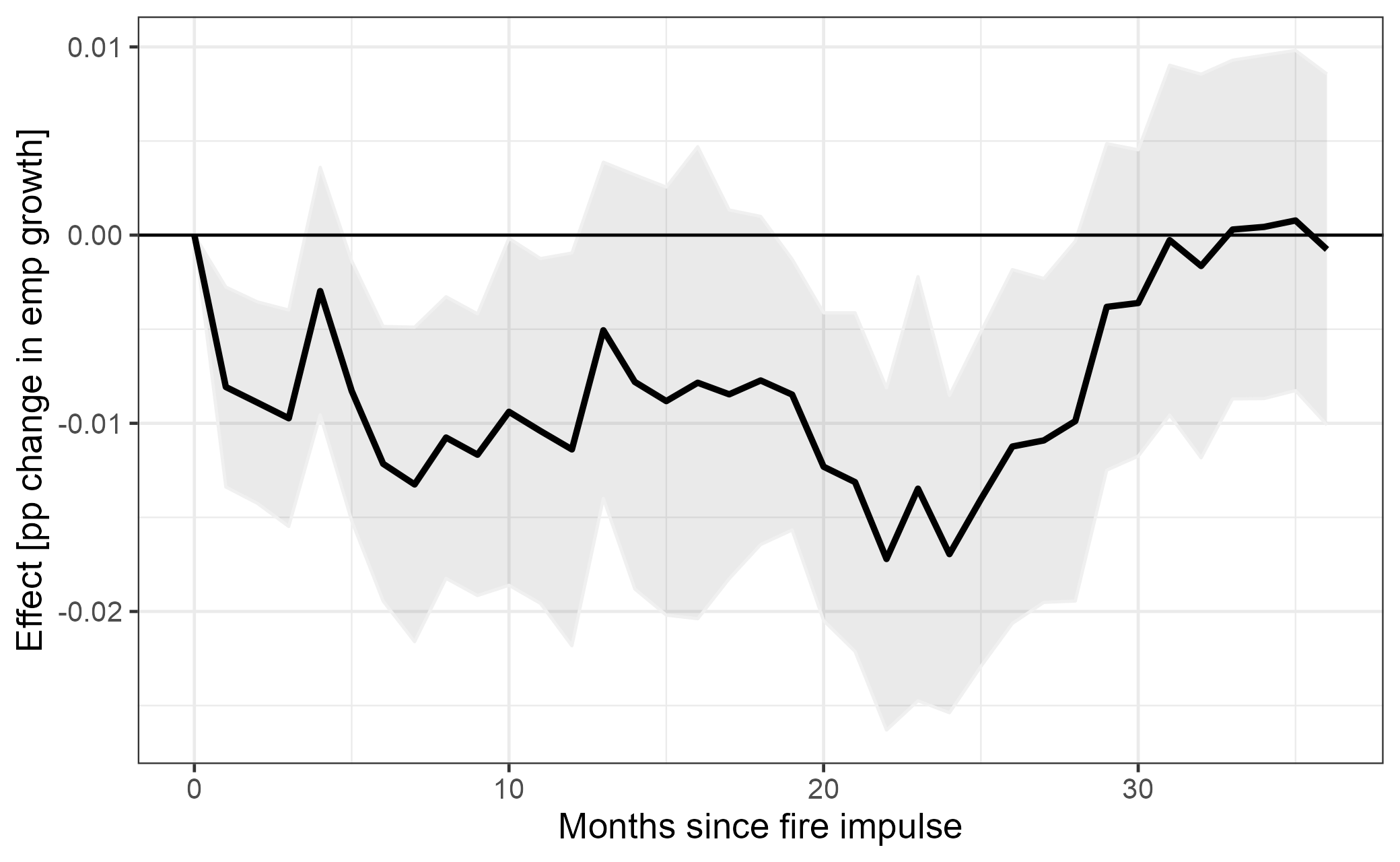}
\caption{West}
\label{fig_west}
\end{subfigure}
\begin{subfigure}[b]{0.49\textwidth}
\includegraphics[width=\textwidth,height=5cm]{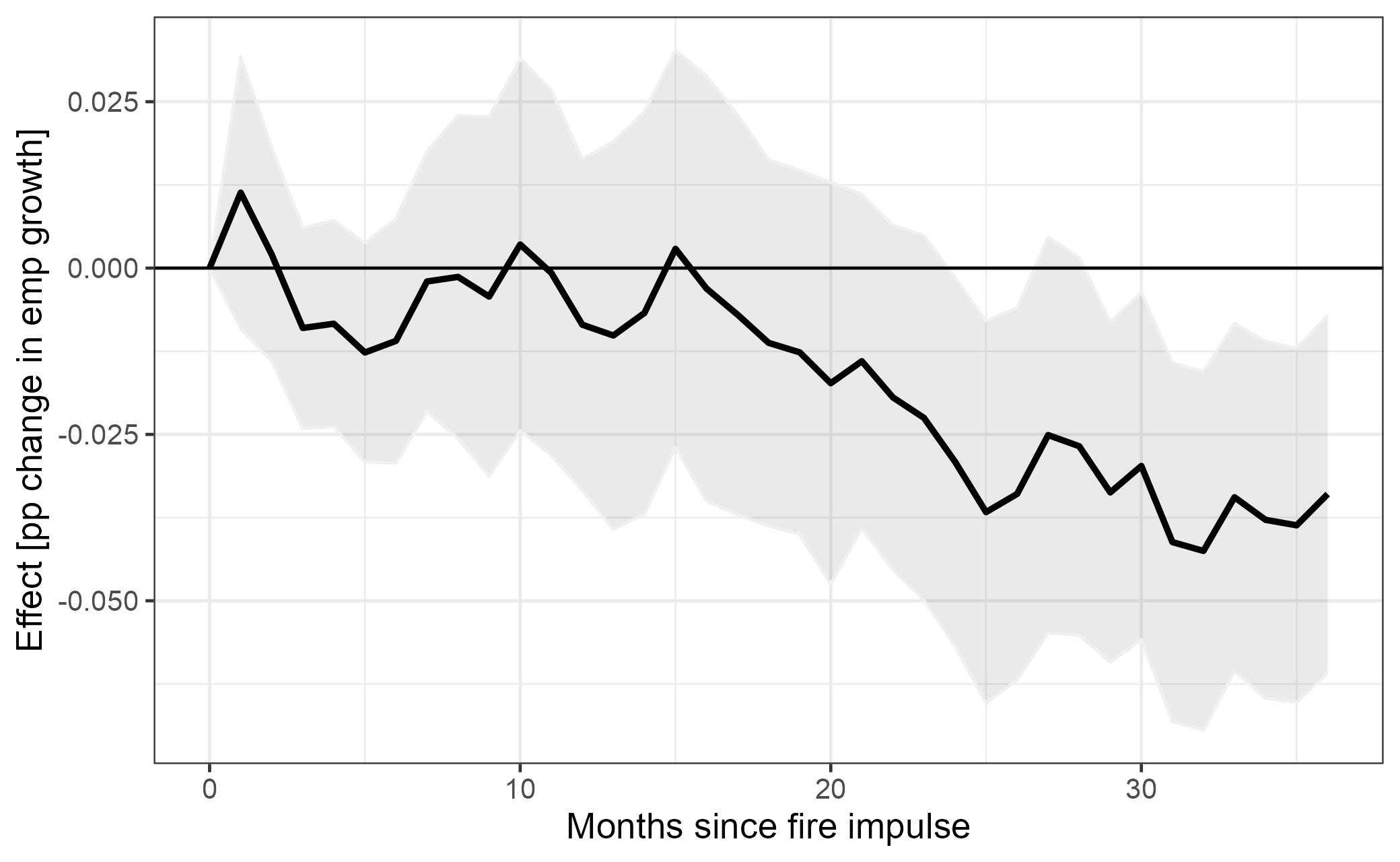}
\caption{Midwest}
\label{fig_midwest}
\end{subfigure}
\begin{subfigure}[b]{0.49\textwidth}
\includegraphics[width=\textwidth,height=5cm]{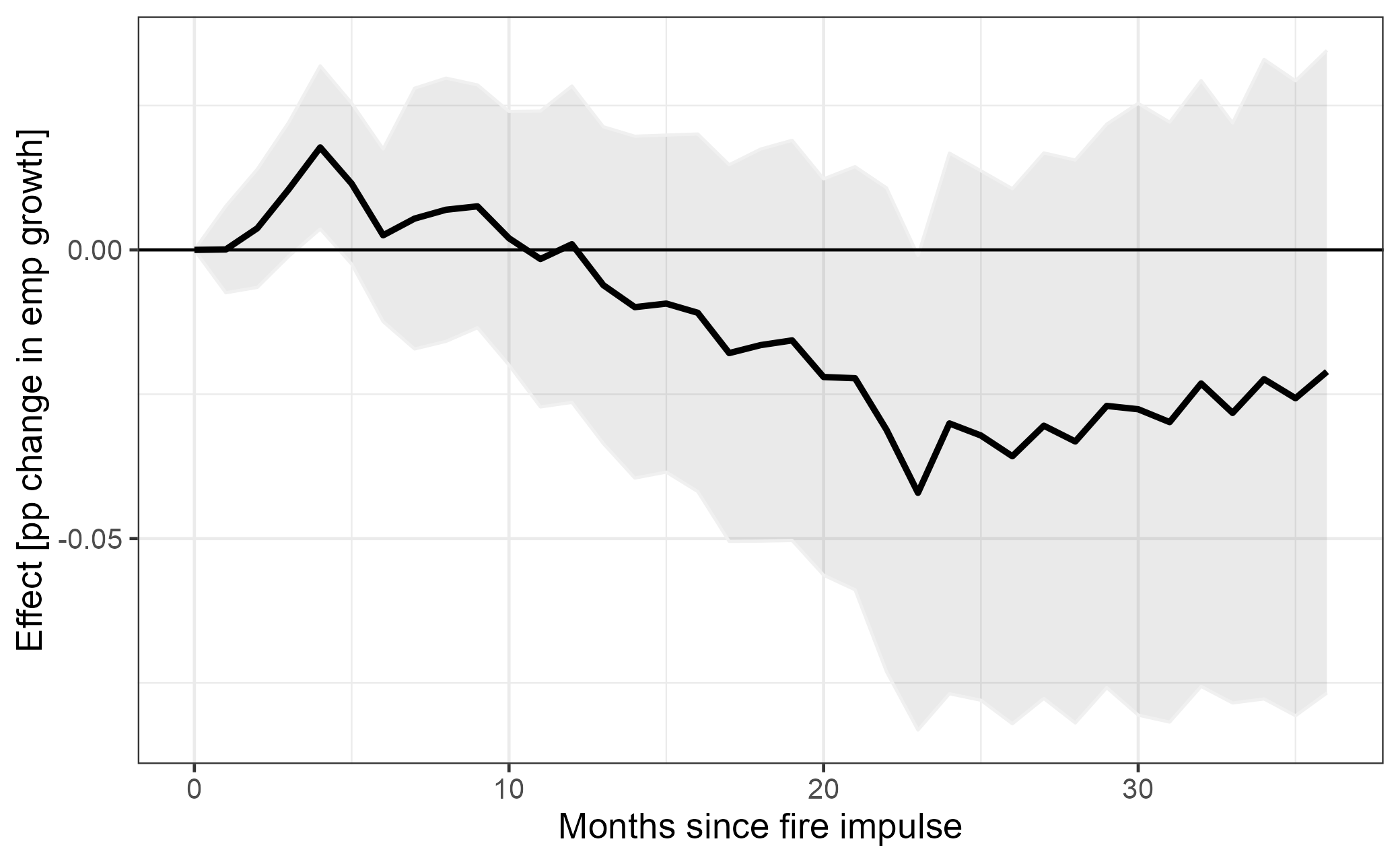}
\caption{South}
\label{fig_south}
\end{subfigure}
\begin{subfigure}[b]{0.49\textwidth}
\includegraphics[width=\textwidth,height=5cm]{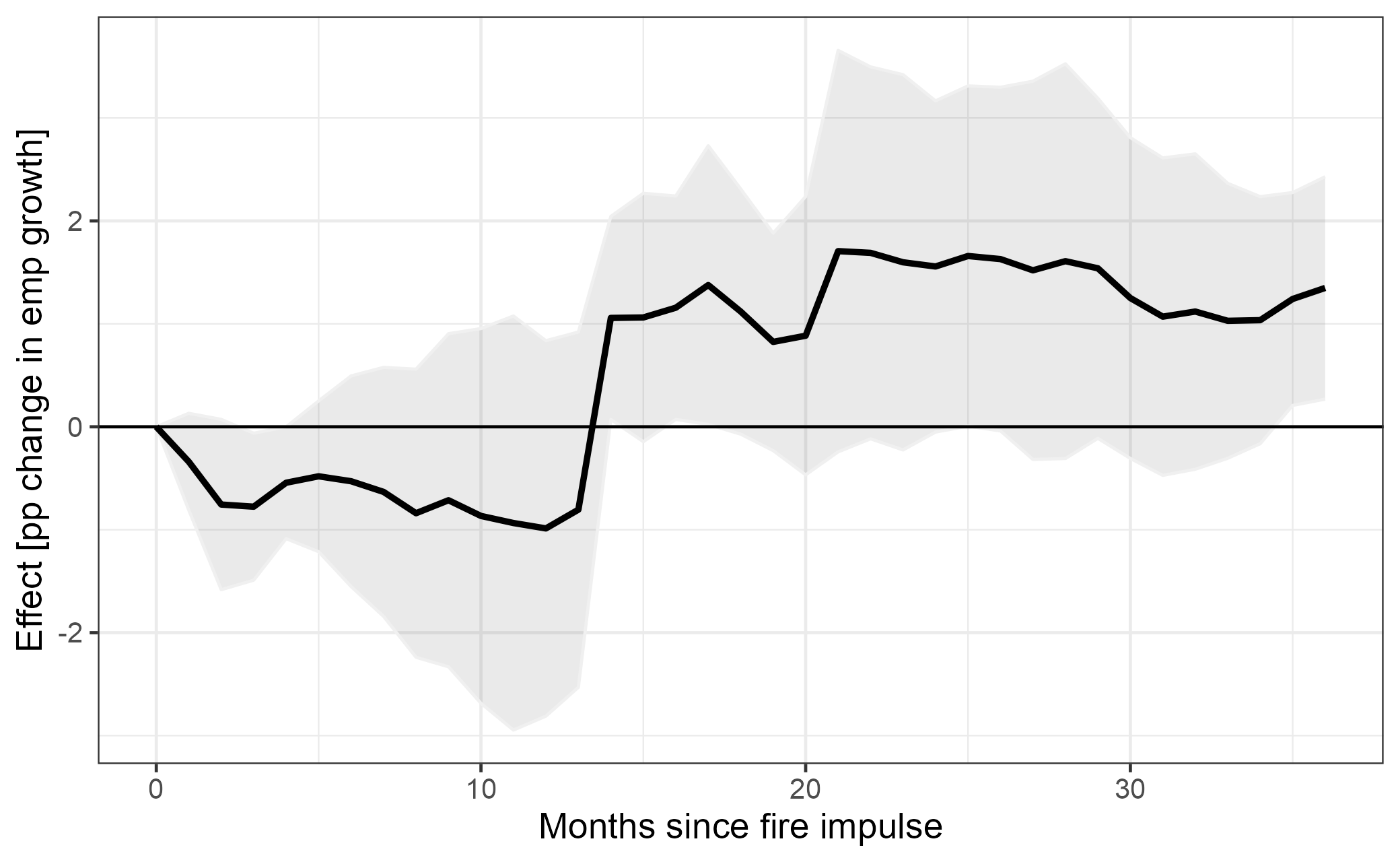}
\caption{Northeast}
\label{fig_northeast}
\end{subfigure}
\raggedright
\begin{singlespace}
\footnotesize{The $y$ axis shows percentage changes in response to a burn area impulse of about 13 km$^2$---the mean burn area in counties that experienced fires. Shaded areas indicate 95\% CIs computed using DK standard errors. Covariates include county and year-month fixed effects and 24 monthly lags of county employment and burn area.} \end{singlespace}
\end{figure}

To investigate this, we estimate equation \ref{eqn:base-model} separately for each Census region. Figure \ref{fig:region-IRFs} shows these results. While the West experiences more fire on average, the same fire impulse leads to a much larger decline in employment in the Midwest and South compared to the West. Two years after a fire, the decline in employment growth is five times larger in the Midwest and South than in the West (-0.1 pp vs. -0.02 pp). The Northeast shows no statistically significant effects at any horizon. The shapes of the regional IRFs indicate that Western fire responses dominate the national IRF shown in figure \ref{fig:base-model-result}.\footnote{
Though the IRFs for the Midwest and Northeast appear to show statistically significant impacts near the end of the horizon, these effects disappear by 40 months after the impulse. There is also no corresponding effect for net out-migration in these regions.} 

\subsection{Simulated historical impacts}

An advantage of using geophysical measures of fire exposure is the ability to simulate the implied employment impacts of historical fire scenarios. To illustrate this we simulate the historical employment impacts of fires in each US Census Region over our sample period. The simulated historical employment impact of a burn sequence $(D_{c,0}, \dots, D_{c,h})$ is the summation of present and lingering employment effects of fires which have occurred in county $c$ over the previous $h$ months. Since the aggregate impulse response fades after three years (Figure \ref{fig:base-model-result}), we truncate the impulse response to zero after 36 months. The simulated historical employment impact in county $c$, $HEI_c$, is shown in equation \ref{eqn:cumulative-impact}:
\begin{equation} 
HEI_c = \sum_{j=0}^{36} \beta_{h-j} D_{c,j}. \label{eqn:cumulative-impact}
\end{equation}

To compute the historical impact in each Census region, we first estimate region-specific impulse responses (shown in Figure \ref{fig:region-IRFs}), then calculate the simulated historical employment impact $HEI_c$ for each county, and finally take the population-weighted average across counties within Census regions. Letting the population in county $c$ of region $R$ be $pop_{c,R}$ and the total population in region $R$ be $pop_R$, the simulated historical impact in an average county in region $R$ is shown in equation \ref{eqn:cumulative-impact-regionavg}:
\begin{equation} 
HEI_R = \sum_{c \in R} \frac{pop_{c,R}}{pop_R} HEI_c. \label{eqn:cumulative-impact-regionavg}
\end{equation}

\begin{figure}[!h] 
\caption{Simulated historical fire-driven employment fluctuations by region} \label{fig:total-effect-regions}
\begin{subfigure}[b]{0.49\textwidth}
\includegraphics[width=\textwidth,height=5cm]{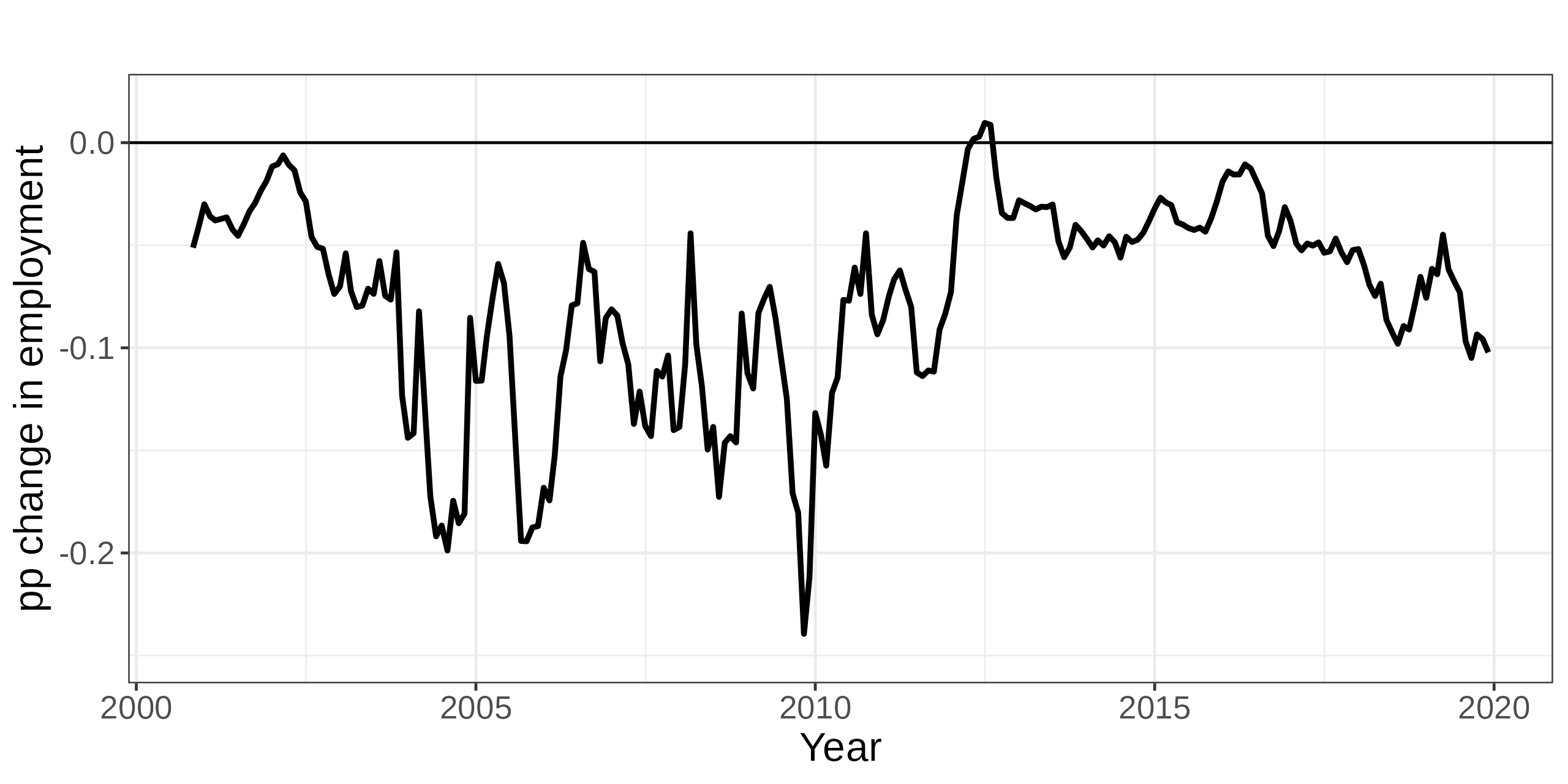}
\caption{West}
\label{fig:total-effect-regions--west}
\end{subfigure}
\begin{subfigure}[b]{0.49\textwidth}
\includegraphics[width=\textwidth,height=5cm]{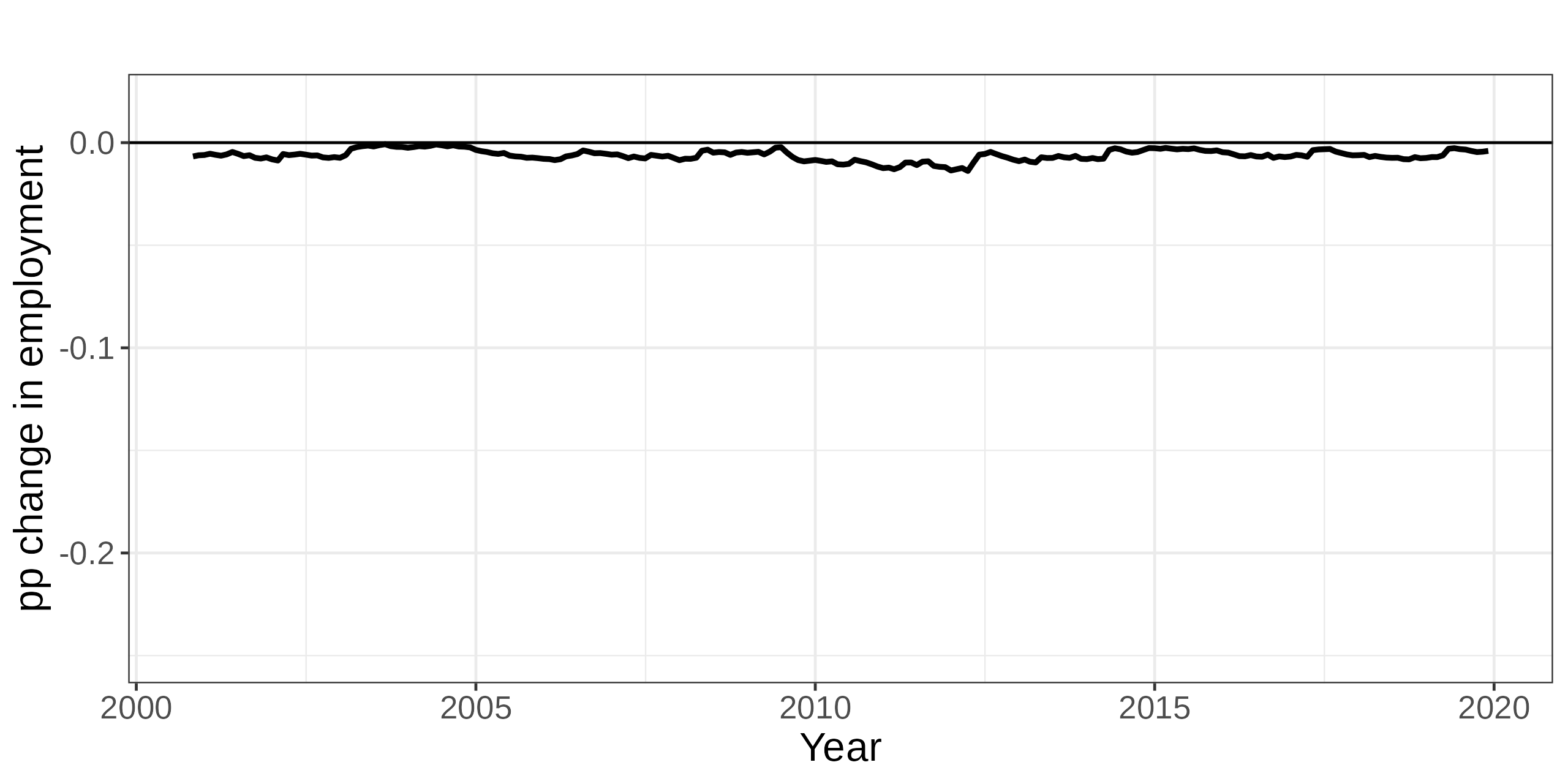}
\caption{Midwest}
\label{fig:total-effect-regions--midwest}
\end{subfigure}
\begin{subfigure}[b]{0.49\textwidth}
\includegraphics[width=\textwidth,height=5cm]{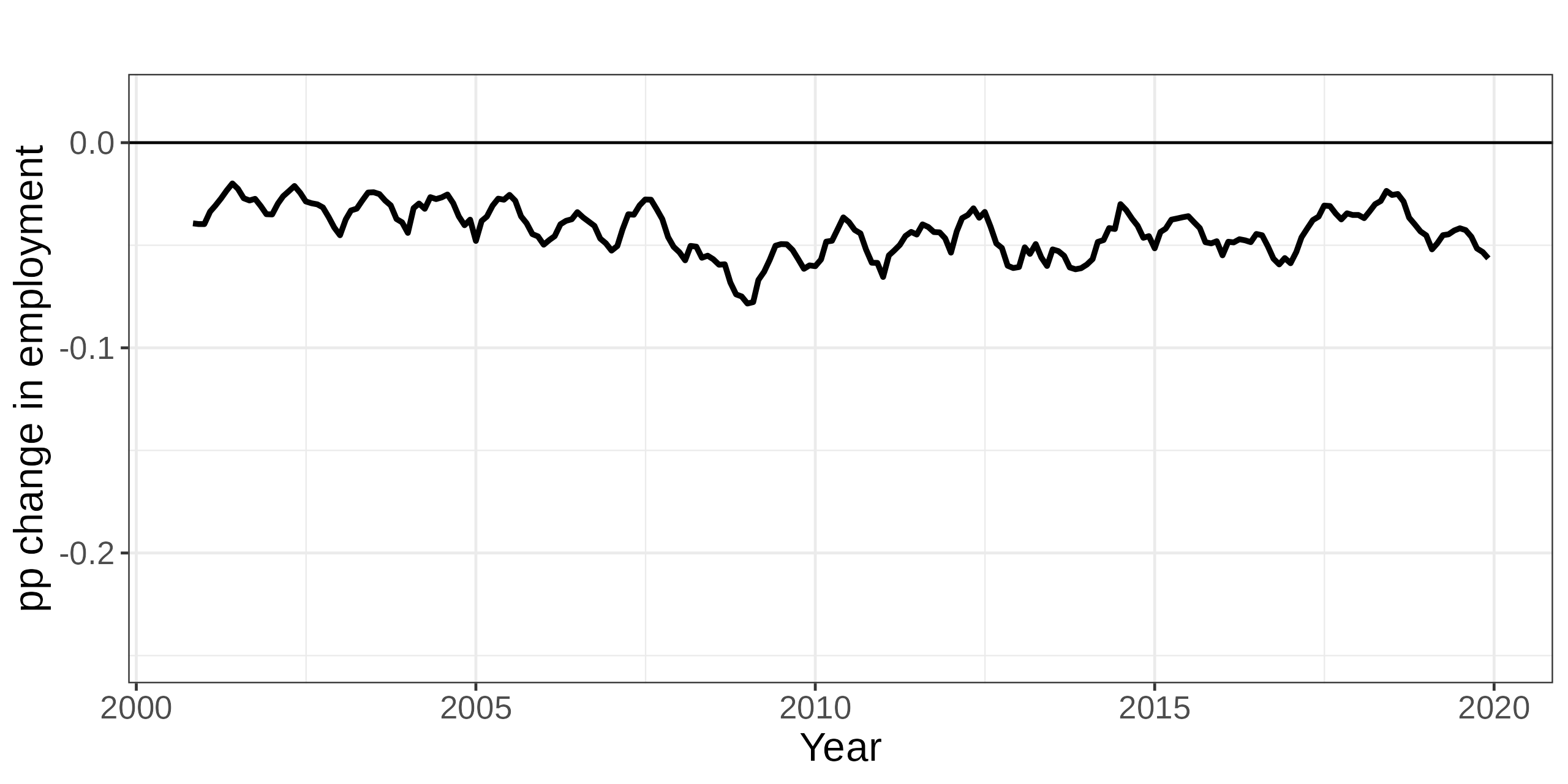}
\caption{South}
\label{fig:total-effect-regions--south}
\end{subfigure}
\begin{subfigure}[b]{0.49\textwidth}
\includegraphics[width=\textwidth,height=5cm]{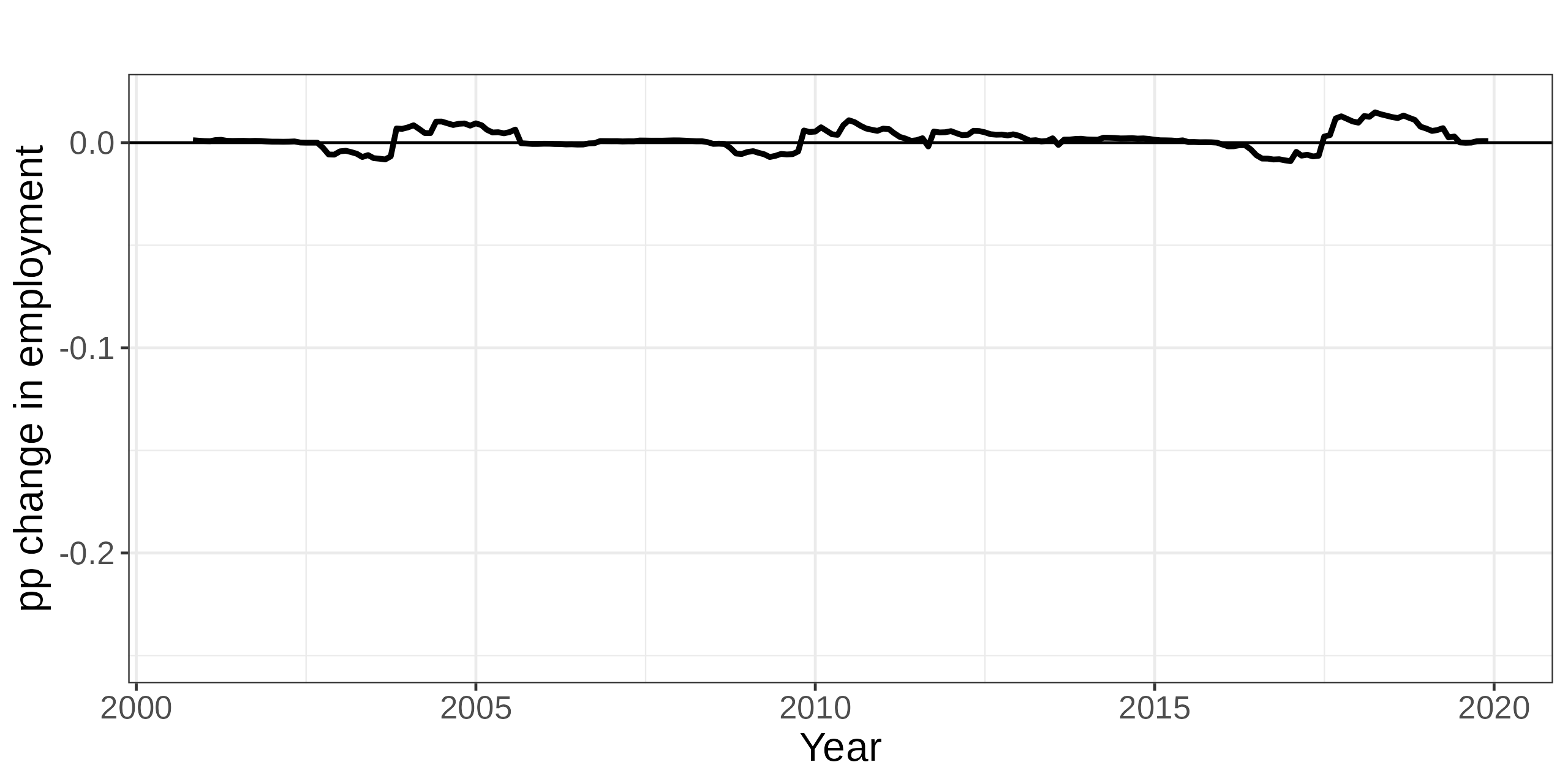}
\caption{Northeast}
\label{fig_northeast}
\end{subfigure}
\raggedright
\begin{singlespace}
\footnotesize{Simulated regional employment growth fluctuations attributable to fire activity over 2000--2021, based on the region's estimated IRF and observed county-level burn sequences.} \end{singlespace}
\end{figure}

The resulting simulated historical fire-driven regional employment fluctuations are shown in Figure \ref{fig:total-effect-regions}. The simulated historical employment impacts reveal the total cost of recurring fire exposure. As might be expected, the West region is most severely impacted on average, followed by the South, the Midwest, and finally the Northeast. Note that the estimated IRFs shown in Figure \ref{fig:region-IRFs} for the South and Northeast are not generally statistically significant. This is unsurprising given the high variation in fire activity across counties in the South and low rate of fire activity overall in the Northeast. 

Still, the magnitude of the effects are substantial. For example, the 2009-2010 fire season in the West is estimated to have cost about 0.25 pp of county-level employment growth on average compared to a no-fire counterfactual. Given that the  average county-level monthly employment in the West over our sample is about 0.07 pp (see Table \ref{tab:empg_summary}), recovering the lost employment growth from the 2009-2010 fire season would have required 3-4 months to be added the year. 

This technique can be extended to forecast fire-driven employment losses in a set of counties given a projected burn area sequence $(\hat{D}_{c,0}, \dots, \hat{D}_{c,h})$. This is enabled by the fact that impulse responses estimated by LP models recover marginal causal effects given the shock's natural evolution \citep{alloza2020dynamic, dube2022local}. Though beyond our scope here, developing such projections is an interesting avenue for future work. 

\subsection{Spatial spillovers}
\label{sec:spatial}
Finally, we conduct robustness checks to assess the degree to which spatial spillovers---employment effects of fires transmitted across county lines---may affect our results. We utilize two strategies to control for such spillovers: controlling for smoke exposure, and controlling for county-level spatial lags of burn area. Each addresses different sources of spatial spillovers. In both cases, we find qualitatively-similar results.

\subsubsection{Smoke and long-range spillovers}

Recent economic literature on wildfires has focused on the effects of smoke on labor market outcomes, finding that smoke exposure reduces earnings and employment \citep{borgschulte2022air}. This poses a potential challenge to identifying the causal effect of fires on labor market activity, as smoke from fires does not remain localized to the county in which it was generated. The thought experiment below describes this issue in more detail. 

Consider two otherwise-identical counties: one which experiences a fire impulse at time $t$ ($A$), and one which does not ($B$). If smoke from the impulse in $A$ at $t$ drifts into $B$ at $t$, then the comparison of fire-related employment outcomes between $A$ and $B$ at $t$ will be distorted as $B$ is no longer a valid control for $A$. Given results from \citet{borgschulte2022air}, this SUTVA violation would bias our estimated effects towards zero. To see this, suppose the entirety of the fire effect was driven by smoke, and wind patterns gave $A$ and $B$ identical smoke exposure ($\mu g / m^3 / day$). The estimated fire IRF would be flat, since there would be no difference in labor market outcomes between the two counties. To the extent that the labor market effects of smoke exposure accumulate or persist beyond one month, a similar issue could arise if $B$ experiences smoke exposure in the months leading up to the fire impulse in $A$. In that case, labor market activity in $B$ would be depressed at the time of the impulse, attenuating the estimated fire effect from $A$. 

\begin{figure}[htpb]
\begin{center}
\caption{Response of employment growth to an increase in burn area with and without smoke lags}  \label{fig:emp_smoke} 
\includegraphics[height=7cm,width=0.8\textwidth]{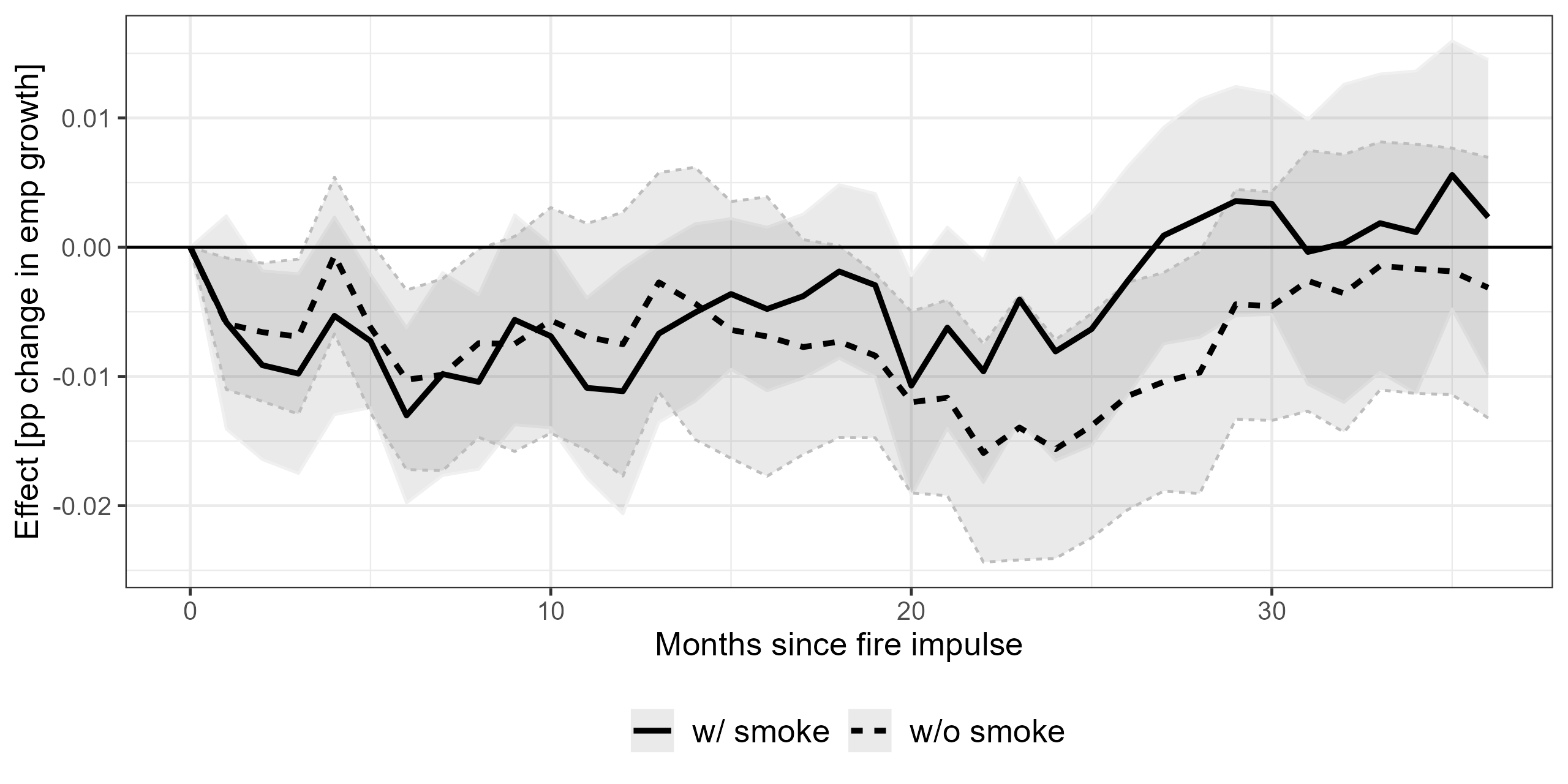}
\end{center}
\vspace{-0.4cm}
\raggedright
\begin{singlespace}
\footnotesize{The $y$ axis shows percentage point changes in employment growth in response to a burn area impulse of about 13 km$^2$---the mean burn area in counties that experienced fires---with and without smoke exposure controls. Shaded areas indicate 95\% CIs computed using DK standard errors. }\end{singlespace}
\end{figure}

To address these issues, we control for contemporaneous and lagged smoke exposure. Contemporaneous exposure addresses the channel described above, while lagged exposure controls for lingering or accumulating effects of earlier smoke exposure.\footnote{Controlling for contemporaneous smoke exposure but not lagged exposure would mechanically distort the path of the IRF: as $B$ recovers at a different pace than $A$, the growing gap between the two would be incorrectly attributed to the fire impulse in $A$ at $t$.} We find qualitatively similar effects in all cases. Figure \ref{fig:emp_smoke} shows the baseline employment IRF from figure \ref{fig:base-model-result} with and without smoke controls. The model with lagged and contemporaneous smoke controls shows a slightly larger decline in employment growth in the short-run and a smaller decline in the medium-run although the difference between the estimated coefficients is not statistically significant. 

\subsubsection{Borders and short-range spillovers}

As Figure \ref{fig:fire_maps} shows, burn area is spatially correlated. In addition, there may be employment spillover effects as people and jobs move across county lines in response to fire damages and suppression or recovery efforts. Such migration may occur if individuals have strong locational preferences (e.g. due to family networks) but wish to reduce fire exposure following a shock, or if firms/establishments relocate locally. 

We model these spillover effects by augmenting equation \ref{eqn:base-model} with controls for fire exposure in counties neighboring $c$ and counties neighboring $c$'s neighbors (i.e. county $c$'s first two spatial lags) and \emph{their} temporal lags. The spatial lags are constructed as the product of a spatial weight matrix $W$ with county fire exposures $D_{c,t}$. A typical element in $W$ is 1 if two counties share a border and 0 otherwise, so $W D_{c,t}$ measures fire exposure in counties neighboring $c$. Our specification becomes 
\begin{equation} \label{eqn:spatial_model}
    y_{c,t+h} - y_{c,t-1} =  \beta_{0,h} D_{c,t} + W D_{c,t} \beta_{1,h} + W^2 D_{c,t} \beta_{2,h} + X'_{c,t} \gamma_h + \alpha_{c,h} + \mu_{t,h} + \epsilon_{c,t+h},
\end{equation} 

where $W^2$ is the lagged weight matrix reflecting second-degree adjacencies (elements are 1 if two counties share a neighbor and 0 otherwise, with the diagonal normalized to 0).\footnote{Formally, \[W^2=W \times W.\] The diagonal may contain values larger than 1 due to loops from $c$ to its neighbors and back to $c$, so is removed to leave only the off-diagonal elements reflecting second-degree adjacencies.} $\beta_{0,h}$ is the effect of fires in $c$ on employment growth in $c$ (the ``own effect''), $\beta_{1,h}$ is the effect of fires in $c$'s neighbors on employment growth in $c$ (the ``first-degree neighbor effect''), and $\beta_{2,h}$ is the effect of fires in $c$'s neighbors' neighbors on employment growth in $c$ (the ``second-degree neighbor effect''). The set of controls $X_{c,t}$ in equation \ref{eqn:spatial_model} includes 24 lags of burn area in first- and second-degree neighbors to control for spatiotemporal correlations in fire activity. 

\begin{figure}[!htbp]
\begin{center}  
\caption{Response of employment growth to an increase in burn area with and without two spatial lags} \label{fig:spat_spillover_own_effects}
\includegraphics[height=7cm,width=0.8\textwidth]{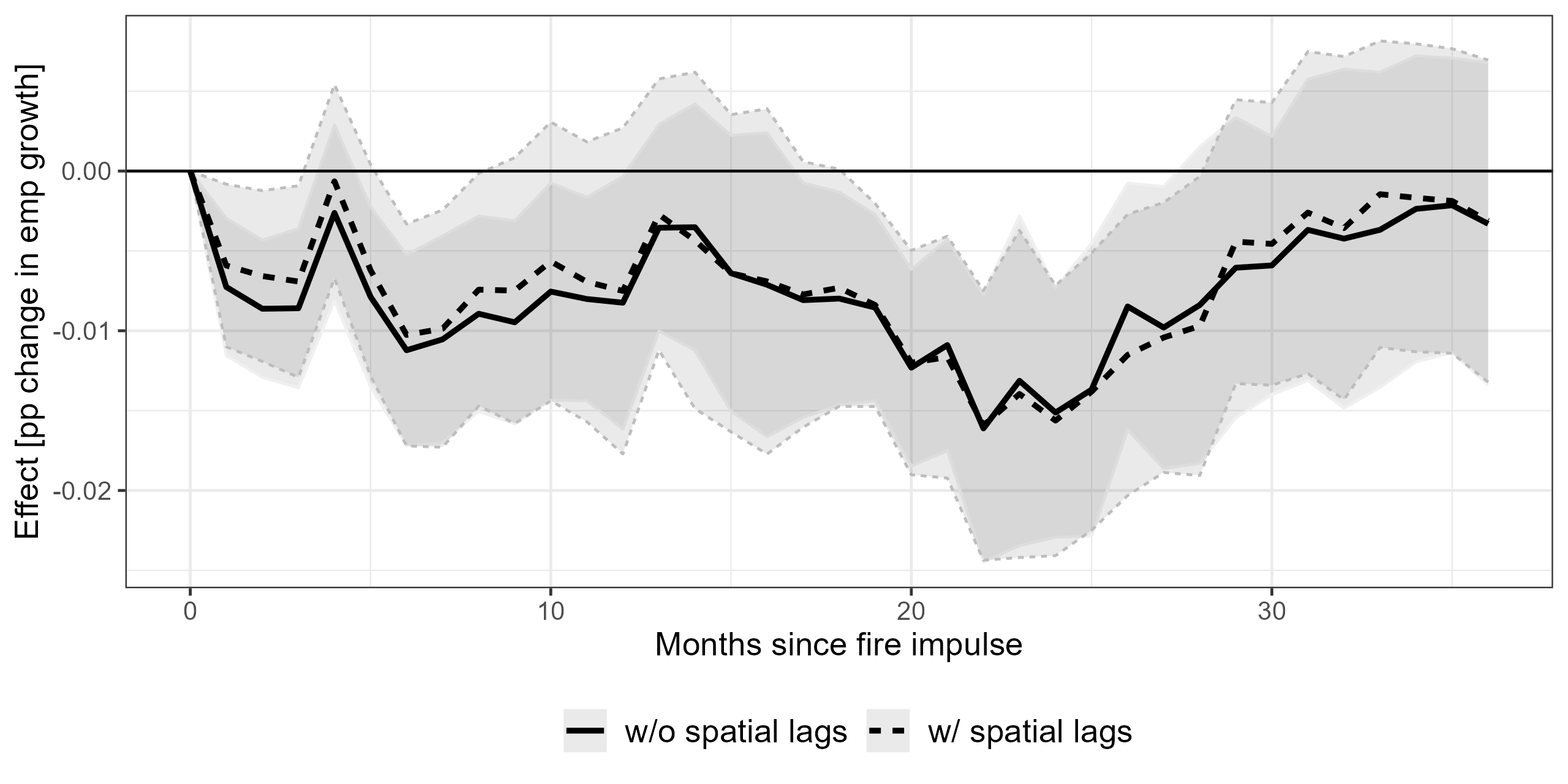}
\end{center}
\vspace{-0.4cm}
\raggedright
\begin{singlespace}
 \footnotesize{Impulse responses of fires in a county on its employment with and without controls for fires in neighboring counties. The $y$ axis shows percentage changes in response to a burn area impulse of about 13 km$^2$---the mean burn area in counties that experienced fires. Shaded areas indicate 95\% CIs computed using DK standard errors.} \end{singlespace}
\end{figure}

With spatial lags included, $\beta_{0,h}$ now represents the effect of fire impulses only in county $c$ holding fire exposure in $c$'s first- and second-degree neighbors constant. Figure \ref{fig:spat_spillover_own_effects} shows the own effect with and without controlling for fires in neighboring counties. Comparison reveals that spillovers from fires in neighboring counties appear to slightly offset the short-run losses of fires in a given county without much effect on the medium-run losses. The latter suggests that migration responses to fires do not involve individuals fine-tuning their fire risk exposure by relocating to nearby counties. 

\begin{figure}[!htbp]
\centering
\caption{Response of employment to an increase in fire exposure in first- and second-degree neighbors} \label{fig:spat_spillover_neighbor_effects} 
\begin{subfigure}[b]{0.49\textwidth}
\includegraphics[width=\textwidth,height=5.5cm]{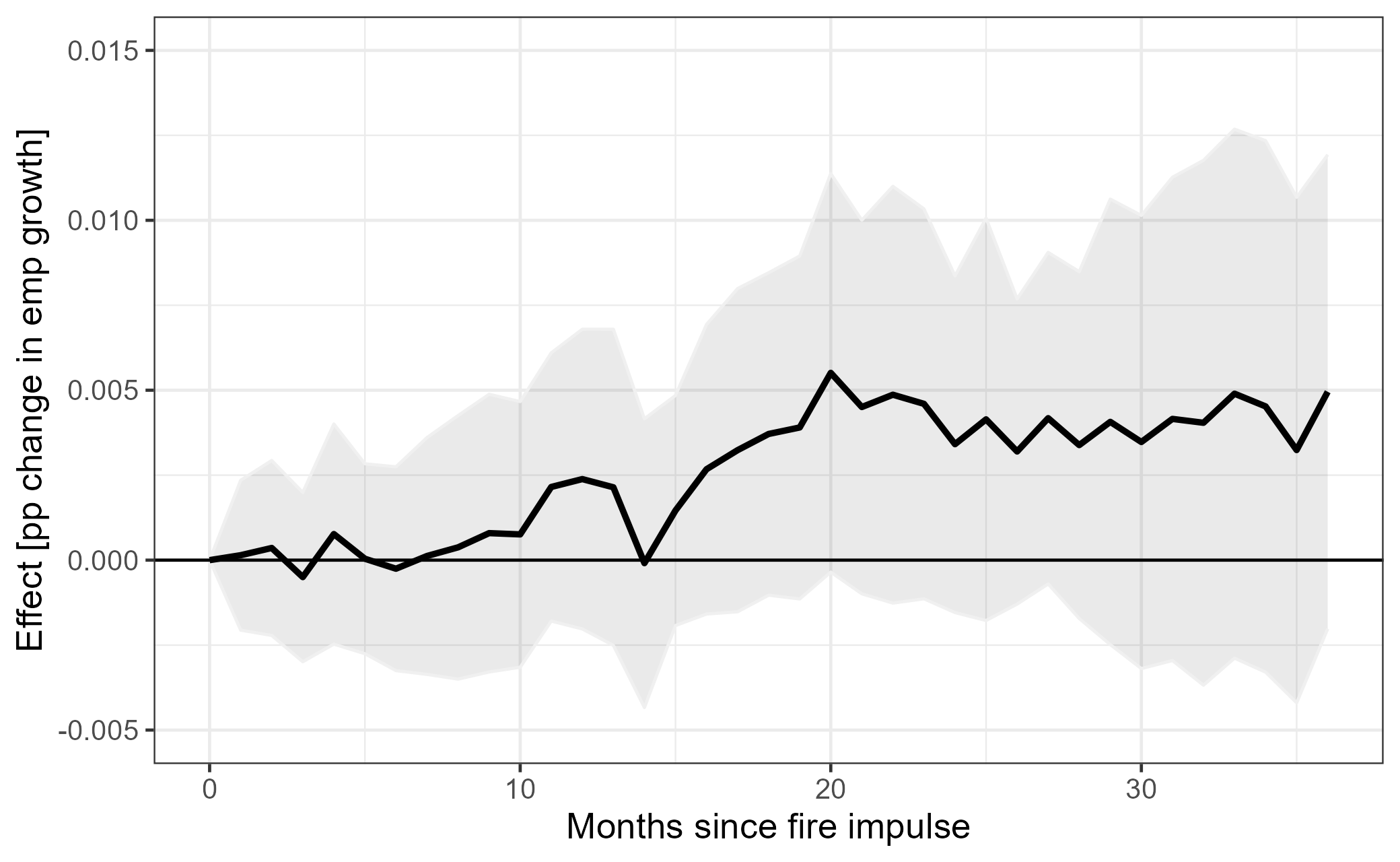}
\caption{First-degree neighbor}
\label{fig_firstneigh}
\end{subfigure}
\begin{subfigure}[b]{0.49\textwidth}
\includegraphics[width=\textwidth,height=5.5cm]{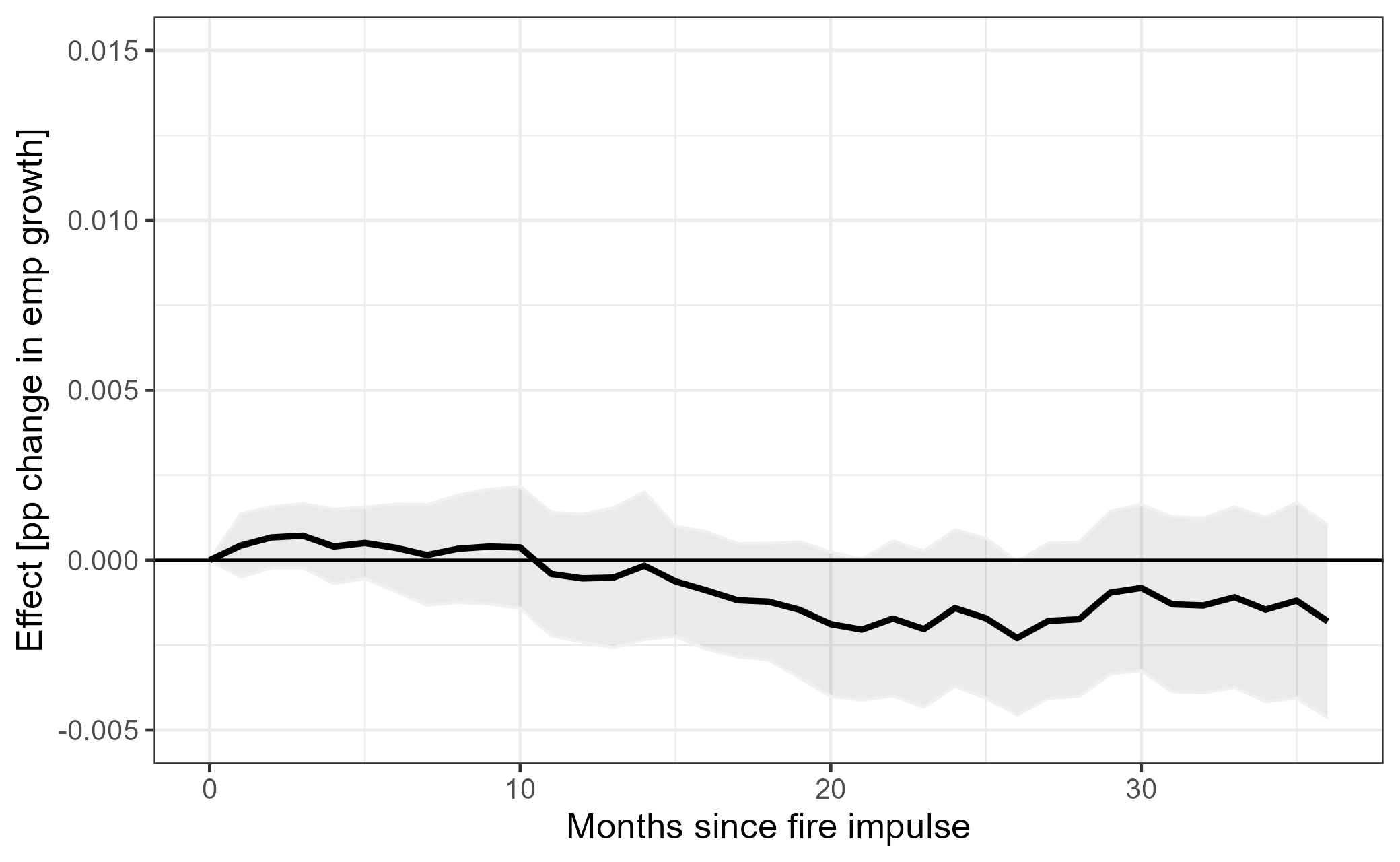}
\caption{Second-degree neighbor}
\label{fig:secneigh}
\end{subfigure}
\raggedright
\begin{singlespace}
 \footnotesize{Impulse responses of fires in a county on employment in its neighbors (first-degree) and neighbors' neighbors (second-degree). The $y$ axis shows percentage changes in response to a burn area impulse of about 13 km$^2$---the mean burn area in counties that experienced fires. Shaded areas indicate 95\% computed using DK standard errors. Covariates include county and year-month fixed effects, 24 monthly lags of county employment and burn area, as well as twelve monthly lags of burn area in first- and second- degree neighbors.} \end{singlespace}
\end{figure}

Figure \ref{fig:spat_spillover_neighbor_effects} shows the effects of fires in a county $c$'s first- and second-degree neighbors on monthly employment growth in $c$. The effects are not statistically significant at the 5\% level over a three-year post-fire horizon and grow noisier over time. The second-degree neighbor effects appear to be more precisely estimated ``zeros''.  

\section{Conclusion}
\label{sec:conclusion}
In this paper we measure the dynamic monthly impact of fires on county-level labor markets across the United States using a novel and granular geophysical measure of fire exposure. We show that employment growth decreases immediately for around a quarter after a fire impulse, and decreases again beginning around a year after the impulse, bottoming out roughly two years later. The magnitude of the effect can be significant: a 13 km$^2$ increase in burn area---the average burn area in counties that experience fires---can erase around 7\% of monthly employment growth on impact and around 17\% of monthly employment growth two years later. The dynamics of this effect may be explained by both short-run ($\approx$1-7 months after the impulse) net job destruction and medium-run ($\approx$2 years after the impulse) net out-migration. These effects typically fade in the long run ($\approx$3 years after the impulse). 

We document that the employment growth responses vary systematically by county education levels, industrial concentration, and the state of the county's business cycle. Counties with lower levels of high school diploma attainment, greater industrial concentration, or in high-slack states at the time of the fire impulse experience more significant negative effects. Counties with higher levels of high school diploma attainment, lower industrial concentration, or in low-slack states at the time of the fire impulse experience no significant negative effects and may even see positive effects. These results suggest that the structure and state of a county's labor market is a critical determinant of its economic resilience to shocks. An alternate, perhaps complementary, explanation is that counties which do not experience significant effects receive greater amounts of recovery and relief funding or are better able to deploy such funds.

Though our results appear robust to likely threats to identification, the mechanisms through which these changes to employment growth occur could be investigated further. The impulse responses of monthly employment growth and annual net out-migration are suggestive of short-run job destruction and medium-run flight from fires, they are not conclusive. Establishment-level data on jobs and vacancies or higher-frequency migration data (ideally covering individuals who do not submit tax returns as well) would help in determining when and the degree to which job destruction or out-migration are responsible for the employment growth effects we identify. Similarly, granular data on housing transactions or new developments would help clarify the role of housing markets in these responses---e.g. \citet{mccoy2018fire} find $\approx$3-year-long declines in housing prices in Colorado following fires, while \citet{black2023wildfire} find reduced housing development following increases in wildfire risk salience. Additionally, though our fire exposure data is very spatially detailed, we lack similarly-detailed population concentration measures. Such measures could be used to estimate the per-capita effects of exposure to fire in regions smaller than counties, complementing existing analyses of the effects of smoke exposure on labor market outcomes \citep{borgschulte2022air}. Lastly, fiscal aid to states in response to fire exposure varies over time, and distribution and deployment within states varies across the country. The effects we measure are net of such funding deployments and do not isolate the effects of fires on employment in the absence of such spending. This packaging of effects limits the extent to which these results can predict the effects of marginal fire aid allocations. Detailed data on federal aid flows and utilization could address this. 

Our use of a geophysical fire measure facilitates integrating estimated impulse responses with burn area projections (e.g. from coupled weather-fire models) to project regional economic activity and its response to different fire patterns. Applied to the Nov 2000-May 2022 period, the results suggest that fires may have appreciably reduced overall employment growth in particularly fire-stricken regions, such as the West and South. Our effect magnitudes also suggest that government spending on fire prevention, suppression, and recovery efforts could have large fiscal multipliers, particularly when targeted to counties likely to experience larger negative effects. 

Measuring economic activity is among the oldest and most-central issues in economics, and fires are among the oldest types of shocks to human civilization. Climate change threatens to make fires both more frequent and more severe for ever-greater fractions of the population, making them an increasingly important source of economic fluctuations. Understanding how fires affect economic activity is an important step in adapting to a warming planet.

\newpage

\bibliographystyle{chicago}
\bibliography{lib}

\newpage

\begin{appendices}

\section*{Appendices}

\section{Summary statistics}
\label{apdx:summary-stats}

 \begin{figure}[htpb] 
 \caption{Total burn area in US counties split by industrial concentration level. Darker colors indicate larger burn areas within counties.} \label{fig:fire_maps_hhi}
 \centering
 \begin{subfigure}[b]{0.49\textwidth} \includegraphics[height=4cm,width=\textwidth]{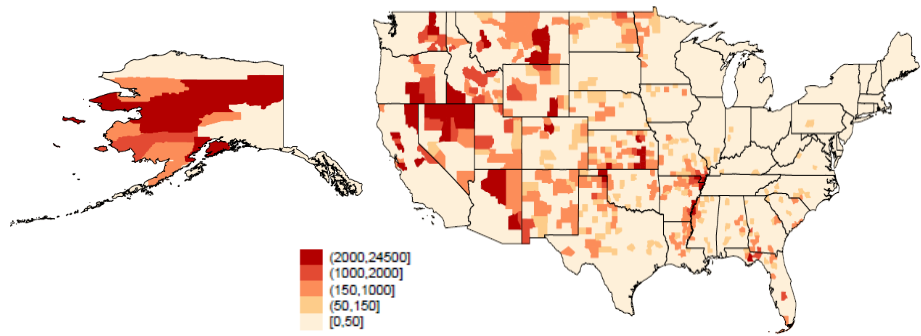}  
 \caption{High concentration}
 \end{subfigure}
 \begin{subfigure}[b]{0.49\textwidth}
\includegraphics[height=4cm,width=\textwidth]{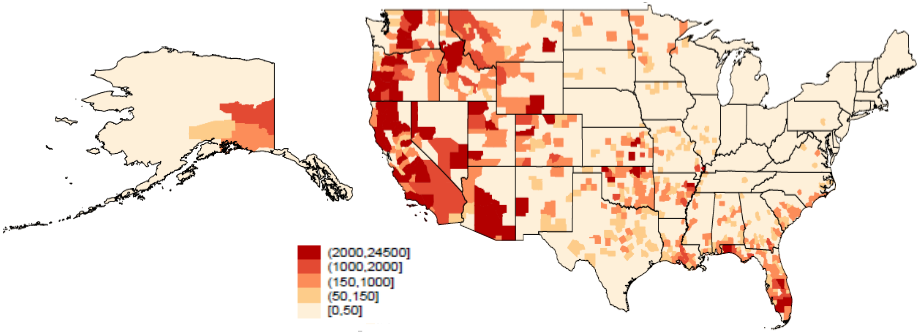}  
 \caption{Low concentration}
 \end{subfigure}
 \end{figure}
\vspace*{-0.5cm}
\begin{singlespace}
\noindent \footnotesize{The figure shows the total fire burn area in 500m$^2$ over the 2017--2021 period for high concentrated counties (left panel) and low concentrated counties (right panel).}
\end{singlespace}

\hspace{3cm}

\begin{figure}[htpb] 
 \caption{Total burned area in US counties split by education level. Darker colors indicate larger burned areas within counties.} \label{fig:fire_maps_edu}
 \centering
 \begin{subfigure}[b]{0.49\textwidth}
\includegraphics[height=4cm,width=\textwidth]{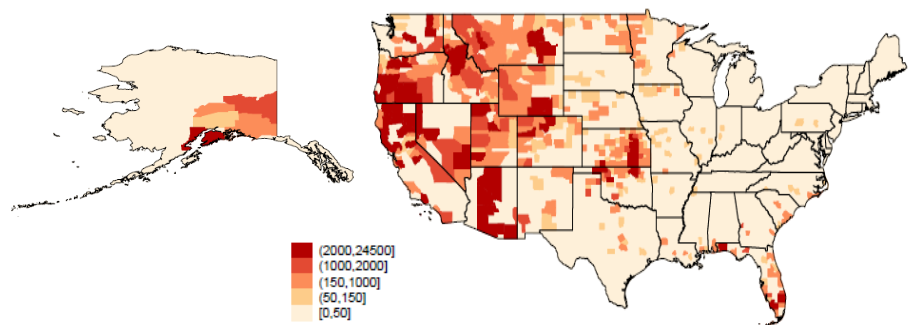}  
 \caption{High education}
 \end{subfigure}
 \begin{subfigure}[b]{0.49\textwidth}
\includegraphics[height=4cm,width=\textwidth]{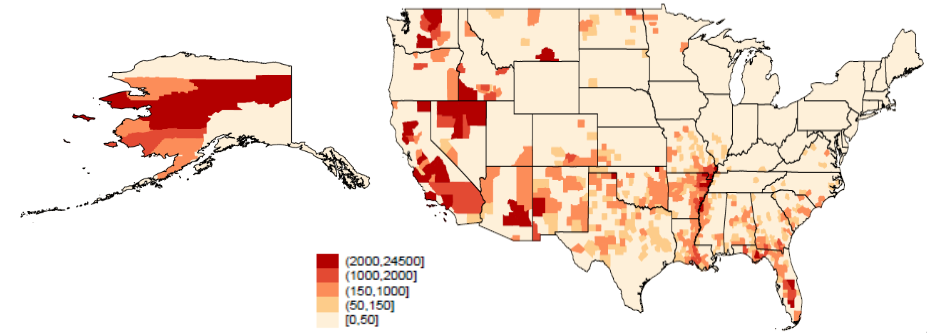}  
 \caption{Low education}
 \end{subfigure}
 \end{figure}
\vspace*{-0.5cm}
\begin{singlespace}
\noindent \footnotesize{The figure shows the total fire burn area in 500m$^2$ over the 2017--2021 period for high educated counties (left panel) and low educated counties (right panel).}
\end{singlespace}

\newpage
\begin{table}[!htpb] 
\caption{Industry composition by county characteristics}
\label{table:indcomp}
\centering
\hspace*{-1.5cm}
\begin{tabular}{lccccccc} 
  \hline
Sample & Natural & Goods  & Trade, trans.  & Professional  & Leisure and & Education     \\ 
 & resources & & and utilities & services & hospitality & and health \\
  \hline
\multicolumn{7}{l}{Share w/o HS degree} \\
\hspace{0.25cm} above median & 7.5\% & 24.0\% & 19.1\% & 7.4\% & 6.2\% & 22.1\% \\
\hspace{0.25cm} below median & 7.5\% & 22.0\% & 18.2\% & 10.6\% & 7.1\% & 21.8\% \\
 & & & & & & \\
Industrial concentration \\
\hspace{0.25cm} above median & 9.4\% & 22.7\% & 17.4\% & 8.0\% & 6.2\% & 23.7\% \\
\hspace{0.25cm} below median & 5.3\% & 23.3\% & 19.9\% & 10.0\% & 7.1\% & 20.3\% \\
    \hline
\end{tabular}
\end{table}
\vspace*{-0.5cm}
\begin{singlespace}
\noindent \footnotesize{The table shows the average share of employment across industry types by county characteristics. Natural resources includes agriculture and mining (sectors 11 and 21). `Goods' refers to goods-producing industries, which are construction and manufacturing (sectors 23, 31-33).}
\end{singlespace}

\begin{table}[htpb]
\caption{Herfindahl-Hirschman Index} \label{tab:HHI-summary}
\begin{tabular}{cll@{\hskip 0.5in}ccll} \hline 
\multicolumn{3}{c}{Low   concentration (bottom 10)} &  & \multicolumn{3}{c}{High concentration (top 10)} \\ \hline
1     & Storey County (NV)  & 0.110    &  & 1    & Petroleum County (MT)       & 0.886  \\
2     & Park County (CO)       & 0.123 &  & 2    & Clark County (ID)  & 0.801  \\
3     & Lyon County (NV)       & 0.124    &  & 3    & Los Alamos County (NM)        & 0.779  \\
4     & Cameron Parish (LA)       & 0.125    &  & 4    & Loving County (TX)   & 0.704  \\
5     & Logan County (OK)         & 0.128    &  & 5    & Slope County (ND)         & 0.693  \\
6     & Jefferson County (ID)      & 0.128    &  & 6    & Carter County (MT)       & 0.692 \\
7     & Lee County (TX)        & 0.129    &  & 7    & King County (TX)        & 0.611  \\
8     & Amador County (CA)         & 0.130    &  & 8    & Reagan County (TX)        & 0.603  \\
9     & Currituck County (NC)       & 0.130    &  & 9    & Kenedy County (TX)      & 0.592  \\
10    & Adams County (CO)    & 0.130    &  & 10   & Lander County (NV)   & 0.580 \\ \hline 
\end{tabular}
\end{table}
\vspace*{-0.5cm}
\begin{singlespace}
\noindent \footnotesize{The table shows the Herfindahl-Hirschman Index for the bottom ten (left panel) and top ten (right panel) counties.}
\end{singlespace}

\begin{table}[!htpb] 
\caption{Average outcomes of interest by county characteristics}
\label{table:descstat}
\centering
\hspace*{-1.5cm}
\begin{tabular}{lcccccc} 
  \hline
Sample & Fire  & Emp-to-pop  & Population  & Income  & Urban    \\ 
 & [km$^2$] & [\%] & [people] & [\$] & [\%] \\
  \hline
Full sample & 723,845 & 45.5\% & 92,639 & 46,726.7 & 42.5\% \\
 & (22.50) & (7.1\%) & (299,025) & (98,885.7) & (32.0\%) \\
 & & & & & \\
\multicolumn{6}{l}{Share w/o HS degree} \\
\hspace{0.25cm} above median & 354,344 & 42.2\% & 66,200 & 35,402.8 & 36.2\% \\
\hspace{0.25cm} below median & 369,501 & 48.8\% & 119,637 & 58,873.5 & 49.2\% \\
 & & & & & \\
Industrial concentration \\
\hspace{0.25cm} above median & 322,294 & 45.4\% & 73,947 & 39,261.1 & 39.0\% \\
\hspace{0.25cm} below median & 401,551 & 45.6\% & 111,201 & 54,206.3 & 46.0\% \\
    \hline
\end{tabular}
\end{table}
\vspace*{-0.5cm}
\begin{singlespace}
\noindent \footnotesize{Total fire burn area (in km$^2$) and average employment-to-population ratio, number of people (rounded to the nearest integer), GDP per capita, and percentage of population living in an urban area by county characteristics. Standard errors are in parentheses. Industrial concentration is measured using the Herfindahl-Hirschman Index (HHI).}
\end{singlespace}

\begin{table}[!htpb] 
\caption{Correlation matrix of county characteristics}
\label{table:correlation}
\centering
\hspace*{-1.5cm}
\begin{tabular}{lccccc} 
  \hline
 & HHI & Pop w/o HS degree  & Rural   & Income  & Population    \\ \hline 
HHI & 1.00 & & & & \\
Pop w/o HS degree & 0.00 & 1.00 & & & \\
Rural & 0.17 & 0.28 & 1.00 & & \\
Income & -0.08 & -0.13 & -0.41 & 1.00 & \\
Population & -0.07 & -0.10 & -0.40 & 0.99 & 1.00 \\ \hline 
\end{tabular}
\end{table}
\vspace*{-0.5cm}
\begin{singlespace}
\noindent \footnotesize{All values are rounded to two decimal places. Industrial concentration is measured using the Herfindahl-Hirschman Index (HHI).}
\end{singlespace}

\begin{table}[htpb]
\caption{Percentage of population without a high school diploma} \label{tab:edu-summary}
\begin{tabular}{cll@{\hskip 0.5in}lcll} \hline 
\multicolumn{3}{c}{Low   education (bottom 10)} &  & \multicolumn{3}{c}{High education (top 10)} \\ \hline 
1     & Kalawao County (HI)       & 60.5    &  & 1    & Douglas County (CO)       & 3.1  \\
2     & Maverick County (TX)       & 57.9    &  & 2    & Los Alamos County (NM)  & 3.6  \\
3     & Presidio County (TX)       & 55.3    &  & 3    & Pitkin County (CO)        & 3.7  \\
4     & Hudspeth County (TX)       & 53.9    &  & 4    & Routt County (CO)         & 4.7  \\
5     & Reeves County (TX)         & 53.4    &  & 5    & Johnson County (KS)         & 5.1  \\
6     & Owsley County (KY)      & 50.7    &  & 6    & Hamilton County (IN)       & 5.8  \\
7     & Clay County (KY)        & 50.6    &  & 7    & Banner County (NE)        & 5.8  \\
8     & Brooks County (TX)         & 50.2    &  & 8    & Gilpin County (CO)        & 5.9  \\
9     & La Salle County (TX)       & 49.9    &  & 9    & Gunnison County (CO)      & 5.9  \\
10    & Magoffin County (KY)    & 49.9    &  & 10   & Washington County (MN)   & 5.9 \\ \hline 
\end{tabular}
\end{table}
\vspace*{-0.5cm}
\begin{singlespace}
\noindent \footnotesize{The table shows the percentage of the population without a high school diploma for the bottom ten (left panel) and top ten (right panel) counties.}
\end{singlespace}

\section{Clean controls and negative weighting}

\label{apdx:robustness}

\begin{figure}[!htbp] 
\begin{center}
\caption{Response of employment to an increase in fire exposure}  \label{fig:clean_control} 
\includegraphics[height=7cm,width=0.8\textwidth]{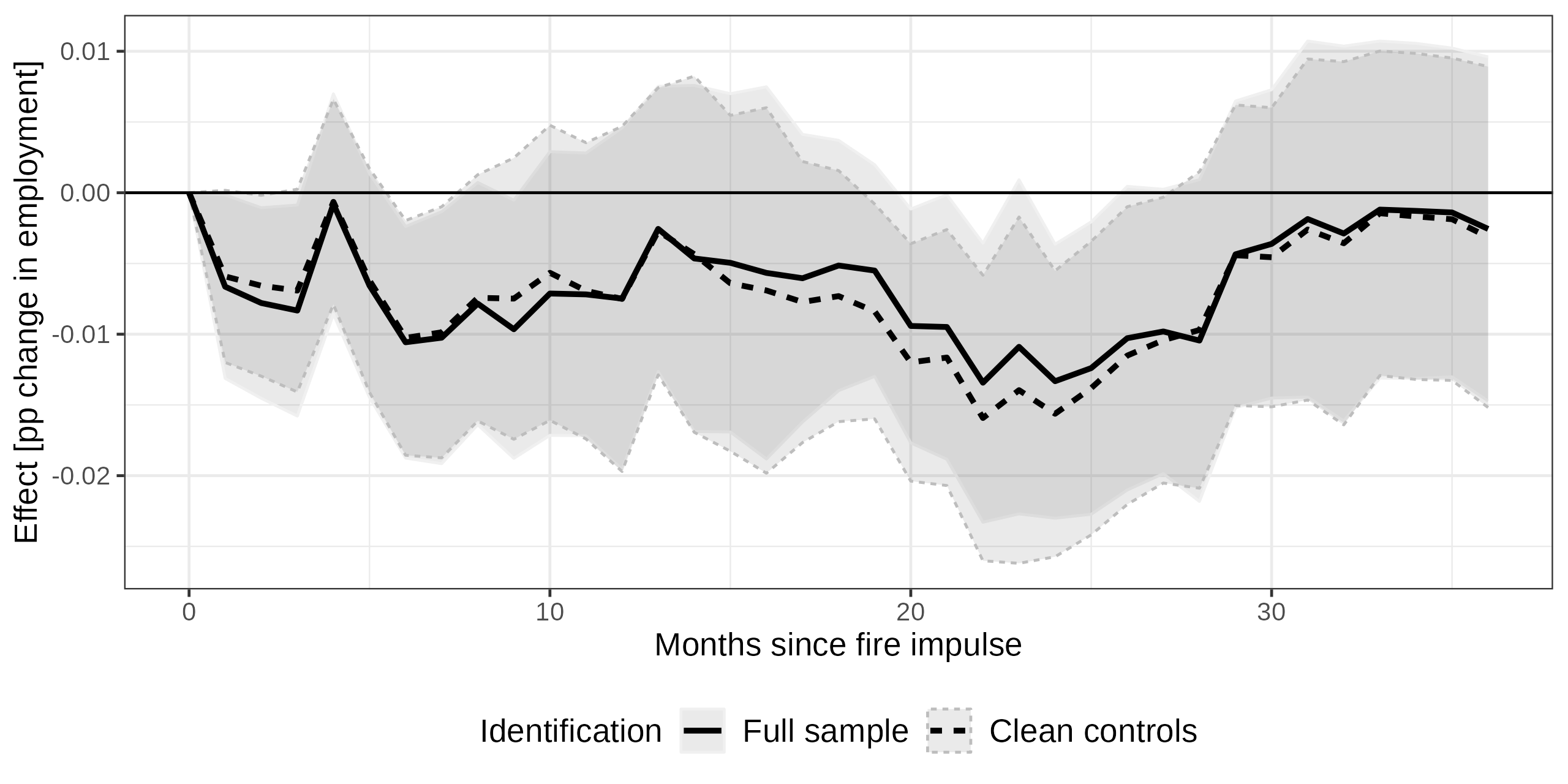}
\end{center}
\vspace{-0.4cm}
\raggedright
\begin{singlespace}
\footnotesize{The $y$ axis shows percentage changes in response to a burn area impulse of about 13 km$^2$---the mean monthly burn area in counties that experienced fires. Shaded areas indicate 95\% CIs.} \end{singlespace}
\end{figure}

To determine whether negative weighting issues are biasing our estimates, we construct ``clean control'' groups from counties which have not experienced a fire in the past 36 months and do not experience fires 36 months after a given impulse. This ensures that our identification does not come from any ``forbidden comparisons'', i.e. we do not compare treated counties to counties recovering from treatment or which receive treatment during the treated county's recovery period \citep{cengiz2019effect, dube2022local}.Figure \ref{fig:clean_control} shows the estimated IRF using only clean controls to the IRF estimated using the full sample. The results are similar under either identification strategy. \\

\section{Calculating the cumulative fire impulse effect}
\label{apdx:cumulative-effect}

Consider a county with $y_t$ people employed at time $t$. At time $0$ there is an impulse $dx$, and the growth rates from time $1$ onwards are altered by a period-specific amount $\beta_t$. Let $g_t$ denote the growth rate of $y_t$ relative to time $-1$, i.e. $g_t = \log(y_{t}) - \log(y_{-1}) $. Table \ref{tab:growth-rates} summarizes the observed (with shock) and counterfactual (without shock) growth rate.

\begin{table}[H]
\centering
\renewcommand{\arraystretch}{1.5}
\begin{tabular}{|c|c|c|c|c|}
\hline
\multirow{2}{*}{\textbf{Month}} & \multicolumn{2}{c|}{\textbf{Growth Rates}} & \multicolumn{2}{c|}{\textbf{Employment Level}} \\
\cline{2-5}
& \textbf{with shock} & \textbf{without shock} & \textbf{with shock} & \textbf{without shock} \\
\hline
$1$ & $g_{1} + \beta_1$ & $g_{1}$ & $y_t(g_1 + \beta_1)$ & $y_t g_1$ \\
\hline
$2$ & $g_{2} + \beta_2$ & $g_{2}$ & $y_t(g_2 + \beta_2)$ & $y_t g_2$ \\
\hline
$3$ & $g_{3} + \beta_3$ & $g_{3}$ & $y_t(g_3 + \beta_3)$ & $y_t g_3$ \\
\hline
\end{tabular}
\caption{Growth rates and employment levels with and without shock}
\label{tab:growth-rates}
\end{table}

The cumulative difference in the employment level due to the shock over months 1-3 is
\begin{align}
    \Phi(3) &= \left(y_t g_1 - y_t(g_1 + \beta_1)\right) + \left(y_t g_2 - y_t(g_2 + \beta_2)\right) + \left(y_t g_3 - y_t(g_3 + \beta_3)\right) \\
    &= y_t \beta_1 + y_t \beta_2 + y_t \beta_3,
\end{align}

which can be expressed in terms of the differential rate of employment growth due to the shock by dividing out $y_t$:
\begin{equation}
    \phi(3) = \beta_1 + \beta_2 + \beta_3.
\end{equation}

Suppose the change in months $t-1$ and $t$ is 0 by definition (shock hadn't happened at $t-1$, assume shock effect is only measurable after period $t$ when it occurs). Then, in general, the ``cumulative fire impulse effect'' out to horizon $H$---expressed as the cumulative change in the employment growth rate attributed to the shock---is:
\begin{equation}
    \phi(H) = \sum_{h=1}^H \beta_h. \label{eqn:fire-multiplier}
\end{equation}

We are interested in the sensitivity of this quantity to the sample definition. Is the cumulative fire impulse effect's magnitude due to a few influential counties? To address this, we compute the mean and standard deviation of $\phi(H)$ over random subsets of counties (holding the time dimension fixed) as
\begin{align}
    E[\phi(H)] &= \sum_{h=1}^H E[\beta_h] ,\\
    V[\phi(H)] &= \sum_{h=1}^H V[\beta_h] + 2 \sum_{i=1}^N \sum_{j=i+1}^{N} Cov[\beta_i,\beta_j].
\end{align}

Calculating the mean of the cumulative fire impulse effect is straightforward---equation \ref{eqn:fire-multiplier} offers an unbiased estimator. Though the variance terms $V[\beta_h]$ could similarly be calculated from the local projections, the covariance terms $Cov[\beta_i,\beta_j]$ are more challenging. We therefore utilize a ``block jackknife'' approach similar to the general weighted jackknife described in \citet{wu1986jackknife} with uniform weights over counties.\footnote{While \citet{wu1986jackknife} recommends weighting observations proportional to the determinant of the Fisher information matrix of the subset, it is not obvious how to extend this weighting scheme to county blocks. Given temporal dependence between observations, dropping observations as though they were IID would be inappropriate.} The procedure for a single sequence of local projection estimates $\{\beta_t\}_{t=1}^H$ is as follows:

\begin{enumerate}
\item Drop 5\% of counties uniformly at random and estimate the local projection. Label these estimates $\{\beta^{(1)}_t\}_{t=1}^H$.
\item Repeat this process $K$ times for different randomly-selected 95\% subsets.
\item Using the set of $K$ local projection estimates, estimate the covariance matrix $Cov[\beta_i,\beta_j]$ as $Cov_K[\beta^{(k)}_i,\beta^{(k)}_j]$ (i.e. the sample covariance of the block jackknife estimates $\{\{\beta^{(k)}_t\}_{t=1}^H\}_K$).
\end{enumerate}

The resulting covariance matrix identifies sensitivity of the cumulative fire impulse effect to exclusion of any subsample of 5\% of counties \citep{kaffine2017multi}.\footnote{Given the large size of our data, the number of horizons we consider, and the set of control variables we use---$H=36$, approximately 850,000 observations, 24 horizon-specific lags of employment and burn area, and horizon-specific county and year-month fixed effects---a stacked regression approach (e.g. as described in \citet{dube2022local}) is computationally infeasible and does not directly address the question of sensitivity to influential counties. The block jackknife procedure is much less computationally demanding and directly addresses the question of sensitivity to influential counties.} We compute the cumulative fire impulse effect directly from the full-sample local projection estimates. We use 1000 jackknife draws for the covariances, and compute the standard deviation of $\phi(H)$ from the covariance matrix of the $\{\{\beta^{(k)}_t\}_{t=1}^H\}_K$ draws.

\section{Additional figures and tables}
\label{apdx:additional-figures-tables}

In this section we present additional figures and tables referenced in the text. These results offer more insight into the nature of the effects we identify.

\subsection{Fire size}

\begin{figure}[!htpb] 
\begin{center}
\caption{Response of employment to an increase in fire exposure by fire size} \label{fig:fire-size-heterogeneity-model}
\begin{subfigure}[b]{0.31\textwidth}
\includegraphics[width=\textwidth,height=4.5cm]{images/Paper/fire_top_1_dk_24_l_emp_fire.png}
\caption{Top 1\%}
\label{fig_top1}
\end{subfigure}
\begin{subfigure}[b]{0.31\textwidth}
\includegraphics[width=\textwidth,height=4.5cm]{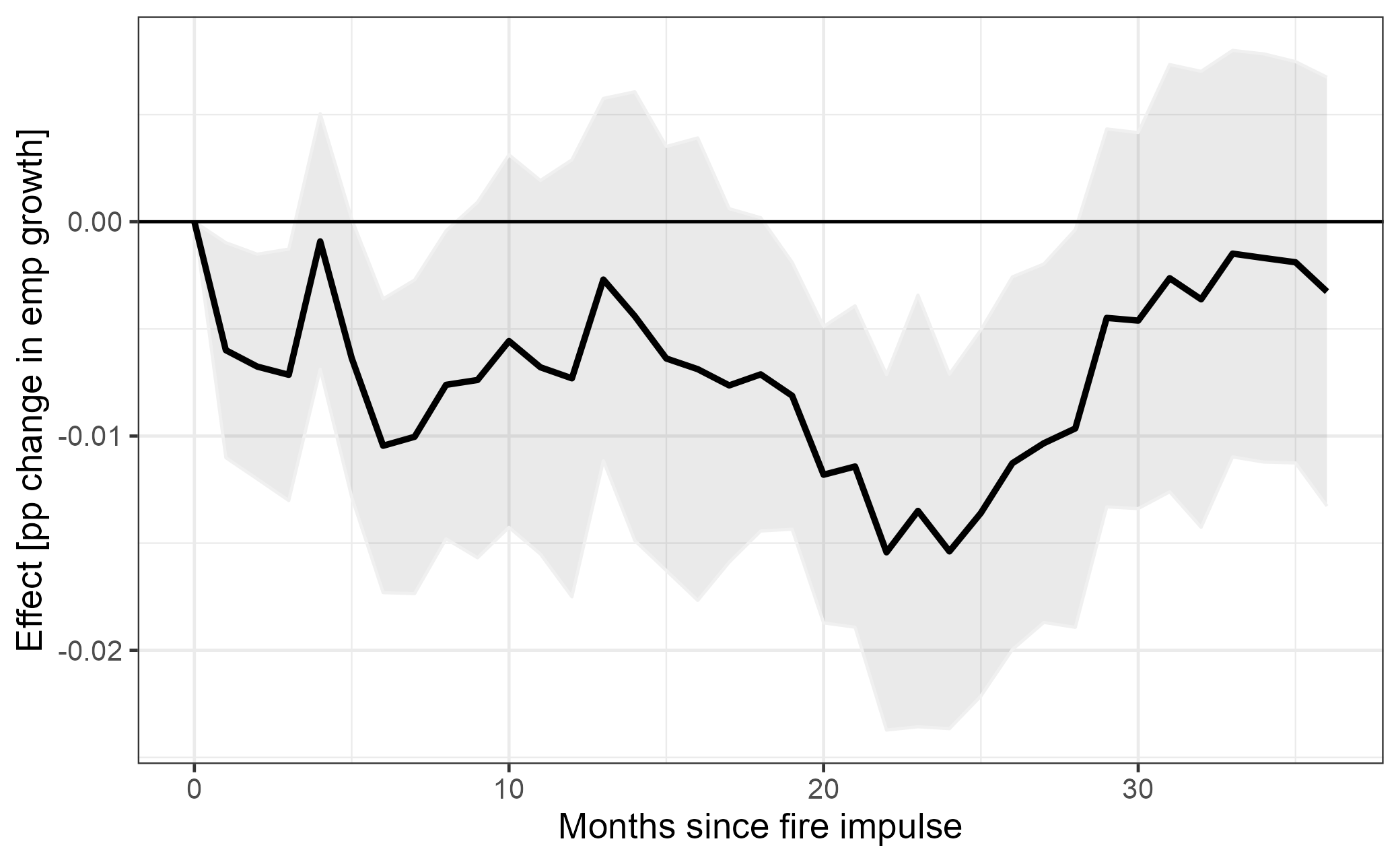}
\caption{Top 5\%}
\label{fig_top5}
\end{subfigure}
\begin{subfigure}[b]{0.31\textwidth}
\includegraphics[width=\textwidth,height=4.5cm]{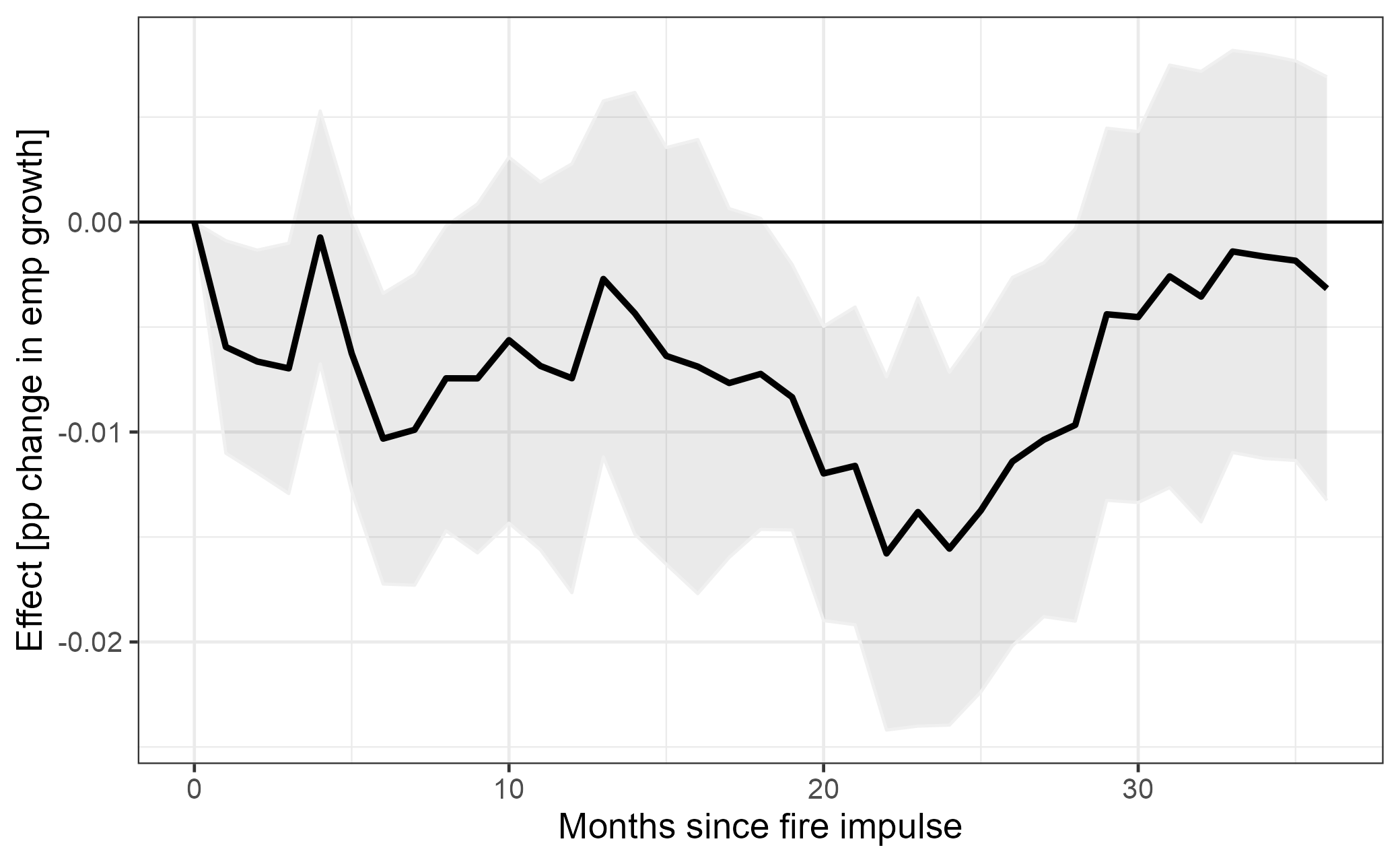}
\caption{Top 10\%}
\label{fig_top10}
\end{subfigure}
\begin{subfigure}[b]{0.31\textwidth}
\includegraphics[width=\textwidth,height=4.5cm]{images/Paper/fire_bottom_99_dk_24_l_emp_fire.png}
\caption{Bottom 99\%}
\label{fig_bottom99}
\end{subfigure}
\begin{subfigure}[b]{0.31\textwidth}
\includegraphics[width=\textwidth,height=4.5cm]{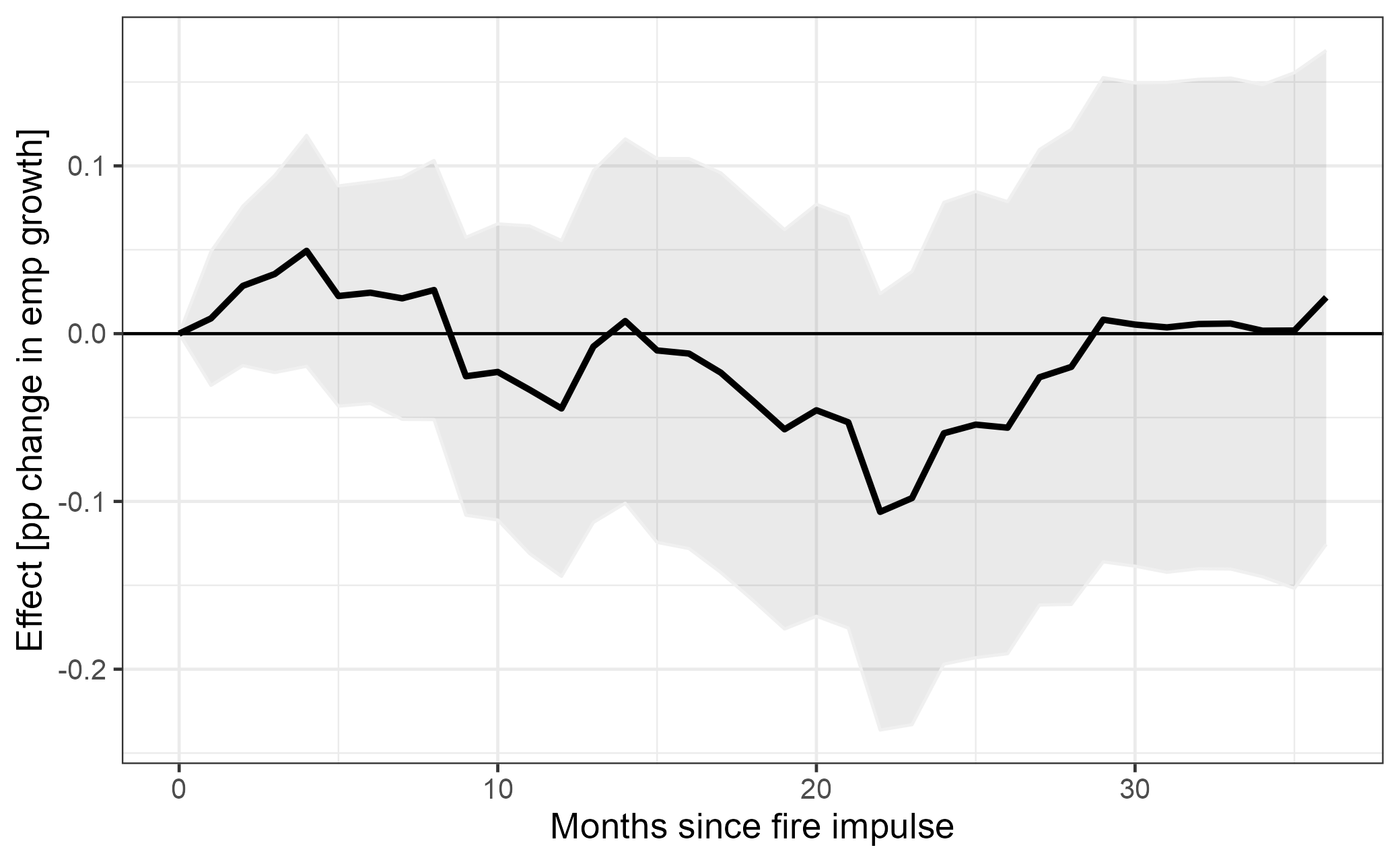}
\caption{Bottom 95\%}
\label{fig_bottom95}
\end{subfigure}
\begin{subfigure}[b]{0.31\textwidth}
\includegraphics[width=\textwidth,height=4.5cm]{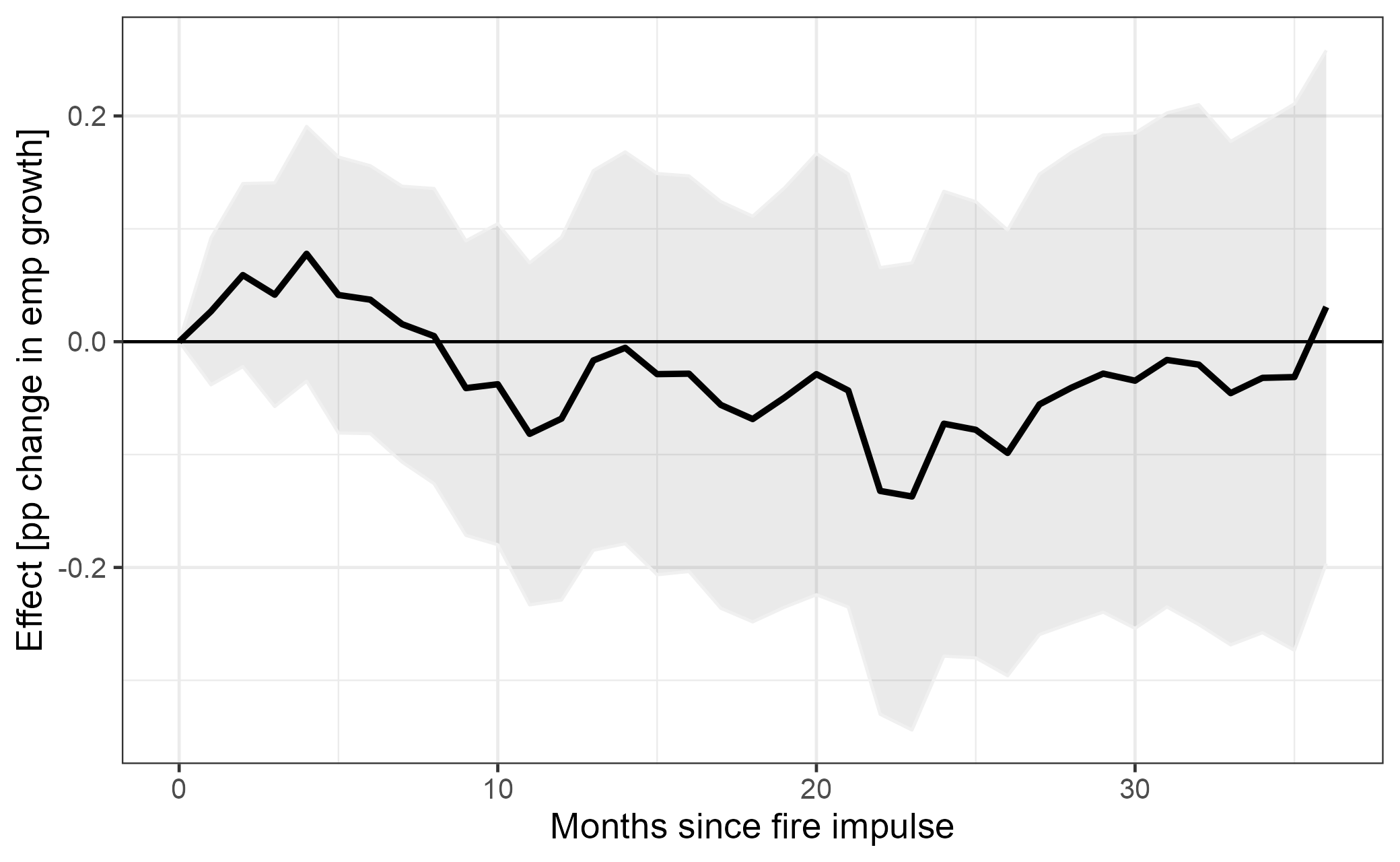}
\caption{Bottom 90\%}
\label{fig_bottom90}
\end{subfigure}
\end{center}
\raggedright
\begin{singlespace}
\footnotesize{The $y$ axis shows percentage changes in response to a burn area impulse of about 13 km$^2$---the mean burn area in counties that experienced fires. Shaded areas indicate 95\% CIs computed DK standard errors. Covariates include county and year-month fixed effects and 24 monthly lags of county employment and burn area.} \end{singlespace}
\end{figure}

\subsection{Migration responses}

\begin{figure}[!htpb] 
\begin{center}
\caption{Migration responses to an increase in fire exposure} \label{fig:mig-app}
\begin{subfigure}[b]{0.45\textwidth}
\includegraphics[width=\textwidth,height=5cm]{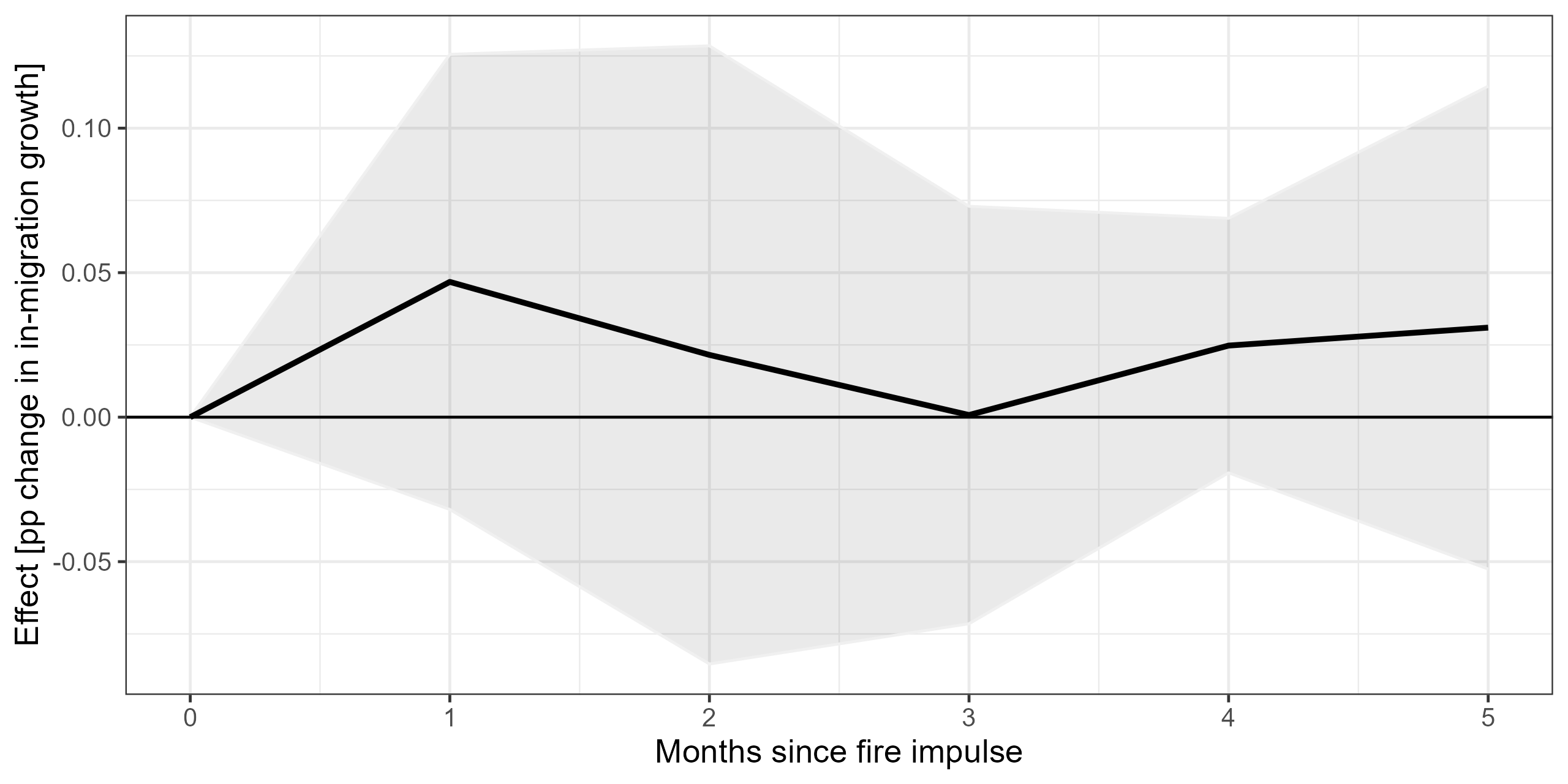}
\caption{In-migration}
\label{fig_inmig}
\end{subfigure}
\begin{subfigure}[b]{0.45\textwidth}
\includegraphics[width=\textwidth,height=5cm]{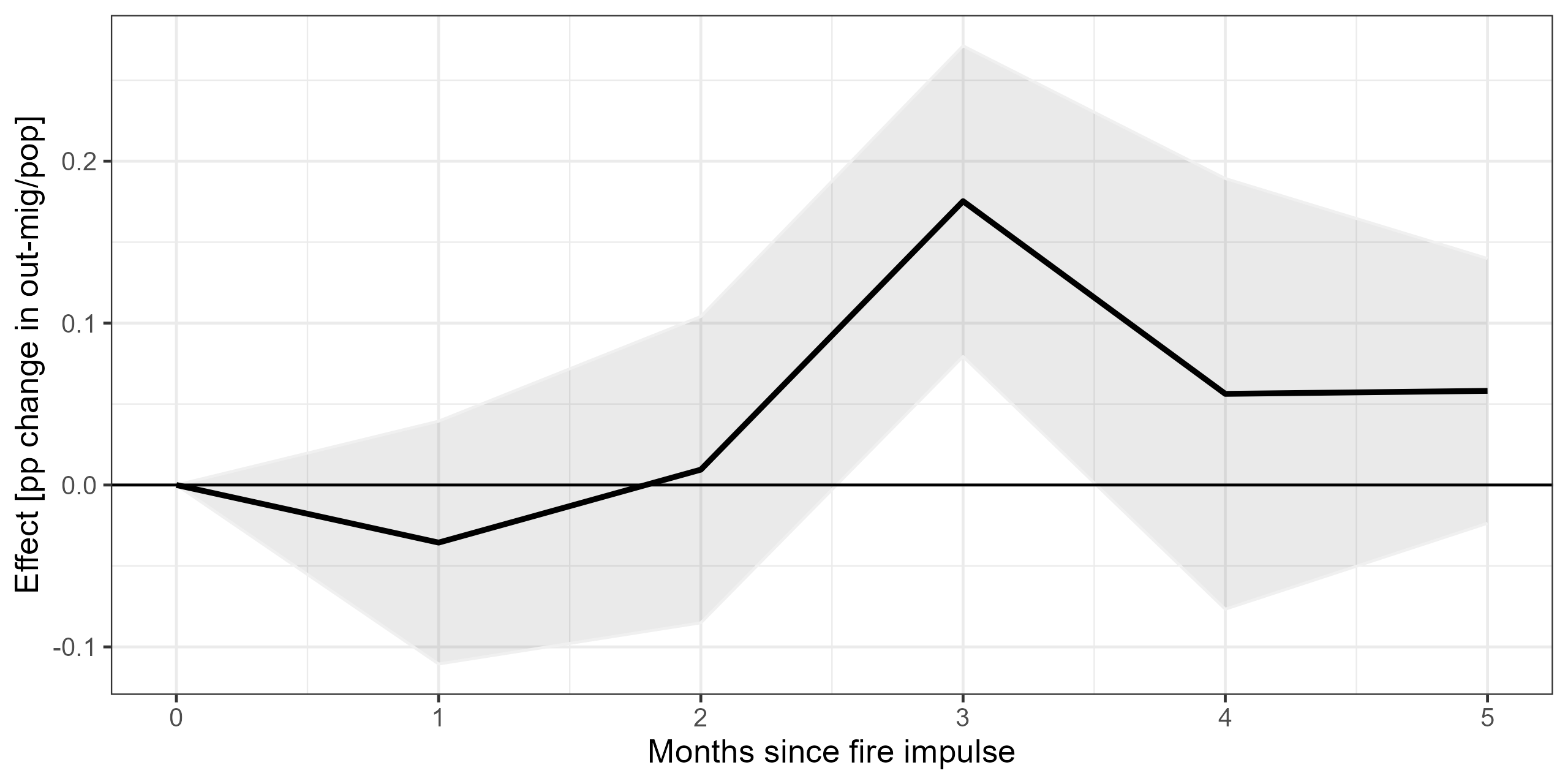}
\caption{Out-migration}
\label{fig_outmig}
\end{subfigure}
\end{center}
\raggedright
\begin{singlespace}
\footnotesize{The $y$ axis shows percentage changes in response to a burn area impulse of about 13 km$^2$---the mean burn area in counties that experienced fires. Shaded areas indicate 95\% CIs computed using DK standard errors.} \end{singlespace}
\end{figure}

\begin{figure}[!htbp]
\begin{center}
\caption{Response of employment to an increase in fire exposure split by share of rural population} \vspace{0.35cm} \label{fig:rural}
\includegraphics[width=\textwidth]{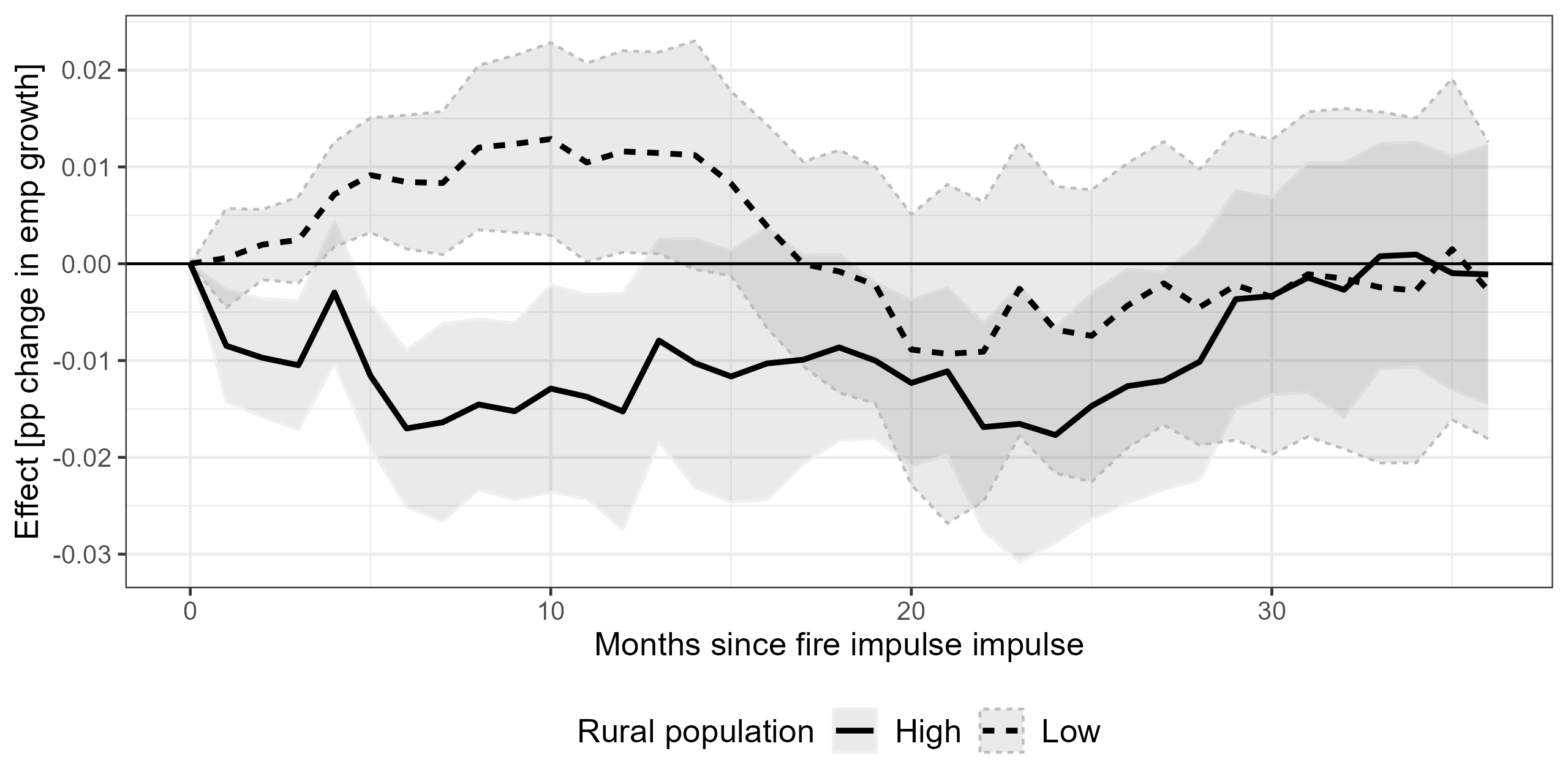}    
\end{center}
\raggedright
\begin{singlespace}
 \footnotesize{The $y$ axis shows percentage changes in response to a burn area impulse of about 13 km$^2$---the mean burn area in counties that experienced fires. Shaded areas indicate 95\% CIs computed DK standard errors. Covariates include county and year-month fixed effects and 24 monthly lags of county employment and burn area.} \end{singlespace}
\end{figure}

\subsection{Separating counties by other variables}

\begin{figure}[!htbp]
\begin{center}
\caption{Response of employment growth to an increase in fire exposure split by population size} \vspace{0.35cm} \label{fig:pop}
\includegraphics[width=\textwidth]{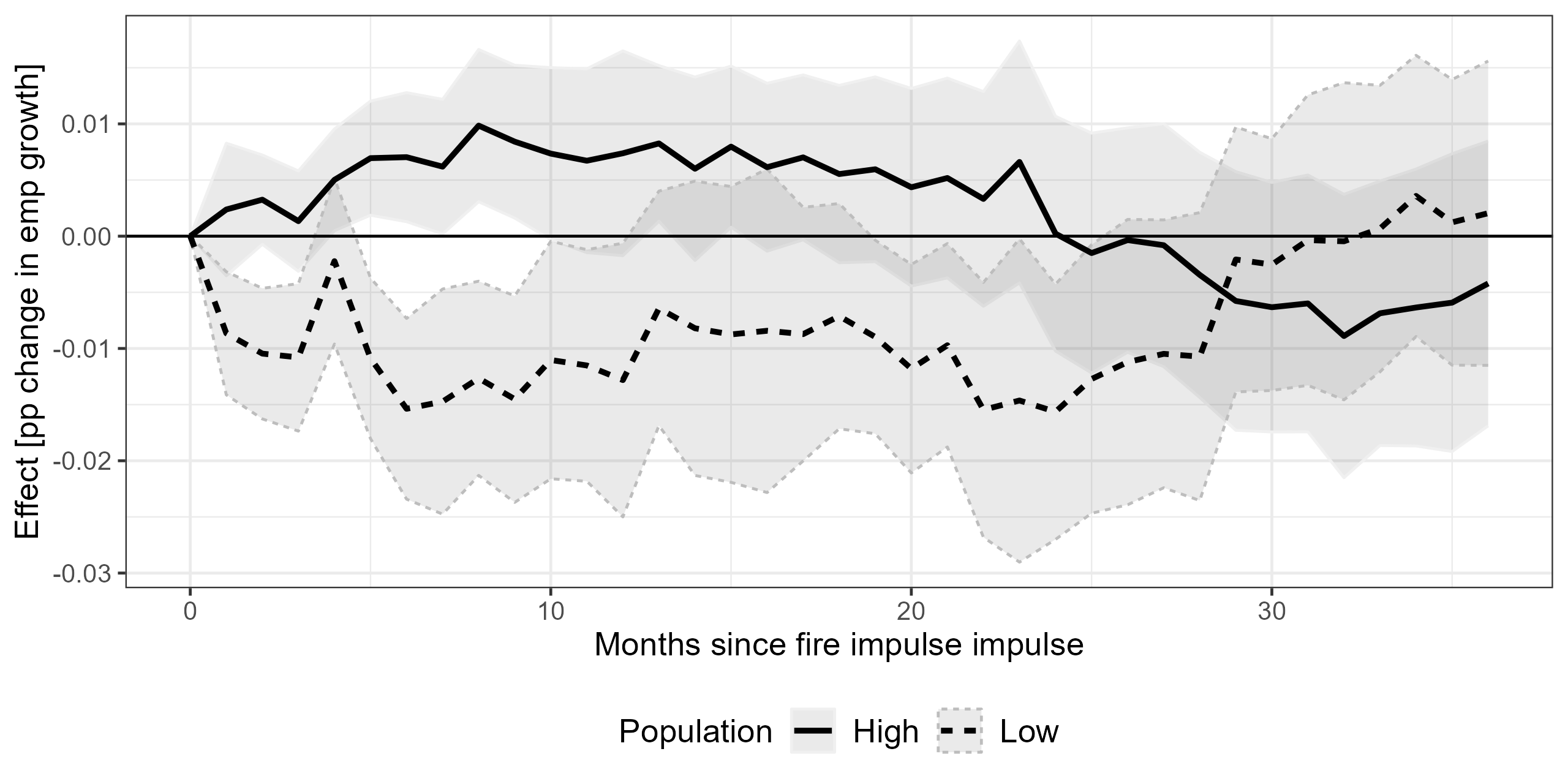}    
\end{center}
\raggedright
\begin{singlespace}
 \footnotesize{The $y$ axis shows percentage changes in response to a burn area impulse of about 13 km$^2$---the mean burn area in counties that experienced fires. Shaded areas indicate 95\% CIs computed DK standard errors. Covariates include county and year-month fixed effects and 24 monthly lags of county employment and burn area.} \end{singlespace}
\end{figure}

\begin{figure}[!htbp]
\begin{center}
\caption{Response of employment growth to an increase in fire exposure split by income} \vspace{0.35cm} \label{fig:income}
\includegraphics[width=\textwidth]{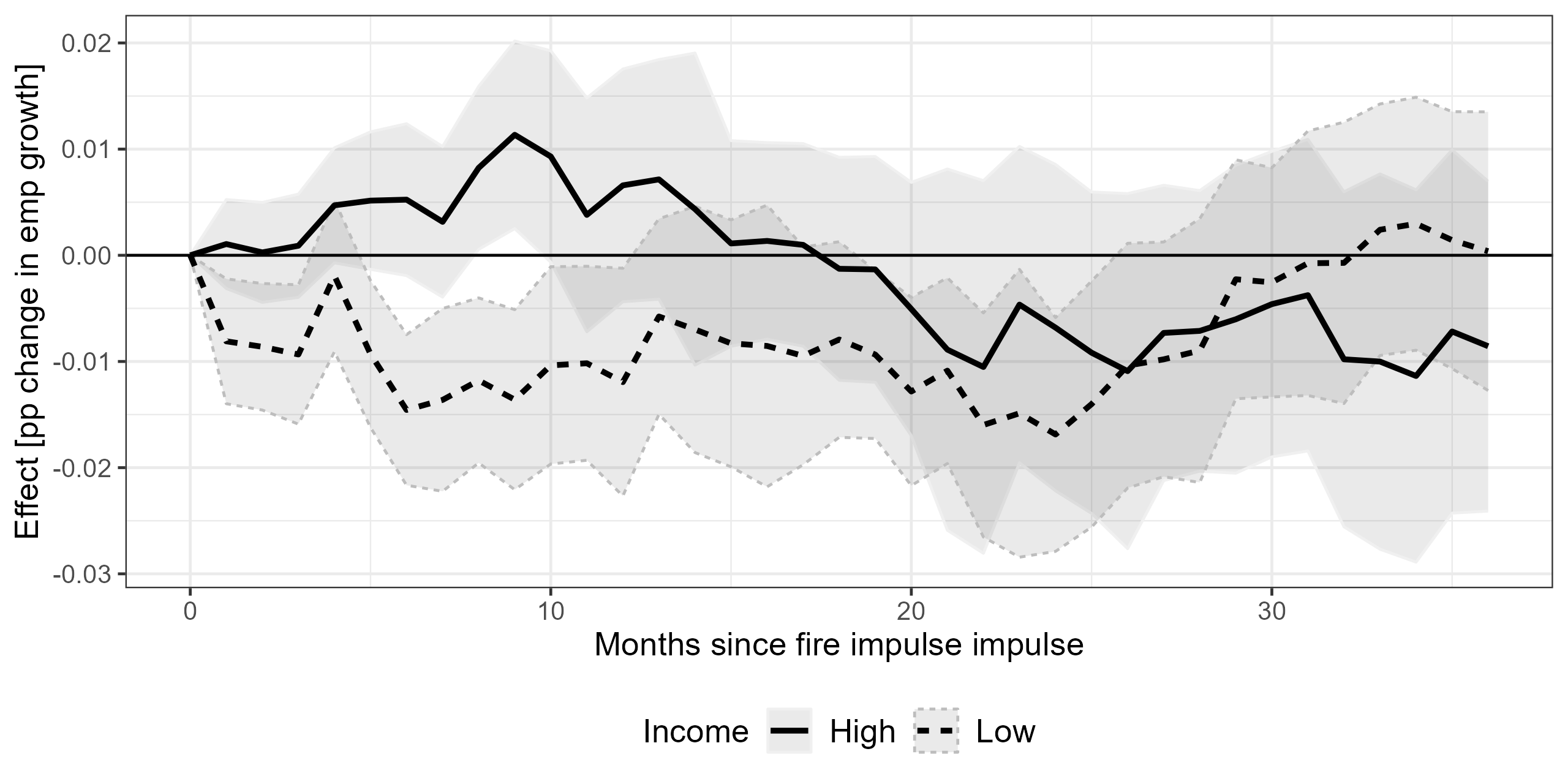}    
\end{center}
\raggedright
\begin{singlespace}
 \footnotesize{The $y$ axis shows percentage changes in response to a burn area impulse of about 13 km$^2$---the mean burn area in counties that experienced fires. Shaded areas indicate 95\% CIs computed DK standard errors. Covariates include county and year-month fixed effects and 24 monthly lags of county employment and burn area.} \end{singlespace}
\end{figure}

\end{appendices}

\end{document}